\documentclass[aps,prd,twocolumn,nobalancelastpage,superscriptaddress,showpacs]{revtex4}

\usepackage[english]{babel}
\usepackage[usenames]{color}
\usepackage{color,graphicx,epsfig,amsmath,amssymb}   

\usepackage{lscape}
\usepackage{natbib}
\usepackage{rotating}

\newcommand{\beq}{\begin{equation}}
\newcommand{\eeq}{\end{equation}}
\newcommand{\ben}{\begin{eqnarray}}
\newcommand{\een}{\end{eqnarray}}
\newcommand{\bi}{\begin{itemize}}
\newcommand{\ei}{\end{itemize}}

\newcommand{\alpham}{{\ifmmode \alpha_M \else $\alpha_M$\fi}}
\newcommand{\mchi}{{\ifmmode m_{\chi} \else $m_{\chi}$\fi}}
\newcommand{\rhosun}{{\ifmmode \rho_{\odot} \else $\rho_{\odot}$\fi}}
\newcommand{\sigvdm}{{\ifmmode \langle \sigma_{\rm ann} v \rangle \else 
    $\langle \sigma_{\rm ann} v \rangle$\fi}}

\newcommand{\Msun}{{\ifmmode M_{\odot} \else $M_{\odot}$\fi}}
\newcommand{\Rsun}{{\ifmmode R_{\odot} \else $R_{\odot}$\fi}}
\newcommand{\vecxsun}{{\ifmmode \vec{x}_{\odot} \else $\vec{x}_{\odot}$\fi}}
\newcommand{\xsun}{{\ifmmode x_{\odot} \else $x_{\odot}$\fi}}
\newcommand{\ysun}{{\ifmmode y_{\odot} \else $y_{\odot}$\fi}}
\newcommand{\zsun}{{\ifmmode z_{\odot} \else $z_{\odot}$\fi}}
\newcommand{\Mearth}{{\ifmmode M_{\oplus} \else $M_{\oplus}$\fi}}
\newcommand{\Rearth}{{\ifmmode R_{\oplus} \else $R_{\oplus}$\fi}}
\newcommand{\vecxearth}{{\ifmmode \vec{x}_{\oplus} \else $\vec{x}_{\oplus}$\fi}}
\newcommand{\xearth}{{\ifmmode x_{\oplus} \else $x_{\oplus}$\fi}}
\newcommand{\yearth}{{\ifmmode y_{\oplus} \else $y_{\oplus}$\fi}}
\newcommand{\zearth}{{\ifmmode z_{\oplus} \else $z_{\oplus}$\fi}}
\newcommand{\gev}{{\ifmmode {\rm GeV} \else ${\rm GeV}$\fi}}

\newcommand{\vconv}{{\ifmmode V_{\rm conv} \else $V_{\rm conv}$\fi}}
\newcommand{\pbar}{{\ifmmode \overline{p} \else $\overline{p}$\fi}}
\newcommand{\gtilde}{{\ifmmode \tilde{\cal G} \else $\tilde{{\cal G}}$\fi}}
\newcommand{\gtildepos}{{\ifmmode \gtilde^{\; e^+} \else 
    $\gtilde^{\; e^+}$\fi}}
\newcommand{\gtildepbar}{{\ifmmode \gtilde^{\; \pbar} \else 
    $\gtilde^{\; \pbar}$\fi}}

\newcommand{\lcdm}{{\ifmmode \Lambda{\rm CDM} \else $\Lambda{\rm CDM}$\fi}}

\begin{document}

\title{Antimatter cosmic rays from dark matter annihilation:\\
  First results from an N-body experiment}

\author{J.~Lavalle}
\affiliation{Dipartimento di Fisica Teorica,
  Universit\`a di Torino \& INFN,
  via Giuria 1,
  10125 Torino - Italia}
\email{lavalle@to.infn.it}
\author{E.~Nezri}
\affiliation{Laboratoire d'Astrophysique de Marseille, 
  CNRS \& Universit\'e Aix-Marseille I, 
  2 place le Verrier, 
  13248 Marseille Cedex 4 - France}
\email{Emmanuel.Nezri@oamp.fr, lia@oamp.fr}
\author{F.-S.~Ling}
\affiliation{Service de Physique Th\'eorique,
  Universit\'e Libre de Bruxelles,
  boulevard du Triomphe - CP225,
  1050 Bruxelles - Belgique}
\email{fling@ulb.ac.be}
\author{E.~Athanassoula} 
\affiliation{Laboratoire d'Astrophysique de Marseille, 
  CNRS \& Universit\'e Aix-Marseille I, 
  2 place le Verrier, 
  13248 Marseille Cedex 4 - France}
\email{name@oamp.fr}
\author{R.~Teyssier}
\affiliation{Service d'Astrophysique,
  Commissariat \`a l'\'Energie Atomique,
  Orme des Merisiers,
  91191 Gif sur Yvette - France}
\email{romain.teyssier@cea.fr}

\begin{abstract}
While the particle hypothesis for dark matter may be very soon investigated at 
the LHC, and as the PAMELA and GLAST satellites are currently taking new data 
on charged and gamma cosmic rays, the need of controlling the 
theoretical uncertainties affecting the possible indirect signatures of dark 
matter annihilation is of paramount importance. The uncertainties which 
originate from the dark matter distribution are difficult to estimate because 
current astrophysical observations provide rather weak dynamical constraints, 
and because, according to cosmological N-body simulations, dark matter is 
neither smoothly nor spherically distributed in galactic halos. Some 
previous studies made use of N-body simulations to compute the $\gamma$-ray 
flux from dark matter annihilation, but such a work has never been performed 
for the antimatter (positron and antiproton) primary fluxes, for which 
transport processes complicate the calculations. We take advantage of the 
galaxy-like 3D dark matter map extracted from the HORIZON Project results to 
calculate the positron and antiproton fluxes from dark matter annihilation, in 
a model-independent approach as well as for dark matter particle
benchmarks relevant at the LHC scale (from supersymmetric and 
extra-dimensional theories). We find that the flux uncertainties 
arise mainly from fluctuations of the local dark matter density, and 
are of $\sim$ 1 order of magnitude. We compare our results to analytic 
descriptions of the dark matter halo, showing how the latter can well 
reproduce the former. The overal antimatter predictions associated with our 
benchmark models are shown to lie far below the existing measurements, and 
in particular that of the positron fraction recently reported by PAMELA, and 
far below the background predictions as well. Finally, we stress the limits of 
the use of an N-body framework in this context.
\end{abstract}


\maketitle

\begin{flushleft}
Preprint: DFTT-21/2008
\end{flushleft}

\section{Introduction}
\label{sec:intro}

The idea that dark matter is made of exotic weakly interacting massive 
particles (WIMP) is very appealing in the sense that the related existing 
frameworks in particle physics beyond the standard model (BSM) could solve (i) 
problems inherent to elementary particle theory, e.g. the force unification 
scheme and/or the stabilization of the theory against quantum corrections, 
which would thus confer a more fundamental meaning to this theory; and in the 
sametime (ii) the dark matter issue as characterized in astrophysics and 
cosmology (see e.g.~\cite{review_dm_murayama_07} for a recent review). Indeed, 
the cold dark matter (CDM) paradigm seems to feature a powerful and 
self-consistent theory of structure formation, which at present is able to 
reproduce and explain most of large and mid scale observations (see 
e.g.~\cite{2007NuPhS.173....1P} and references therein).

While the Large Hadron Collider at CERN (LHC) is about to start hunting BSM 
particle physics and may soon provide new insights about the dark matter 
particle hypothesis, high energy astrophysics experiments are also of strong 
interest to try to determine the actual nature of dark matter. Indeed, if dark 
matter is made of self-annihilating particles, a property which is found in 
many BSM models and which provides a natural mechanism to explain the dark 
matter cosmological abundance as observed today, there should be traces of 
annihilations imprinting the cosmic ray spectra (indirect detection of dark 
matter, see e.g.~\cite{susy_dm_jungman_etal_96,review_dm_bergstrom_00,
  review_dm_bertone_etal_05,review_dm_carr_etal_06}). Therefore, satellite 
experiments such as PAMELA and GLAST, which are dedicated to large field of 
view observations of charged cosmic rays and $\gamma$-rays, respectively, in 
the MeV-TeV energy range, should be able to yield additional constraints or 
smoking guns very soon in this research field~\cite{2006JCAP...12..003M}. The 
most promising astrophysical messengers that could trace the annihilation 
processes are indeed gamma-rays and antimatter cosmic rays, which have been 
first investigated in~\cite{1978ApJ...223.1015G} and 
in~\cite{1984PhRvL..53..624S} respectively. Confinement and diffusion on 
magnetic turbulences limit the origin of the latter to our Galaxy, whereas the 
former, which only experiences the common $r^{-2}$ flux dilution, may be 
observed even when emitted from extra-galactic regions.

Nevertheless, predictions of these exotic signals are affected by many 
uncertainties coming from (i) the underlying WIMP model (ii) the distribution 
of dark matter in the relevant sources (iii) the propagation of charged 
cosmic rays in the Galaxy in the case of searches in the antimatter 
cosmic ray spectra. The first point is encoded in the WIMP mass, the 
annihilation cross-section and the annihilation final states which fully 
define the injected cosmic ray spectra: this has been widely investigated for 
years. The second point is currently surveyed by the state-of-the-art N-body 
experiments, but the use of N-body dark matter maps in the context of indirect 
detection has only been performed for $\gamma$-ray 
predictions~\cite{2003MNRAS.345.1313S,2007ApJ...657..262D,2008arXiv0801.4673A,
  2008arXiv0805.4416K}. Besides, there has been lots of efforts to 
estimate the effect of varying the dark matter distribution by means of
analytical calculations, allowing for instance to study the effects of 
sub-halos on the $\gamma$-ray~\cite{1999PhRvD..59d3506B,
  2002PhRvD..66l3502U,berezinsky_etal_03,berezinsky_etal_06,
  2008MNRAS.384.1627P} as well as on the antimatter primary
fluxes~\cite{2007A&A...462..827L,2008A&A...479..427L}. Such 
analytical studies mostly rely on the statistical information supplied by 
N-body simulations. However, though they provide very nice theoretical bases 
to unveil the salient effects of any change in the relevant parameter space, 
and permit to go beyond the current numerical resolution limits, they are 
usually bound to simplifying hypotheses, such as the sphericity of sources. 
Especially, no study of the antimatter signatures has been directly performed 
in the frame of N-body environments up to now. Finally, point (iii), referring 
to cosmic ray propagation, has also widely been investigated for 
antiprotons~\cite{2004PhRvD..69f3501D,2005JCAP...09..010L} and 
positrons~\cite{2008A&A...479..427L,2008PhRvD..77f3527D}, and 
while the transport processes depend on sets of still degenerate parameters 
within different propagation models, the origins of uncertainties are now 
rather well understood.

In this paper, we make use of a realistic (in the sense of \lcdm\ cosmology) 
3D dark matter distribution of a galactic-sized halo extracted from the 
HORIZON Project simulation results to study the antimatter signals. Our 
main purposes are to characterize and quantify the uncertainties coming from 
dark matter inhomogeneities (not necessarily clumps) and departure from 
spherical symmetry in a virialized Milky-Way-like object; and finally 
make predictions for different dark matter particle candidates. Despite 
the rather low resolution of our numerical data compared with the highest-level 
artillery on the market, as portrayed, e.g., by the Via Lactea 
simulations~\cite{2007ApJ...657..262D,2008arXiv0805.1244D}, we study 
for the first time the production of antimatter cosmic rays from dark matter 
annihilation in an N-body framework. The resolution issue is however less 
important for charged cosmic rays than for gamma-rays, because the signal is 
strongly diluted by diffusion effects, which therefore tends to smooth the 
yields from high density fluctuations, provided the latter are not too close to 
the observer. Nevertheless, as very local density fluctuations are expected to 
affect the antimatter signals, we will quantify and correct the consequence 
of loss of resolution by analytically extrapolating our results. For the WIMP 
candidates, we will use a model-independent approach as well as typical models 
of BSM particle theories, some of them being observable at the LHC.

This article is parted as follows: we recall the salient features of the 
HORIZON N-body experiment in Sect.~\ref{sec:horizon}; then, in 
Sect.~\ref{sec:crs}, we briefly review the positron and antiproton 
propagation before sketching the method we adopt to connect the propagation 
to the N-body source terms; we describe the WIMP models in 
Sect.~\ref{sec:wimp}; we finally present and discuss our results and 
predictions of the primary positron and antiproton fluxes in 
Sect.~\ref{sec:res} (where the expert reader is invited to go directly) 
before concluding.

\section{The HORIZON framework}
\label{sec:horizon}

The framework of this study is a simulation of the Horizon collaboration. The 
run used the Adaptive Mesh Refinement code  RAMSES~\cite{2002A&A...385..337T} 
with an effective number of particles of $N_p=1024^3$ in a box of size 
$L=20h^{-1}$ Mpc where the initial conditions were fixed by the WMAP3 results 
($\Omega_m=0.24$, $\Omega_\Lambda=0.76$, $\Omega_b=0.042$, $n=0.958$, 
$H_0=73$, $\sigma_8=0.77$).

A Milky-Way sized halo at $z=0$ was selected and refined using the so-called
``zoom'' technique. We obtained a maximum linear resolution of about 
$200$ pc and the particle mass is $M_p=7.46\ 10^5$ in solar mass ($\Msun$) 
units (we refer the reader to ref.~\cite{2008arXiv0801.4673A} for more details 
on the characteristics and the analysis of the simulation.).

A non biased calculation of the dark matter density around a simulation point 
$i$ is given by the algorithm of Casertano and Hut~\cite{1985ApJ...298...80C}, 
namely:
\begin{equation}
\rho^i_j=\frac{j-1}{V(r_j)}M_p\;;
\end{equation}
where $V(r_j)=4\pi/3 r_j^3$ is the volume of the smallest sphere around the 
particle $i$ that includes $j$ neighbors.

\begin{figure*}[t]
\begin{center}
\includegraphics[width=0.65\columnwidth, clip]{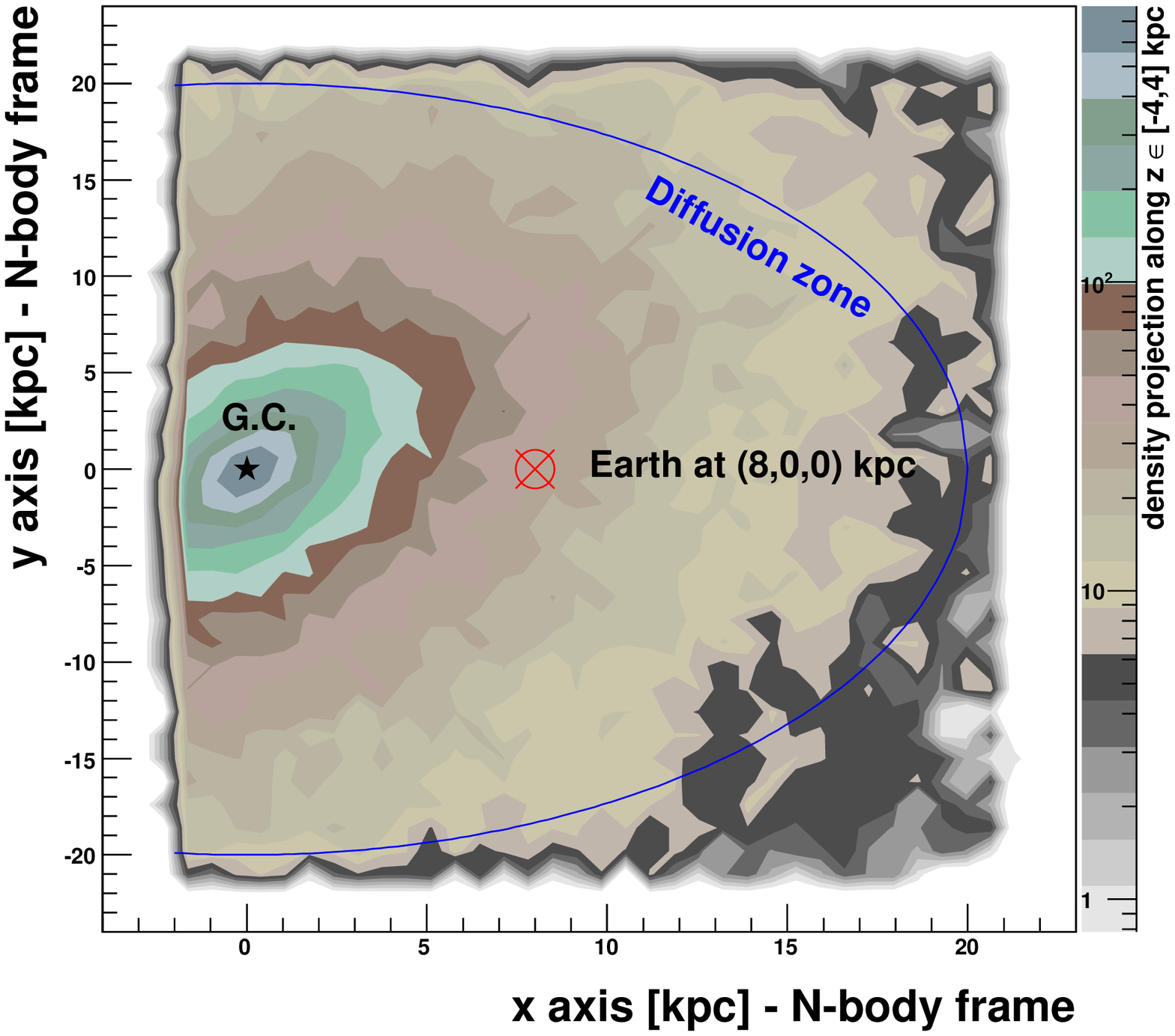}
\includegraphics[width=0.65\columnwidth, clip]{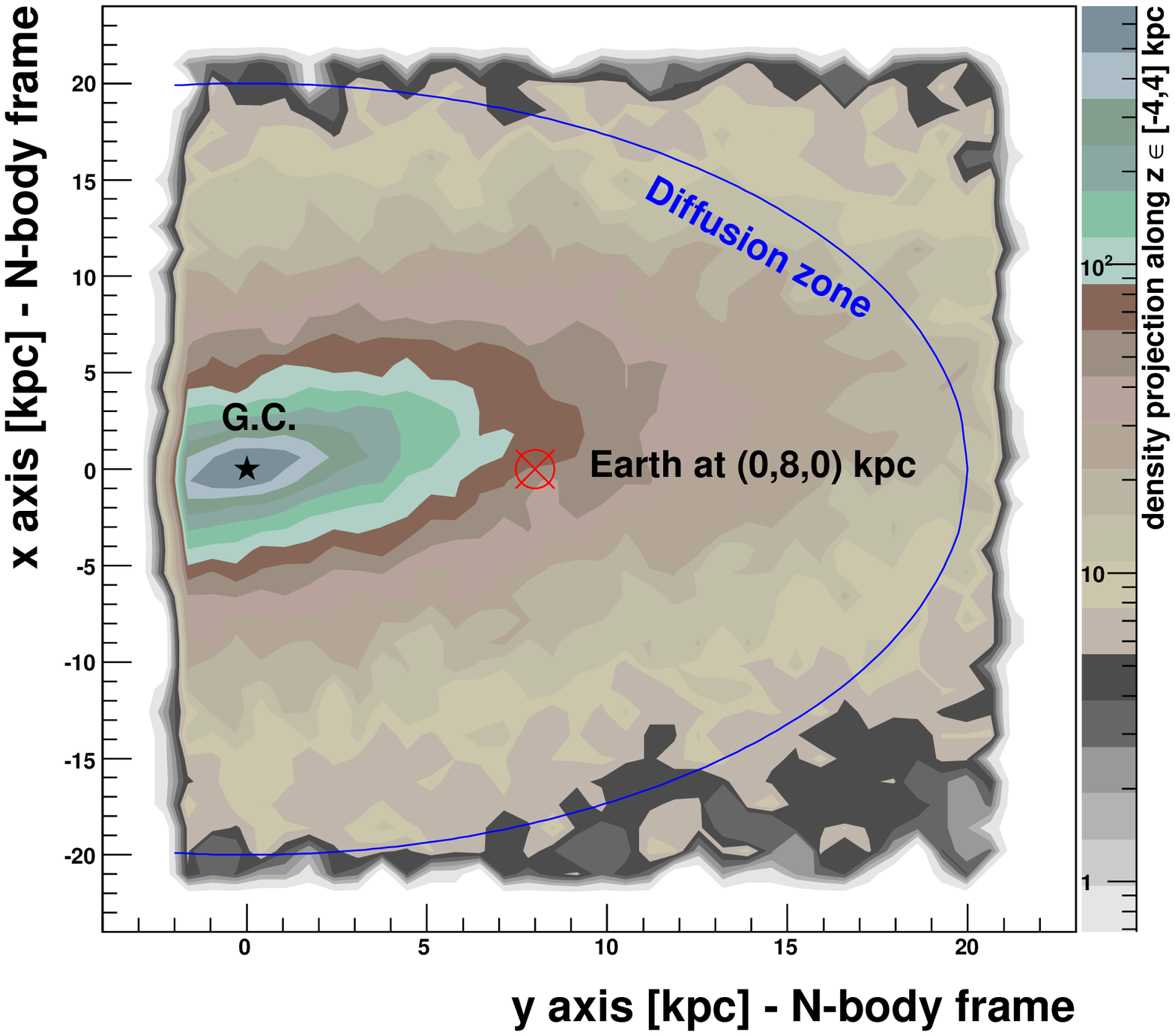}
\includegraphics[width=0.65\columnwidth, clip]{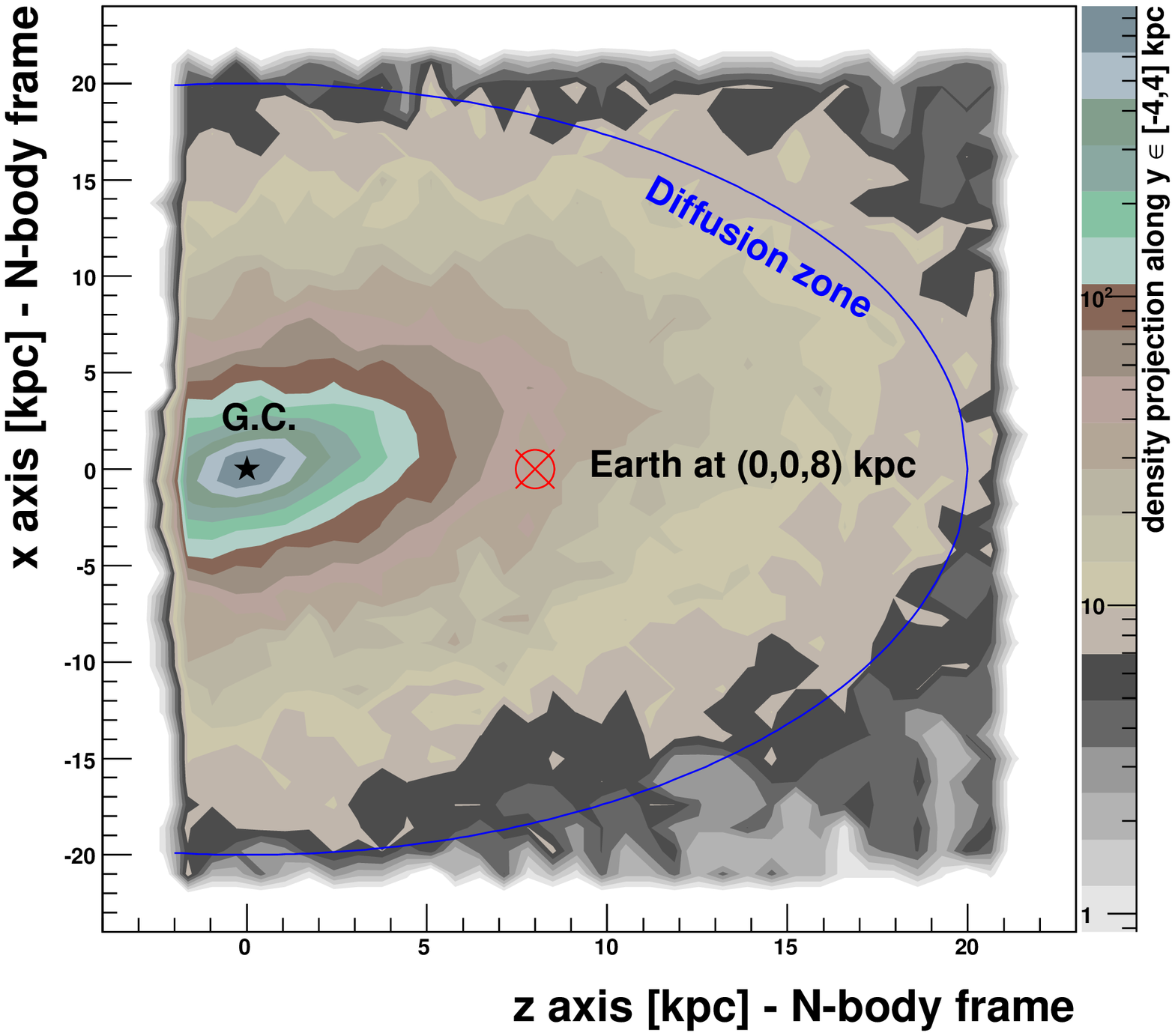}
\caption{\small Projected dark matter density in the cosmic ray diffusion 
plane, for different choices of the Earth location (Cartesian coordinates). 
Left: Earth at $x=8$ kpc $(\rhosun = 0.23 \;{\rm GeV.cm^{-3}})$, diffusion 
plane in $Oxy$; middle: Earth at $y=8$ kpc $(\rhosun = 0.28\;
{\rm GeV.cm^{-3}})$, diffusion plane in $Oxy$; right: Earth at $z=8$ 
kpc $(\rhosun = 0.13\;{\rm GeV.cm^{-3}})$, diffusion zone in $Oxz$. The 
diffusion slab contour is explicitly drawn (visually ellipsoidal instead of 
spherical due to different horizontal and vertical scales).}
\label{fig:dm_density_map}
\end{center}
\end{figure*}

The virial radius (defined as $\rho(r_{vir})=200\rho_c$) of our halo is equal 
to 253 kpc, corresponding to an enclosed mass of  $6.05 \times 10^{11} {\rm
  \Msun}$ or $8.1\times 10^5$ particles. Within this radius, the dark matter 
halo density can be fitted by the usual spherical parameterization:
\begin{equation}
\label{eq:profile}
  \rho(r) = \rho_{0} ~\left(\frac{r}{r_0}\right)^{-\gamma} \left[
  \frac{1+\left(r_0/a\right)^{\alpha}}
  {1+\left(r/a\right)^{\alpha}}\right]^{\frac{\beta-\gamma}{\alpha}},
 \end{equation}
where  $\rho_0=\rhosun$ is the local density in the solar neighborhood and
$r_0=\Rsun=8$ kpc is the distance from the Sun to the Galactic center (GC). 
Density maps, after projections in the $(x,y)$ and $(z,x)$ planes, are
shown on Fig.~\ref{fig:dm_density_map}, where the Earth has been located at 
three different positions: at $x = 8$, $y = 8$ and $z = 8$ kpc, respectively 
(see Sect.~\ref{subsec:dm_source} for 
further details). When considering the radial distribution of our data, the 
best fit gives $(\alpha,\beta,\gamma)=(0.39,3.72,0.254)$,
and $a=13.16$ kpc, but a NFW-like profile $(1,3,1)-(a=10\;{\rm kpc})$ 
and a cored profile $(0.5,3.3,0)-(a=4.5\;{\rm kpc})$ are also in good 
statistical agreement due to the resolution limits, especially in the 
central region. The averaged density at a radius of 8 kpc is 
$\rhosun = 0.25\pm 0.19$ GeV/cm$^3$. 108 sub-halos have been identified with 
masses between $2 \times 10^7 \Msun$ and $2.4 \times 10^{10} \Msun$ with 
a mass distribution following $N(M_{sub})\propto M_{sub}^{-1}$ above 
$5 \times 10^8 \Msun$. Inside those substructures, the dark matter density 
scales like a universal power law $\rho_{cl}(r)\propto r^{-2.5}$ for the outer 
part. Regarding the inner part, the logarithmic slope is highly speculative 
due to the resolution limit. Furthermore, all clumps are found to be rather 
far from the Galactic center and will actually not modify our coming results. 
Anyway, though we will further mimic higher resolutions 
thanks to analytical extrapolations, this N-body environment is still relevant 
for the punch line of this study, which is to quantify the importance of 
solar neighborhood dark matter density fluctuations.

\section{Antimatter cosmic rays}
\label{sec:crs}

Antimatter cosmic rays are very interesting messengers to look for 
dark matter annihilation traces, because they are seldom produced in standard 
astrophysical processes. The astrophysical secondary background at the Earth 
is generically created out of the spallation of cosmic rays off the 
interstellar medium (ISM), and is expected to be highly power-law suppressed 
at energies above $\sim 10$ GeV.
\begin{widetext}
For charged Galactic cosmic rays in the GeV-TeV regime, the relevant (steady 
state) propagation equation which characterizes the transport, for any 
species, of the cosmic ray number density per unit of energy 
${\cal N}_{\rm cr}\equiv dn_{\rm cr}(E)/dE$ reads:
\ben
  \vec{\nabla} \left[ K(E)\vec{\nabla} {\cal N}_{\rm cr} -
    \vec{V}_{\rm conv} {\cal N}_{\rm cr} \right]
  +\frac{\partial}{\partial E}\left[b(E){\cal N}_{\rm cr} + 
    K_{EE} \frac{\partial}{\partial E}  {\cal N}_{\rm cr} \right]
  + \Gamma(E) {\cal N}_{\rm cr} + {\cal Q}
   =  0\;; 
  \label{eq:propag_crs}
\een
where $K(E)$ encodes the spatial diffusion due to scattering off magnetic 
turbulences, \vconv\ is the velocity of the convective wind that drifts cosmic 
rays out of the 
Galactic disk, $b(E)$ stands for energy losses, $K_{EE}$ characterizes 
diffusion in momentum space, $\Gamma(E)$ features the spallation reactions 
and ${\cal Q}$ is the (stationary) cosmic ray source of interest.
\end{widetext}

Though the above equation stands for any cosmic ray species, some of the 
processes can be safely neglected in the GeV regime, depending on the species. 
For positrons, the main processes are spatial diffusion and energy losses, 
so that we will neglect convection and reacceleration. Such an approximation 
is very often used in the literature, and its relevance is explained in great 
details in ~\cite{2008arXiv0809.5268D}. For antiprotons, the dominant processes 
beside diffusion are convection and spallation, which are more efficient at 
low energies, typically $\lesssim 10\; {\rm GeV}$, whereas reacceleration 
and energy losses do not play a major 
role~\cite{2001ApJ...563..172D,2004PhRvD..69f3501D}. We will therefore 
neglect those latter effects for antiprotons in the following.

The diffusion equation can be solved either fully numerically (see 
e.g.~\cite{1998ApJ...509..212S}) or semi-analytically in typical energy 
regimes and/or geometries of the diffusion zone 
(see e.g.~\cite{berezinsky_book_90}). We will use the latter 
approach in this paper. Generic solutions for point-like sources 
$\propto \delta^4(E_S,\vec{x}_S)$ (4-D Dirac function) can be expressed in terms of Green 
functions ${\cal G}(E,\vec{x}_{\odot}\leftarrow E_S,\vec{x}_S)$ that 
characterize the probability for cosmic rays of energy $E_S$ injected 
at position $\vec{x}_S)$ to be detected at the Earth with energy $E$. Hence, 
for a WIMP particle of mass $\mchi$ and of thermally averaged annihilation 
cross-section $\sigvdm$, the primary exotic flux at the Earth is merely given 
by:
\ben
\label{eq:flux}
\frac{d\phi_{\rm cr}(E)}{dE} &\equiv& 
\frac{v_{\rm cr}}{4\pi}\times {\cal S}\nonumber\\
& & \times \int_{\rm slab}d^3\vec{x}_S 
     \gtilde(E;\vec{x}_{\odot}\leftarrow \vec{x}_S) \times 
            {\cal Q}(\vec{x}_S)
     \;,
\een
where $v_{\rm cr}$ is the cosmic ray velocity, ${\cal S}\equiv  (\sigvdm /2)
\times (\rhosun/\mchi)^2$ allows a convenient normalization to the local 
WIMP-annihilation rate, ${\cal Q}$ is the spatial source term that we will 
further discuss in Sect.~\ref{subsec:dm_source}, and 
$\gtilde\equiv \int dE_S {\cal G} dN/dE_S$ is the Green function convoluted 
with the injected cosmic ray spectrum $dN/dE_S$.

In the following, we present the main features of our propagation model for 
both antiprotons and positrons. We will end this section by discussing the 
source term ${\cal Q}$ as set in our N-body framework.

\subsection{Propagation model}
\label{subsec:propag}
The propagation model that we adopt in this paper is characterized by 
a slab geometry with the origin located at the Galactic center, a radial 
extent $R_{\rm slab}$ and a vertical half-height $L$. This defines the cosmic 
ray confinement zone (see Fig.~\ref{fig:dm_density_map}), thanks to which we 
impose Dirichlet boundary 
conditions to Eq.(\ref{eq:propag_crs}), that is 
${\cal N}(R_{\rm slab},z) = {\cal N}(r,L) = 0$. The Earth is located on a 
circle in the Galactic disc with cylindrical coordinates $(r,z)=(8,0)$ kpc 
(the polar angle is left free for the moment). As regards the transport 
processes, we take a spatial independent diffusion coefficient of the form 
$K(E)=\beta K_0{\cal R}^{\delta}$ (where $\beta$ and ${\cal R}=pc/Ze$ are the 
particle velocity and rigidity respectively) and 
a constant wind \vconv\ directed outwards the Galactic plane along the 
vertical $z$ axis. The reader is referred to 
\cite{maurin_etal_01} for a detailed presentation of this framework. The 
propagation parameters can be constrained with cosmic ray measurements of 
secondary to primary ratios, and are found to be 
degenerate~\cite{maurin_etal_01,2002A&A...394.1039M,2002A&A...381..539D}. 
Throughout this paper, we will use the \emph{median} set of propagation 
parameters defined in~\cite{2004PhRvD..69f3501D}, which has been constrained 
with the current B/C data and which is recalled in Tab.~\ref{tab:prop}. Once 
the fluxes will be computed, they will be modulated to account for the solar 
modulation. We will use the force field approximation method, with an electric 
potential $\phi = 600$ MV~\cite{1987A&A...184..119P}. 

\begin{table}[t!]
\begin{center}
{\begin{tabular}{c c c c c c}
\hline
\hline
&  $\delta$  & $K_0$ & $L$ & $R_{\rm slab}$ & $\vconv$  \\
&    & (kpc$^{2}$.Myr$^{-1}$) & (kpc) & (kpc)& (km.s$^{-1}$)  \\
\hline
{\rm med} &  0.70  & 0.0112 & 4  & 20 & 12.0   \\
\hline
\end{tabular}}
\caption{
Propagation parameters compatible with B/C data giving the median antiproton 
primary (from dark matter annihilation) fluxes ; 
from~\cite{2004PhRvD..69f3501D}.
\label{tab:prop}}
\end{center}
\end{table}

\subsubsection{Antiproton and positron propagation}
\label{subsubsec:pbars}
The exotic antiproton production and propagation have widely been studied in 
the literature~\cite{2002astro.ph.12111M,2004PhRvD..69f3501D,
  2002A&A...388..676B,2005PhRvD..72f3507B,2007PhRvD..75h3006B}, and this 
also holds for positrons~\cite{1974Ap&SS..29..305B,1999PhRvD..59b3511B,
  2007A&A...462..827L,brun_etal_07,2008A&A...479..427L}. In this paper, we 
will use the Green functions as described in~\cite{2008A&A...479..427L} for 
both the propagations of positrons and antiprotons. Both species experience
the same diffusion processes, but have quite different propagation behaviors: 
positron propagation is mainly affected by Inverse Compton (IC) energy losses, 
while antiprotons, though their energy losses are negligible, are swept 
out of the galactic disc (destroyed) by the convective wind (spallations off 
the interstellar gas) at low energy. Therefore, their respective horizon is 
completely reversed one from the other in terms of the energy dependence. 

The typical propagation length for positrons is given 
by~\cite{2007A&A...462..827L}:
\beq
\lambda_D = \left\{ \frac{4 K_0\tau_E}{1-\delta} \left( \epsilon^{\delta-1}-
  \epsilon^{\delta-1}_S\right) \right\}^{1/2}\;,
\eeq
where $\tau_E\simeq 10^{16}{\rm s}$ is the characteristic timescale of the 
energy loss in the IC regime, $\epsilon\equiv (E/1 {\rm GeV})$ stands for the 
energy at which the positron is detected at the Earth, while 
$\epsilon_S\geq \epsilon$ is the energy at which the positron is injected at 
the source. Given $\epsilon_S$, this is clearly a decreasing function of the 
energy, which is hardly greater than a few kpc.

For antiproton, convection is the relevant quantity that bounds the 
propagation length, and the horizon is set by~\cite{2003A&A...402..971T}:
\beq
\Lambda_D = \frac{K(E)}{\vconv}\;,
\eeq
which is, contrarily to the positron case, an increasing function of the 
energy, and which can quickly get values much greater than the typical size 
of the diffusion zone.

\begin{figure}[t]
\begin{center}
\includegraphics[width=\columnwidth, clip]{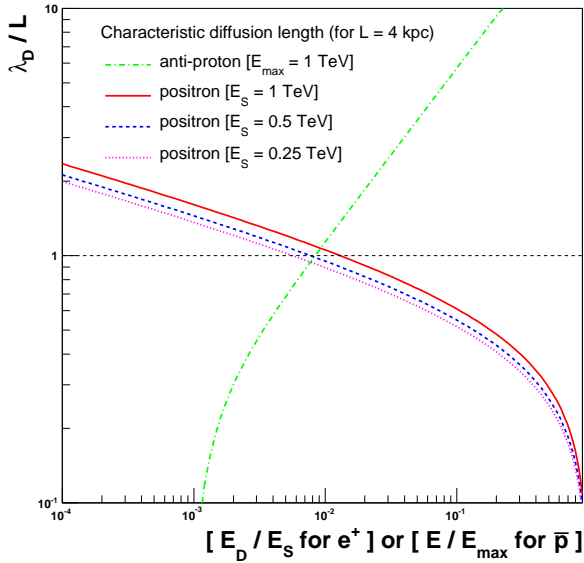}
\caption{\small Propagation scales for positrons and antiprotons as 
functions of energy. For the former, the energy reported on the $x$ axis is 
$E_d/E_S$, that is the detected energy divided by the injected energy 
(positrons loose energy), and for the latter, this is merely $E/E_{\rm max}$. 
Scales are normalized to $L = 4$ kpc, the vertical half-height of the 
diffusion zone.}
\label{fig:prop_lengths}
\end{center}
\end{figure}

The characteristic lengths for both antiprotons and positrons are reported 
in Fig.~\ref{fig:prop_lengths}, as functions of the detected energy at the 
Earth. For positrons, we have considered different injected energies $E_S$ of 
250, 500 and 1000 GeV, and $\lambda_D$ is plotted as a function of the fraction 
of $E_s$ still carried by the particle at the Earth. For antiprotons, we have 
plotted $\Lambda_D$ as a function of the detected energy divided by 1 TeV.

These quantities are very useful to figure out the volume from the Earth we 
are sensitive to when computing the exotic antimatter fluxes. When the positron 
(antiproton, respectively) energy is high (low), contributions to the 
flux are mostly local, and large fluctuations are therefore expected if the 
Earth position is varied inside an N-body environment because the dark matter 
density varies. This is the contrary at low (high) energies, for which 
contributions are integrated inside much larger volumes, and tend to smooth 
the density gradients.

This will be discussed in more details in Sect.~\ref{subsubsec:ellipt} 
and~\ref{subsubsec:inhom}, but the 
reader can already look at Fig.~\ref{fig:comp_xyz}, where it is shown that 
different local distributions of dark matter strongly affect the positron 
and antiproton signals, at high and low energy respectively.

\subsection{Connection to the N-body source term}
\label{subsec:dm_source}
N-body simulation results consist in a set of points with coordinates and 
masses. Though all points correspond to particles of the same mass, the 
algorithm of density computation defined in Sect.~\ref{sec:horizon} allows to 
associate a dark matter density to each point, which actually reflects the 
particle number density in the point neighborhood. Therefore, each point will 
be considered as an individual source term of Eq.~(\ref{eq:propag_crs}). 
In order to compute the antimatter flux at the Earth, we have to fix the 
locations of the Earth and of the diffusion zone with respect to the N-body 
system of coordinates. Due to the cosmological origin our simulation, we have 
no freedom in rescaling the distances, so that the Earth is set at 
$\Rearth=\Rsun=8$ kpc, while the Euler angles are random parameters. 
Once the Earth position is set, we also have to randomly fix the slab 
diffusion zone, where obviously, the Earth itself is located.

As already detailed in~\cite{2008A&A...479..427L}, 
constraints on the local dark matter density are very important for indirect 
detection of dark matter with antimatter cosmic rays, because 
local contributions dominate the signal in energy ranges depending on the 
species (high/low energy for positron/antiproton). Therefore, it is meaningful 
to add a prescription on top of the random procedure described above, so as to 
account for the observational constraints on the local dark matter density. 
According to \cite{2008A&A...479..427L} (and references therein), it 
is reasonable to bracket $\rhosun$ within 
$[0.2,0.4]\; {\rm GeV.cm^{-3}}$. Notice that such a constraint is by itself a 
very strong limit on the possible variations of the expected fluxes. This 
will be discussed in Sect.~\ref{sec:res}.

For visual purpose, we show in Fig.~\ref{fig:dm_density_map} 
projections of the dark matter density $\rho$ in the diffusion slab 
according to three different positions for the Earth, at 8 kpc along each axis 
$x$, $y$ and $z$, respectively (the associated local densities are 
$\rhosun = 0.23,\,0.28,\,0.13\;{\rm GeV/cm^{3}}$). These can be seen as 
different maps of the source term of the diffusion 
equation~(\ref{eq:propag_crs}) (the actual source term is in fact more 
contrasted because scaling like $\rho^2$). Note that those three different 
configurations are typified by very different local environments, with sizable 
density fluctuations, and one can expect some 
variations on the predicted antimatter cosmic ray fluxes 
(to go straight to the results, see Fig.~\ref{fig:comp_xyz}).

\section{WIMP models}
\label{sec:wimp}
Despite its numerous successes, the high energy physics standard model of
elementary particles needs some extensions. Indeed, some questions like the 
hierarchy problem, the force unification or the neutrino masses can
not be answered without advocating new physics. Furthermore in a large variety of
standard model extensions, new particles should be present and some of
them could be WIMPs and candidates to dark matter. Some of those scenarios
have deep theoretical motivations. Supersymmetric models like the
Minimal Supersymmetric Standard Model (MSSM) are inspired
from supergravity or string theory and predict new particles below
the TeV(s) scale. Those theoretical frameworks also require the
existence of extra spatial dimensions, and have inspired the building of 
phenomenological models in which new particles related to those new 
dimensions can arise at energies as low as the TeV scale. Some more effective 
models propose to extend the standard model in more minimal ways, in order to 
address some specific questions and to consider new phenomenologies
without hypotheses on the tricky and unsolved question of the link
with some possible fundamental aspects or inspiring sector. After a model 
independent setting, we will describe some typical
benchmark situations of the MSSM, Kaluza-Klein dark matter, Little Higgs 
Model and Inert Doublet Model, of which we will further survey the 
signatures in the antimatter cosmic ray spectra.

Note that all cosmic ray spectra for the WIMP models discussed in this paper 
have been generated with the PYTHIA Monte Carlo~\cite{2006JHEP...05..026S}, 
and some of the associated relic densities have been computed with the 
micrOMEGAs code~\cite{2006CoPhC.174..577B}.

\subsection{Model-independent setting}
\label{subsec:ind_wimp}
To allow discussion in a very general context, we define here 
two WIMP fiducial models associated with positron and antiproton 
production, respectively. For both models, we will take a WIMP mass of 
200 GeV and a total annihilation cross section of 
$\sigvdm = 3\times 10^{-26}\;{\rm cm^3/s}$.

The first one, associated with positrons, is either a bosonic or a Dirac 
fermionic WIMP and is supposed to provide monochromatic electrons and 
positrons by fully annihilating like $\chi\bar{\chi}\rightarrow e^+e^-$. This 
might, though very roughly, mimic Kaluza-Klein dark matter 
(see Sect.~\ref{subsubsec:kk}). Such a model greatly simplifies the 
calculation for the positron flux since the source term is 
$\propto \delta(E-\mchi)$, and is very convenient to discuss the main 
features of our results.

The second one, associated with antiprotons, is a Majorana fermionic WIMP 
that fully annihilates in $b\bar{b}$. Such a model is generically 
found in SUSY theories (see Sect.~\ref{subsubsec:mssm}).

\subsection{Models observable at the LHC}
\label{subsec:lhc_wimp}

We will show the resulting cosmic ray fluxes from WIMP annihilation in some
popular BSM frameworks relevant at the LHC scale that we describe in the 
next paragraphs.

\subsubsection{The Minimal Supersymmetric Standard Model}
\label{subsubsec:mssm}
The most popular new symmetry of high energy physics is the
supersymmetry (SUSY) \cite{Fayet:1976cr}. The simplest extension in this way is 
known as the Minimal Supersymmetric Standard Model (MSSM) \cite{Haber:1984rc}. 
The gauge group is still $SU(3)\times SU(2)\times U(1)$ and the supersymmetric
transformations associate a scalar (fermion) to each fermion (boson) of the 
standard model. This model unifies the coupling constants of the three forces 
at high energy and nicely achieves the electroweak symmetry breaking through
radiative corrections. It requires two Higgs doublets and we note $\mu$ the 
mass parameter of those (super)fields and $\tan{\beta}$ the ratio of their 
vacuum expected value (vev). This model has numerous (106) parameters but is 
usually simplified assuming at high energy unified values for scalar ($m_0$) 
and fermions ($m_{1/2}$) masses and for trilinear couplings ($A_0$). Thanks to 
the requirement of radiative electroweak symmetry breaking, the last parameter 
is the sign of $\mu$. The set
$ \lbrace m_0,m_{1/2}, \tan{\beta}, sign(\mu) \rbrace $ defines a so called
Constrained Minimal Supersymmetric Standard Model 
\cite{1983PhLB..121..123E,1994PhRvD..49.6173K} (CMSSM or 
mSUGRA) point. In the main part of this parameter space, the dark matter
particle is  the lightest neutralino, $\chi_1\equiv \chi \equiv DM$. The four 
neutralinos $\chi_{1,2,3,4}$) are the mass eigenstates coming from the mixing 
of neutral Higgs and gauge boson superpartners. There were many studies 
dedicated to predictions for antimatter fluxes with SUSY dark matter, see 
e.g.~\cite{1998PhRvD..58l3503B,1999PhRvD..59b3511B,1999ApJ...526..215B,
  2004PhRvD..69f3501D}.

Inspired from the SnowMass Point and Slope \cite{2002EPJC...25..113A} 
definition, we will study in the following some typical benchmark points. We 
have chosen three points as close as possible to SnowMass points but 
respecting the WMAP constraint. For the sake of completeness, we also relaxed 
the GUT scale universality assumptions to obtain two additional points 
providing light neutralinos inspired 
by~\cite{2005PhRvD..72h3518B,2008PhRvD..77k5026B} and potentially observable 
at the LHC (see the main parameters in Tab.~\ref{tab:models}).

\subsubsection{Extra dimensions}
\label{subsubsec:kk}

Initially inspired by the hierarchy problem, large extra dimensions with 
testable ($\lesssim$ TeV) phenomenological consequences have been studied 
during the last ten years. For dark matter phenomenology, two 
kinds of models have been proposed. Either universal large compact extra 
dimensions with flat geometry \cite{2001PhRvD..64c5002A} where all the 
Standard Model fields may propagate, or extra dimensions with warped geometry
 \cite{1999PhRvL..83.4690R,1999PhRvL..83.3370R} where all
fields but the Higgs are in the bulk.

New particles arise as the Kaluza-Klein (KK) modes associated to each 
standard  field. 

In universal extra dimensions (UED), the KK parity (coming from translational 
invariance) implies that the lightest KK particle (LKP) is stable. 
Consequently, the first mode of neutral particles (mainly photon and Z but 
also neutrino) can be interesting WIMP candidates. The tree level masses
are degenerate and $=n/R$, where $R$ is the size of the compact dimension
and $n$ is the mode number.  The radiative corrections add extra terms breaking
the degeneracies and in the gauge boson sector, the electroweak symmetry
breaking induces mixing between the gauge eigenstates. These contributions 
lead to different masses for the Z and photons modes. The dark matter 
particle is most likely a gauge boson. Here, we will consider the LKP 
dark matter model of~\cite{2003NuPhB.650..391S}, which is the first KK 
excitation of the $B^{(1)}$ boson. Previous analyses involving this model in 
the frame of antimatter cosmic rays can be found in 
e.g.~\cite{2003PhRvD..68d4008B,2005PhRvD..72f3507B,brun_etal_07}.

In warped geometry, one has to impose a $Z_3$ symmetry to preserve proton 
stability. Consequently, the lightest $Z_3$ charged particle (LZP) is also 
stable and has to be the KK RH neutrino to be a WIMP candidate 
\cite{2004PhRvL..93w1805A,2005JCAP...02..002A}. The dark matter particle is 
thus a Dirac fermion. The 4-D masses arise from the localization of wave 
function of massless modes along the 5$^{\rm th}$ dimension. This mechanism 
is popular to explain the structure of Yukawa couplings. For previous 
studies of indirect detection involving LZP DM candidates, see 
e.g.~\cite{lzp_hooper_servant_05,2005PhRvD..72f3507B,2007A&A...462..827L,
brun_etal_07}.

We have selected one UED model and another one in warped geometry 
(see Tab.~\ref{tab:models}).

\subsubsection{The Inert Doublet Model}
\label{subsubsec:idm}

The scalar sector is a key point of numerous high energy physics theories, 
especially for the standard model and some of its extensions. Up to now, no 
scalar particle has been discovered and this sector is still unknown.
The Inert Doublet Model \cite{Barbieri:2006dq} is a more ad-hoc framework, but 
quite minimal. It is a two Higgs doublet extension of the standard model with 
a $Z_2$ symmetry under which one of the doublet and the fields of the Standard
Model are even. The $Z_2$ symmetry is not broken preserving flavor changing 
neutral currents and forbidding a vacuum expectation value for the odd 
doublet. This scalar sector provides three additional  odd particles, a
neutral ($H_0 \equiv H \equiv DM$), a pseudo ($A_0$) and a charged scalar 
($H_+)$. In our notation, a point of this model is then defined by the mass of 
the extra scalar particle masses : $M_{H_0},M_{A_0},M_{H+}$, the mass of
the standard model Higgs $m_h$  and two scalar potential parameters related to 
the odd doublet: a mass dimensional parameter $\mu_2$ and a coupling 
$\lambda_2$ (see \cite{LopezHonorez:2006gr} for more details concerning the 
parameterization). The neutral extra scalar particle of the model, $H_0$, is a 
typical candidate of WIMP dark matter with an interesting phenomenology. 
Inspired from an extended study of the dark matter phenomenology of the Inert 
Doublet Model \cite{LopezHonorez:2006gr} we have taken two benchmark 
points respecting the WMAP constraint (see Tab.~\ref{tab:models}).

\subsubsection{Little Higgs Model}
\label{subsubsec:lhm}

Another possibility to solve the hierarchy problem is to build a model such 
that the Higgs particle appears as a Nambu-Goldstone boson. In the Little 
Higgs model with T-parity (see~\cite{arkani_01,schmaltz_05,cheng_03} and 
references therein), a global symmetry $SU(5)$ with 
locally gauged subgroup $[SU(2) \times U(1)]^2$ is spontaneously broken down 
to $SO(5)$. The local subgroup is also broken down to the diagonal subgroup 
$SU(2) \times U(1)$, identified with the SM gauge group. To prevent the 
contributions from the new heavy gauge bosons associated with the enlarged 
symmetry to spoil the electroweak precision measurements, a $Z_2$ symmetry, 
the T-parity, is introduced~\cite{han_03}. Therefore, the lightest T-odd 
particle, which happens to be the heavy photon $A_H$, is a potential candidate 
for dark matter~\cite{birkedal_06}. Other T-odd heavy bosons include the 
heavy gauge bosons $W_H$ and $Z_H$ and a scalar triplet Higgs boson $\Phi$. 
The parameters of a point in this model will be given by the value of the vev 
associated with the symmetry breaking $f$, the masses of the heavy T-odd 
bosons, $m_{A_H}$, $m_{W_H}$, $m_{Z_H}$ and $m_\Phi$, and the mass of the Higgs 
particle $m_h$. The annihilation of the heavy photon leads to the production 
of $W$, $Z$ or $h$. Points that satisfy the WMAP constraint lie on two 
branches in the plane $f - m_h$. For points on the upper branch (U-branch), 
$m_h > 2 m_{A_H}$, while for points on the lower branch (L-branch), 
$m_h < 2 m_{A_H}$~\cite{asano_07}. Therefore, we have selected one benchmark 
point on each branch (see Tab.~\ref{tab:models}).

To summarize our benchmarks, we have chosen five supersymmetric points,
two models with extra dimension, two Inert Doublet points and two
Little Higgs models. Parameters and details can be found in 
Tab.~\ref{tab:models}.

\begin{widetext}
\newpage
\begin{sidewaystable}
\centering
\renewcommand{\arraystretch}{1.5}
\begin{tabular}{|l|c|c|c|l|}
\hline
Benchmark point & $M_{\chi}$ (GeV)&$\Omega h^2$ & $\langle \sigma v
\rangle ({\rm cm^3s^{-1})}$ & Branching ratios \\
\hline
\hline
SUSY light-1 : light neutralino, annihilation through $A$ funnel  & 9.7 & 
0.116 &$2.17\times 10^{-26}$ & $BR(\chi\chi\xrightarrow{} b\bar{b})= 0.873$ \\
(scenario A \cite{2008PhRvD..77k5026B} like) Non universal parameters  & & & &
$BR(\chi\chi\xrightarrow{} \tau^+\tau^-)= 0.125$ \\
\hline
SUSY light-2 : light neutralino, annihilation into $\tau^+\tau^-$
via $\tilde{\tau}$   & 34.1 & 0.129 &$1.48\times 10^{-26}$&
$BR(\chi\chi\xrightarrow{}  \tau^+\tau^-)= 1$ \\
(scenario B \cite{2008PhRvD..77k5026B} like) Non universal parameters & & & &\\
\hline
SUSY-SPS2 : focus point, mixed higgsino-gaugino neutralino  & 198.7  & 0.102 &
$2.30\times 10^{-26}$& $BR(\chi\chi \xrightarrow{} t\bar{t})= 0.719$ \\
(SPS2-like)   & & & &  $BR(\chi\chi \xrightarrow{} Z Z)= 0.0763$\\
$m_0=$2850 GeV, $m_{1/2} =$500  GeV, $A_0$=0 GeV,
$\tan{\beta}$= 10 $\mu>0$ &   &  & & 
$BR(\chi\chi \xrightarrow{} s \bar{s})= 0.00387$ \\
\hline
SUSY-SPS3 : neutralino-stau coannihilation  & 283.8  & 0.102 &
$2.08\times 10^{-29}$ &$BR(\chi\chi \xrightarrow{}b\bar{b})= 0.733$ \\
(SPS3-like)   & & & &  $BR(\chi\chi \xrightarrow{} \tau^+\tau^-)= 0.176$
\\
$m_0=$150 GeV, $m_{1/2} =$680  GeV, $A_0$=0 GeV,$\tan{\beta}$= 10
$\mu>0$ &   &  & & $BR(\chi\chi \xrightarrow{} t\bar{t})= 0.0762$ \\
\hline
SUSY-SPS4 : bino neutralino with main annihilation through Higgs funnel  & 
252.2 & 0.107 &$1.50\times 10^{-26}$&
$BR(\chi\chi\xrightarrow{} b\bar{b})= 0.849$ \\
(SPS4-like)   & & & &  $BR(\chi\chi \xrightarrow{} \tau^+\tau^-)= 0.147$
\\
$m_0=$770 GeV, $m_{1/2} =$600  GeV, $A_0$=0 GeV,
$\tan{\beta}$= 50 $\mu>0$ &   &  & & 
$BR(\chi\chi \xrightarrow{} s \bar{s})= 0.00387$ \\
\hline
KK-1 :  KK right-handed neutrino  & 50  &$\simeq$ 0.1 &
$2.04\times 10^{-26}$ &$BR(\chi\chi \rightarrow b\bar{b})=0.7$\\
(warped-GUT-like).    & & & & 
$BR(\chi\bar{\chi} \rightarrow l^+l^-)= 0.035 \times 3$ \\
KK boson mass scale $M_{KK}=6\;{\rm TeV}$
&   &  & & $BR(\chi\chi \rightarrow \nu\bar{\nu})= 0.065 \times 3$\\
\hline
KK-2 : first excitation of the $B^{(1)}$ gauge boson  & 1000  & $\simeq$ 0.1 &
$1.7\times 10^{-26}$ &$BR(\chi\chi\rightarrow l^+l^-)= 0.19\times 3$ \\
(UED-like)   & & & & $BR(\chi\chi\rightarrow (q\bar{q})_{u})= 0.1 \times 3$  \\
 &   &  & & $BR(\chi\chi \rightarrow \nu\bar{\nu})= 0.04 \times 3$\\
&   &  & & $BR(\chi\chi \rightarrow (q\bar{q})_{d})\simeq 0.01 \times 3$\\
\hline
IDM-1 :   light dark matter, annihilation through $A$
 funnel  & 45  & 0.109 &
$1.23\times 10^{-27}$ &$BR(\chi\chi \xrightarrow{} b\bar{b})= 0.871$\\
$\mu_2=$35 GeV, $\Delta M_{A_0} =$10  GeV,$\Delta M_{H^+}$=50 GeV,
$m_h$= 120 GeV, $\lambda_2=0.1$ & & & & 
$BR(\chi\chi \xrightarrow{} \tau^+\tau^-)= 0.0435$\\
 &   &  & & $BR(H H \xrightarrow{} c \bar{c})= 0.0835$\\
\hline
IDM-2 :   heavy dark matter, annihilation into gauge bosons & 1000  & 0.117 &
$4.04\times 10^{-26}$ &$BR(H H\xrightarrow{} W+W^-)= 0.589$\\
$\mu_2=$1005 GeV, $\Delta M_{A_0} =$5  GeV,$\Delta M_{H^+}$=10 GeV, 
$m_h$= 120 GeV, $\lambda_2=0.1$ & & & & $BR(H H\xrightarrow{} ZZ)= 0.223$\\
 &   &  & & $BR(H H \xrightarrow{} hh)= 0.174$\\
\hline
LHM-1 :   L-branch &  103 & $\sim 0.13$ &$1.4478\times 10^{-26}$ &
$BR(H H \xrightarrow{} W^+W^-)= 0.7343$\\
$f=702$ GeV, $m_{A_H}=103$ GeV, $m_{W_H}=450$ GeV, & & & & 
$BR(\chi\chi \xrightarrow{} ZZ)= 0.2657$\\ $m_{Z_H}=450$ GeV,
$m_\Phi=484$ GeV, $m_h=120$ GeV  &   &  & & \\
\hline
LHM-2 :   U-branch &  162 & $\sim 0.13$ &$1.9\times 10^{-26}$ &
$BR(\chi\chi \xrightarrow{} W^+W^-)= 0.6893$\\
$f=1050$ GeV, $m_{A_H}=162$ GeV, $m_{W_H}=678$ GeV,  & & & & 
$BR(\chi\chi \xrightarrow{} ZZ)= 0.3107$\\
$m_{Z_H}=678$ GeV, $m_\Phi=2410$ GeV, $m_h=400$ GeV  &   &  & & \\
\hline

\hline
\end{tabular}
\renewcommand{\arraystretch}{1}
\caption{Benchmark points used in the calculation for Fig~\ref{fig:all_models}.}
\label{tab:models}
\end{sidewaystable}
\end{widetext}

\newpage

\section{Results and discussion}
\label{sec:res}
We have performed different calculations in order to extract the most 
relevant features in a cosmological N-body framework. We will 
first discuss the theoretical uncertainties on fluxes that we can infer from 
an analysis using WIMP fiducial models, combining the N-body data with 
analytical extrapolations. Afterwards, we will focus on the 
predictions that we can derive out of an averaged description of our 3D 
dark matter density map for different BSM dark matter particle candidates.

\subsection{Effects of the dark matter distribution and inhomogeneities}
\label{subsec:dm_density}

\subsubsection{Modifications of the spherical smooth profile}
\label{subsubsec:smooth}

\begin{figure}[t]
\begin{center}
\includegraphics[width=\columnwidth, clip]{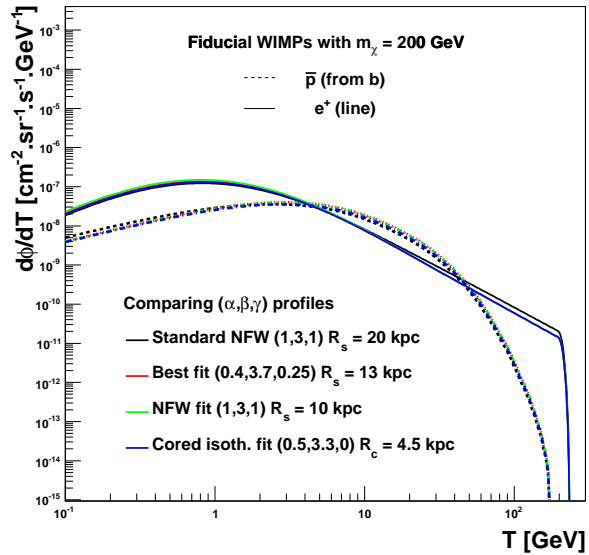}
\caption{\small Comparison of positron and antiproton fluxes obtained with 
 different smooth profiles: three are fitted on the HORIZON 
 simulation (best fit, NFW and cored) and the fourth is the standard NFW fit 
 very often used for the Milky Way.}
\label{fig:comp_smooth}
\end{center}
\end{figure}

Contrarily to dark matter contributions to the Galactic gamma-ray flux, it has 
long been shown that changes in the inner and outer logarithmic slopes of the 
smooth dark matter density profile have no significant effects on the 
charged cosmic ray signatures. This is due to diffusion, which tends to dilute 
the signal when integrated over large volumes (low/high energies for 
positrons/antiprotons). This has been checked for this N-body simulation, 
for which different fits of the density profile give roughly the same 
statistical description of the data. This is illustrated in 
Fig.~\ref{fig:comp_smooth}, where we plot the positron and antiproton fluxes 
corresponding to three different fitted profiles, and where we also plot the 
result obtained with the common NFW model very often used in the
literature to describe our Galaxy. The former three are characterized by an 
averaged density of $\rhosun = 0.25$ GeV/cm$^3$ at 8 kpc from the GC, while 
the latter is taken with its standard value of $\rhosun = 0.3$ GeV/cm$^3$. It 
is noteworthy that there is absolutely no observable difference in the computed 
fluxes between the former three, and that the latter only gives a very small 
additional contribution at high (low) energy for positrons (antiprotons). This 
is merely due to the increase in the local density that translates to a factor 
of $(0.3/0.25)^2$ to fluxes for small propagation lengths. This clearly 
demonstrates that even significant modifications in smooth profiles lead to 
minor effects, provided the local environment is not too much affected.

\subsubsection{Departure from spherical symmetry}
\label{subsubsec:ellipt}

\begin{figure*}[t]
\begin{center}
\includegraphics[width=\columnwidth, clip]{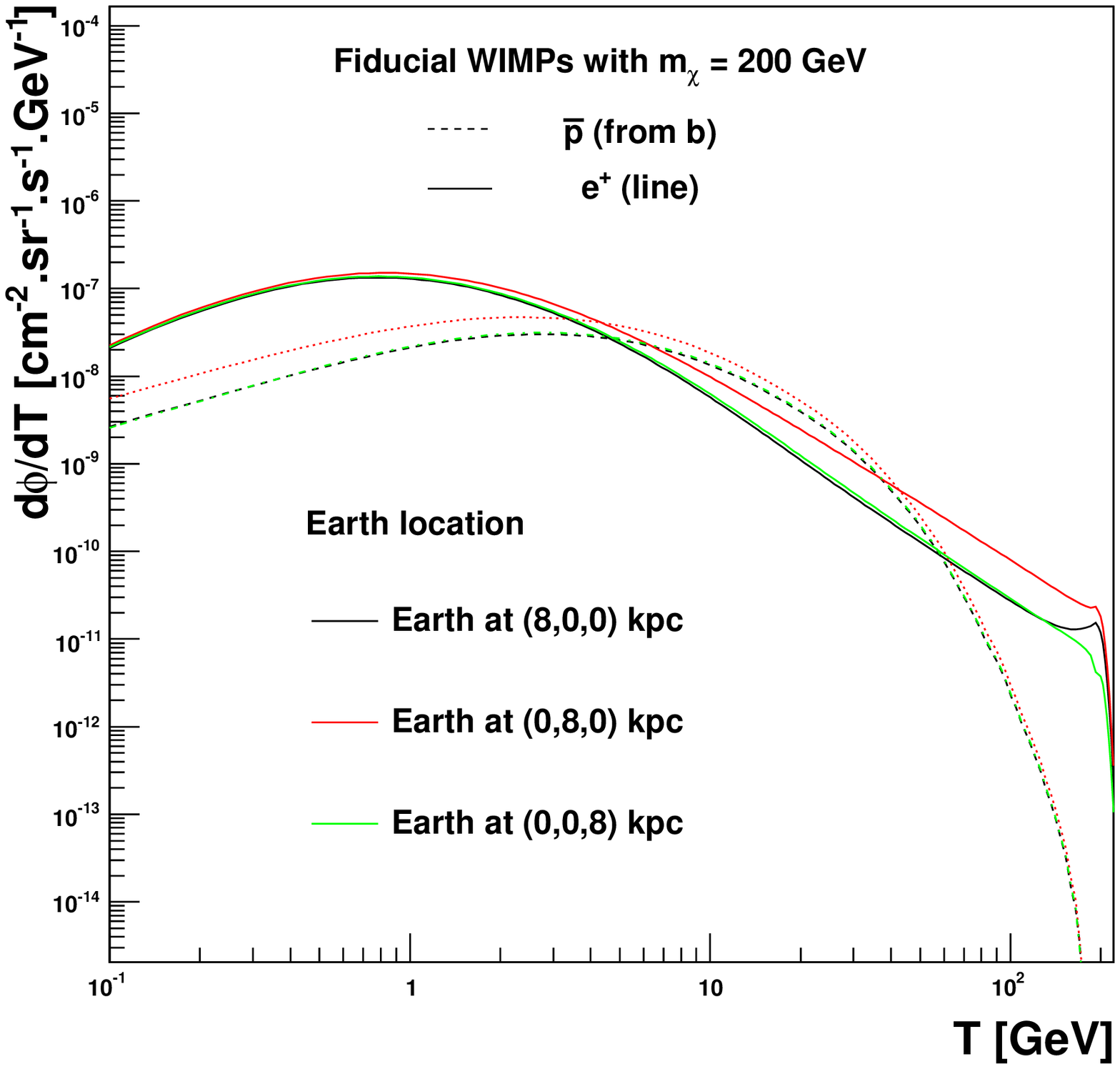}
\includegraphics[width=\columnwidth, clip]{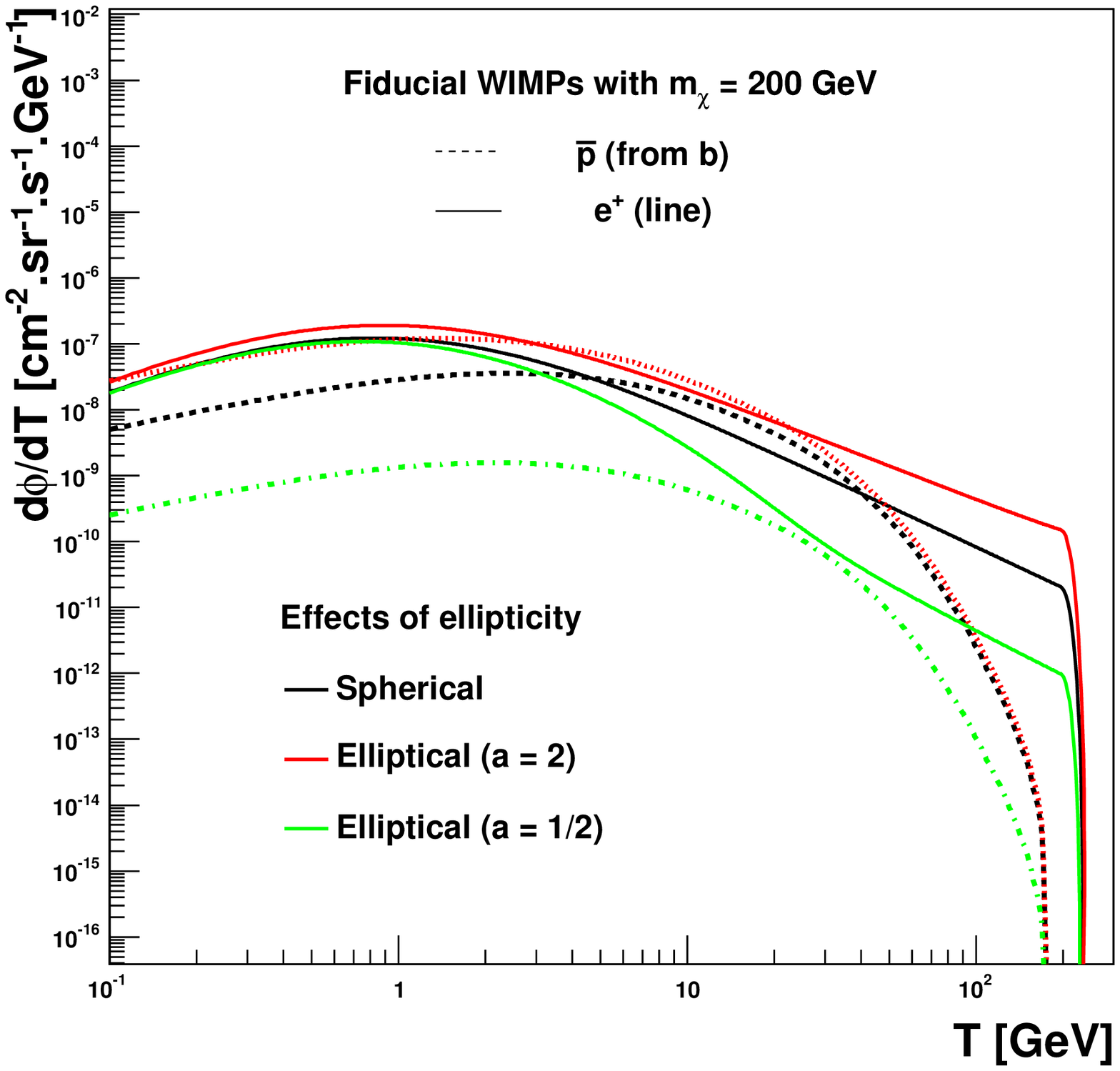}
\caption{\small Positron and antiproton fluxes. Left panel: those 
  corresponding to the three different Earth positions illustrated in 
  Fig.~\ref{fig:dm_density_map}. Right panel: taking into account an 
  elliptical deformation of the dark matter halo (see the details in 
  Sect.~\ref{subsubsec:ellipt}. All fluxes are computed with the fiducial 
  WIMP models (see Sect.~\ref{subsec:ind_wimp}).}
\label{fig:comp_xyz}
\end{center}
\end{figure*}

In the previous paragraph, we have shown that the flux predictions were not 
sensitive to modifications of smooth spherical profiles. However, it turns 
out that though dark matter halos are found to be roughly spherically 
symmetric in N-body simulations in average, they usually do not 
exhibit such a symmetry when 3-D scrutinized. Furthermore, both 
observations and dynamical studies support axisymmetry 
instead~\cite{2002ASPC..275..105D,2002MNRAS.330..591B,bar_lia_07}. The N-body 
framework allows to directly take the effects of departure from spherical 
symmetry into account. This may be seen 
in the left panel of Fig.~\ref{fig:comp_xyz}, where we computed the positron 
and antiproton contributions associated with the three projected density maps 
of Fig.~\ref{fig:dm_density_map} (an elliptic distortion appears in the $xy$ 
plane, along the $y$ axis). We see that for positrons (antiprotons 
respectively) the low (high) energy parts of the spectra are superimposed, 
whereas the high (low) energy parts differ significantly. This comes from that 
low (high) energy contributions correspond to a signal averaged over large 
volumes, erasing information on the source geometry. Conversely, local 
variations of the dark matter density are only relevant at high (low) energies 
for positrons (antiprotons). Therefore, global geometry 
modifications such as departure from sphericity can have an important impact 
if the averaged local dark matter density is changed 
accordingly. This can be further checked by analytically deforming 
a spherical halo to model an elliptical one. The recipe is presented 
in~\cite{non_spher_halo_lia_fu_manu_05}, where an elliptical profile of 
eccentricity $a$ along the $x$ axis in the $xy$ plane can be inferred from the 
spherical dark matter density according to 
$\rho_{\rm elliptical}(x,y,z) \equiv \rho_{\rm spherical}(x/a,ay,z)$. Of course, 
elliptically modifying the halo shape modifies the local dark matter 
density in the same time. Fixing the Earth position at $x=8$ kpc, we have 
considered two cases: $a = 2$ and $a = 1/2$, which results in a modification 
of the local density by a factor of $3.26$ and $0.26$ respectively. Such 
modifications are expected to magnify the differences 
in fluxes observed in the left panel of Fig.~\ref{fig:comp_xyz}. The full 
calculation indeed leads to the right panel of Fig.~\ref{fig:comp_xyz}, where 
the former (latter) case is clearly shown to enhance (decrease respectively) 
the positron and antiproton fluxes by 1 order of magnitude, but mainly at 
energies corresponding to short propagation lengths. The effect 
is much less important at energies corresponding to large propagation lengths, 
where the spherical halo predictions are asymptotically recovered. The 
two previous examples are extreme configurations compared to the current 
observational constraints, and are likely to be attenuated by the 
measured angle of $\phi \sim 20^{\rm o}$ between the GC - Earth axis and the 
semi-major axis~\cite{2002ASPC..275..105D}.

\subsubsection{Density fluctuations and sub-halos}
\label{subsubsec:inhom}

\begin{figure*}[t]
\begin{center}
\includegraphics[width=\columnwidth, clip]{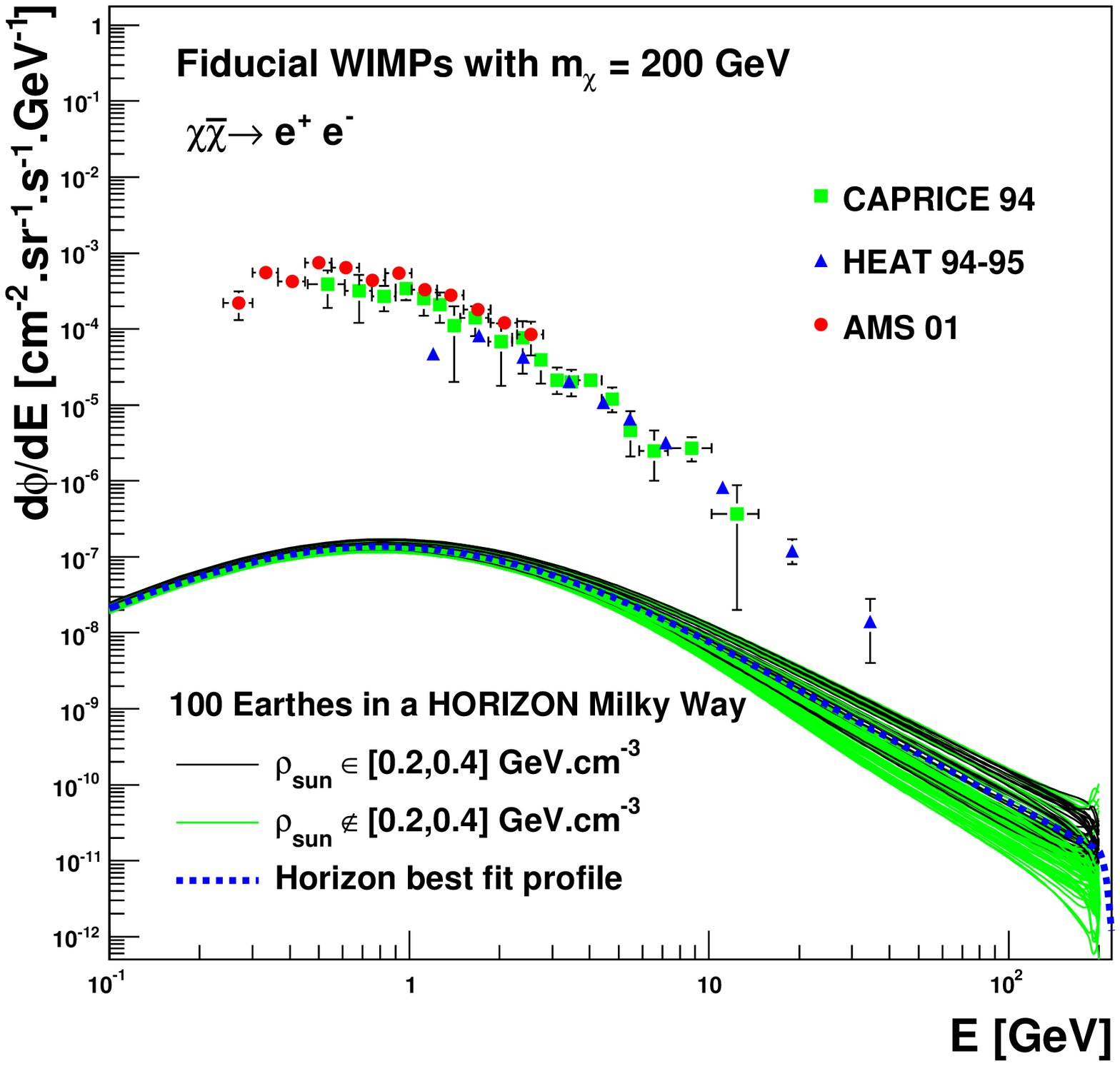}
\includegraphics[width=\columnwidth, clip]{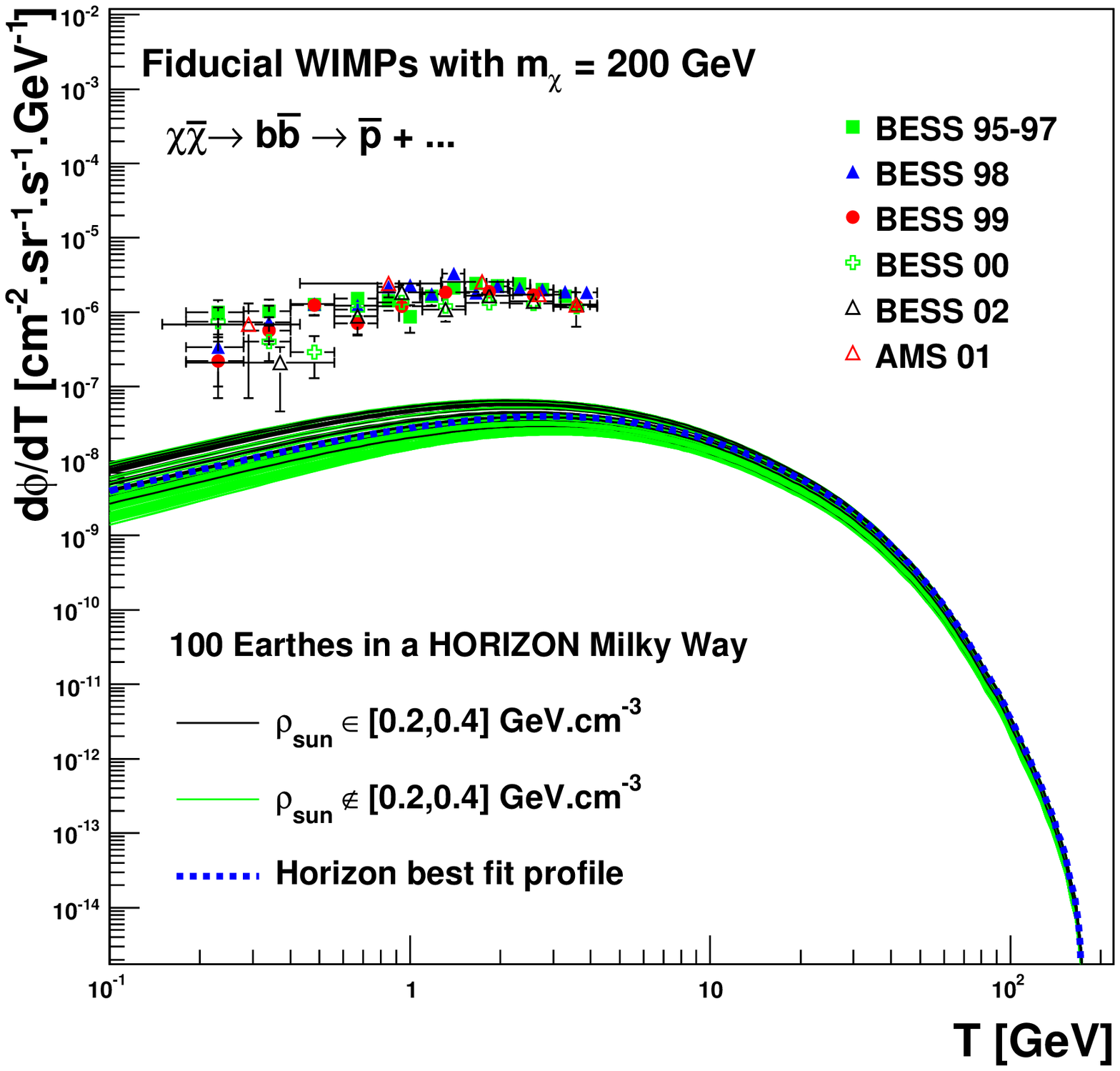}
\caption{\small Varying the Earth position in the N-body simulation. 100 
random positions are sorted and fluxes are computed for positrons (left) and 
antiprotons (right). Fiducial WIMP models detailed in 
Sect.~\ref{subsec:ind_wimp} are used.}
\label{fig:100_earthes}
\end{center}
\end{figure*}

Beside global geometry considerations, dark matter inhomogeneities are 
expected to play a much more important role in characterizing the dark matter 
contribution to the antimatter spectra, but mostly at energies corresponding 
to small propagation scales. Indeed, at energies for which the propagation 
scale is large, the argument is the same as previously, and diffusion erases 
the effect of the source term fluctuations (contributions from the Galactic 
center dominate). On the contrary, the asymptotic limit of fluxes for 
energies $\tilde{E}$ corresponding to very short propagation scales 
only depends on the local dark matter density. As already stressed before, the 
main interest of using an N-body framework is to access a realistic 
inhomogeneous distribution of dark matter. By setting the terrestrial 
observer at different places in the N-body galaxy, we can survey the effect of 
density spatial fluctuations, that is the effect of modifying the local 
environment. This is illustrated in Fig.~\ref{fig:100_earthes}, where we plot 
the flux predictions for positrons (left panel) and antiprotons (right panel) 
associated with 100 different positions of the Earth within our N-body 
grid (the galactocentric radius is fixed at 8 kpc). For comparison, we report 
on the same figures the results when using the best-fit smooth profile, as
well as the existing data 
(from~\cite{2000ApJ...532..653B,2001ApJ...559..296D,2002PhR...366..331A} 
for positrons, and from~\cite{2000PhRvL..84.1078O,2001APh....16..121M,
  2002PhRvL..88e1101A,2005ICRC....3...13H,2002PhR...366..331A} for 
antiprotons). What is interesting is, again, that contributions at energies 
corresponding to large propagation scales (high/low for antiprotons/positrons) 
are very well featured by a smooth description, while fluxes corresponding to 
small propagation scales fluctuate a lot. Such fluctuations are due to 
modifications of the local dark matter density. The larger fluctuations come 
from configurations for which the local density is not constrained to be within 
$[0.2,0.4]$ GeV.cm$^{-3}$ (green curves -- the black ones are 
density-constrained). It is easy to quantify the amplitudes of these 
fluctuations, since the asymptotic flux (small propagation scales) 
merely scales like $\rhosun^2$. The value of $\rhosun =0.25$ GeV.cm$^{-3}$ 
used for the smooth profile allows to have an idea of the local dark matter 
densities corresponding to the extremal fluxes showed in 
Fig.~\ref{fig:100_earthes}, which spread over two orders of magnitude 
(a factor of $\sim 2$ for maximal, and a factor of $\sim 1/4$ for minimal 
local densities).

Although we can nicely deal with density fluctuations in our N-body 
galaxy, the lack of resolution at small scales prevent us to resolve 
sub-halos less massive than $\sim 10^7\Msun$. Moreover, more massive 
sub-halos are only found at the periphery of our simulation data, well beyond 
the cosmic ray diffusion zone, and we do not have evidence for other sub-halos 
in the solar neighborhood that could affect the antimatter flux predictions. 
Sub-halos are expected to have much more important impacts on predictions 
because they are not merely density fluctuations: they are structured objects 
with inner profiles, the first virialized objects of the universe, of which 
the central density could be as or even more cuspy than the galactic center 
itself. The source term of antimatter 
production scaling like the squared density, sub-halos are therefore likely to 
constitute a major component of the exotic primary fluxes. Of course, some 
better simulations have already achieved resolutions good enough to resolve 
clumps as light as $\sim 10^5\Msun$ (e.g. the Via Lactea simulation described 
in~\cite{2008arXiv0805.1244D}), but they are still very far from the typical 
minimal masses $\sim 10^{-6}\Msun$ set by the free streaming scales inferred 
from particle physics. Consequently, the intrinsic resolution limits of 
simulations make the use of analytical extrapolations unavoidable in order to 
study the effects of the smallest dark matter structures on predictions. 
Besides and fortunately, the case of antimatter cosmic rays is well suited for 
such approaches, because contrarily to gamma-rays, the precise location and 
internal features of any source are by no means observable with such 
messengers. Furthermore, an analytical and statistical recipe for calculating 
any clump configuration contribution to the antimatter primaries has been very 
well defined and described in~\cite{2007A&A...462..827L} 
and~\cite{2008A&A...479..427L}. The latter reference provides 
an exhaustive and detailed method to deal with sub-halos as characterized 
in $\Lambda-$CDM cosmology. Armed with these analytical device, we can 
extrapolate our results to different sub-halo configurations, all 
typified by a $M^{-\alpham}$ mass distribution, inner profiles and 
concentrations, and eventually by a spatial distribution. 

\begin{figure*}[t]
\begin{center}
\includegraphics[width=\columnwidth, clip]{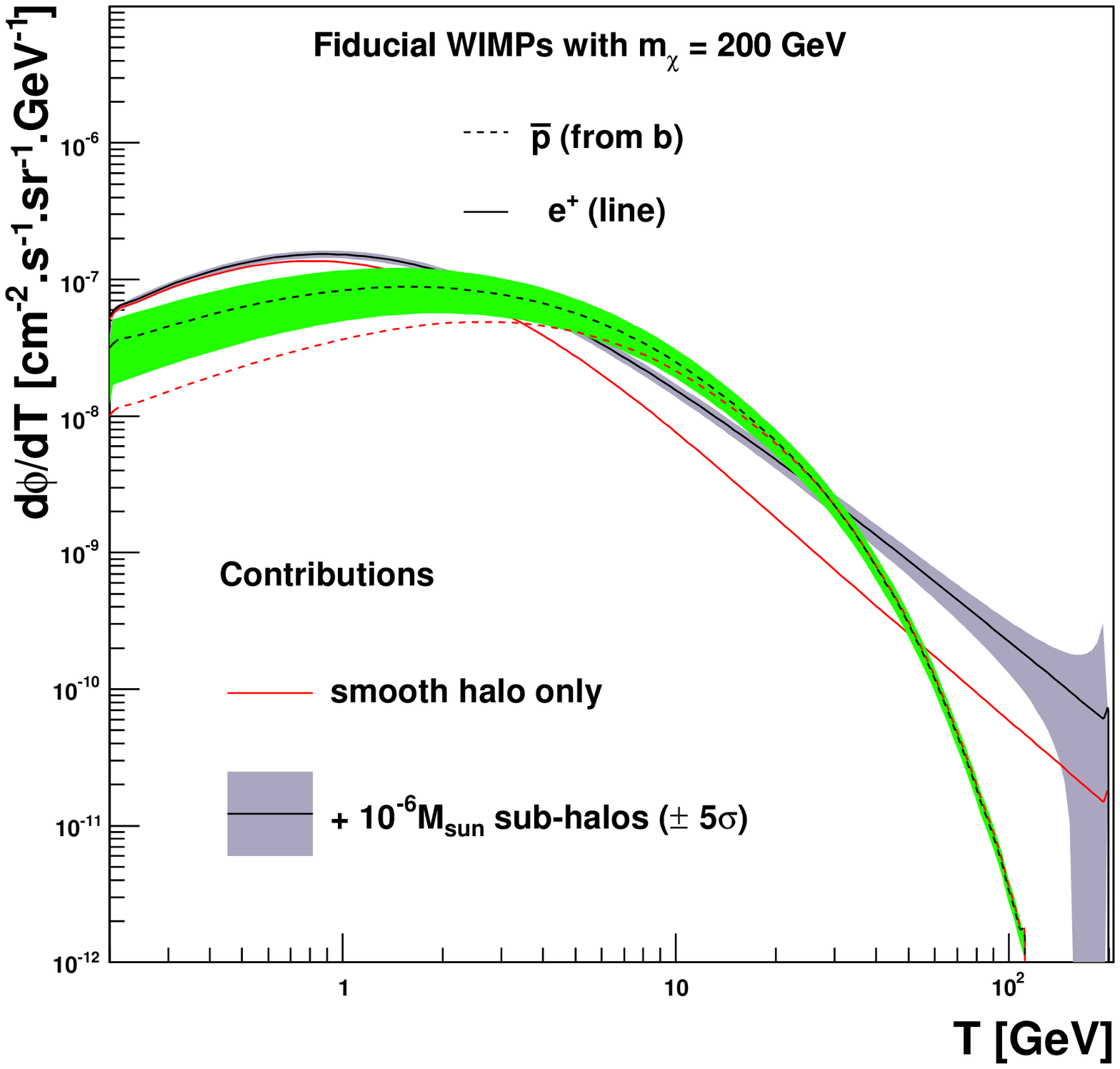}
\includegraphics[width=\columnwidth, clip]{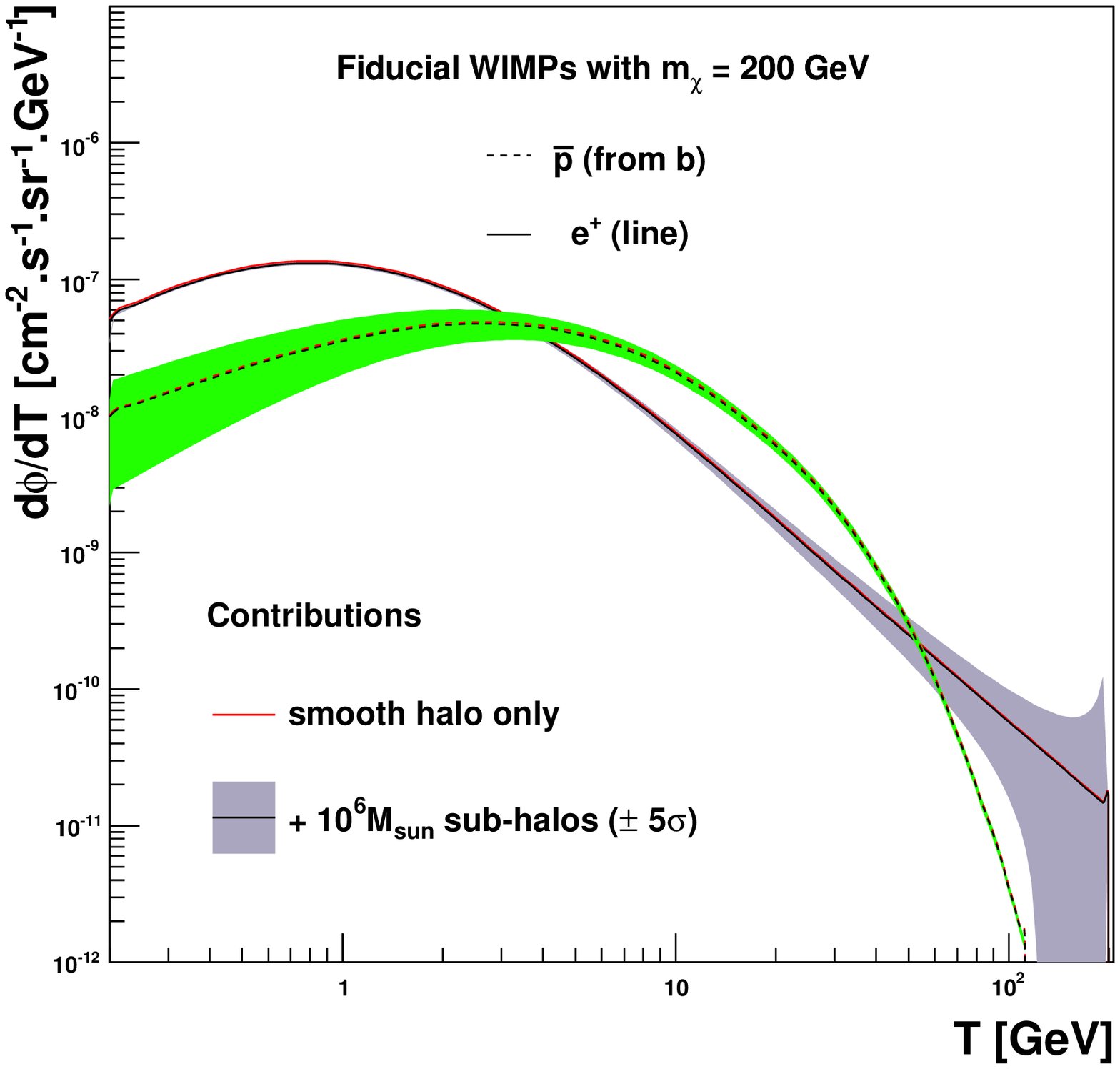}
\caption{\small Analytical predictions for the positron and antiproton fluxes 
when considering sub-halos in the Horizon framework. Left panel: 
\emph{maximal} configuration, where $10^{-6}\Msun$ NFW clumps are involved. 
Right panel: \emph{Via-Lactea-like} configuration, where we consider 
$10^6\Msun$ NFW sub-halos. Fiducial WIMP models detailed in 
Sect.~\ref{subsec:ind_wimp} are used.}
\label{fig:clumps}
\end{center}
\end{figure*}

We have considered two specific cases. The first one is an optimistic 
\emph{maximal} case in which clumps are as light as $10^{-6}\Msun$, with 
$\alpham = 2$, inner NFW 
profiles, and obey the so-called Bullock -- B01 -- concentration-mass 
relation~\cite{bullock_etal_01}; in this configuration, we make sub-halos 
spatially track the smooth profile. The second one is approximately reminiscent 
from what the authors of~\cite{2008arXiv0805.1244D} found in their Via Lactea 
simulation, and will be referred to as the \emph{Via-Lactea-like} 
configuration. It is featured by a minimal clump mass of $10^6\Msun$, 
$\alpham = 1.9$ -- i.e. $\sim 10^{12}$ times less clumps than in the 
\emph{maximal} case --, inner NFW profiles, B01 concentrations, and a spatial 
distribution antibiased with respect to the smooth profile 
(see~\cite{2007ApJ...667..859D} for more details on the \emph{spatial 
antibias}). The reader is referred 
to~\cite{2008A&A...479..427L} for more insights on the impact of 
the clump parameters on predictions for the antimatter cosmic ray fluxes. We 
plot the positron and antiproton fluxes for both configurations in 
Fig.~\ref{fig:clumps}, which somehow quantifies the consequences of our lack 
of spatial resolution, and might even allow to by-pass it. These fluxes are 
characterized by mean values and associated statistical variances (shown as 
contours) taking into account probabilities for clumps to be located inside 
the relevant (energy dependent) diffusion volume and to contribute to the 
signals; in both configurations, the smooth halo is described by our best fit 
parameters (see Sect.~\ref{sec:horizon}). The \emph{maximal} configuration 
(left panel) unveils a flux enhancement (so-called \emph{boost factor}) with 
respect to the contribution of the smooth dark matter component which depends 
on the energy, and saturates at energies corresponding to small propagation 
scales. Indeed, the smooth central halo contribution strongly dominates the 
signals against clumps at large propagation scales, whereas clumps are likely 
to overcome at small propagation scales if numerous enough in the local 
environment~\cite{2007A&A...462..827L,2008A&A...479..427L}. 
The maximal boost factor is $\sim 4$, which 
gives an idea of what we could expect from such a small scale clumpiness. 
Taking a $r^{-1.5}$ profile inside clumps instead of NFW would provide an 
additional factor of $\sim 10$. The \emph{Via-Lactea-like} configuration 
does not provide any flux enhancement in average, but the statistical variance 
of signals is still relevant at short propagation scales, for which a very 
close object, a very improbable situation, could increase the signals. Notice 
that the sub-halo effects yield uncertainties of the same order of magnitude 
as the density fluctuations and departure from spherical symmetry previously 
considered.

\subsection{Predictions for different dark matter particle candidates}
\label{subsec:dm_models}
After having tried to quantify the effects coming from possible dark matter 
density fluctuations with a WIMP model-independent setting, it is useful to 
restore those predictions in a more constrained particle physics scheme. 
This really allows to define and compare the flux orders of magnitude that 
WIMP searches do have to deal with. To this aim, we have used an Earth 
location in the N-body frame which provides intermediate fluxes as illustrated 
in Fig.~\ref{fig:100_earthes}, with a local density of $\rhosun\simeq 0.25$ 
GeV.cm$^{-3}$.

\begin{figure*}[t]
\begin{center}
\includegraphics[width=\columnwidth, clip]{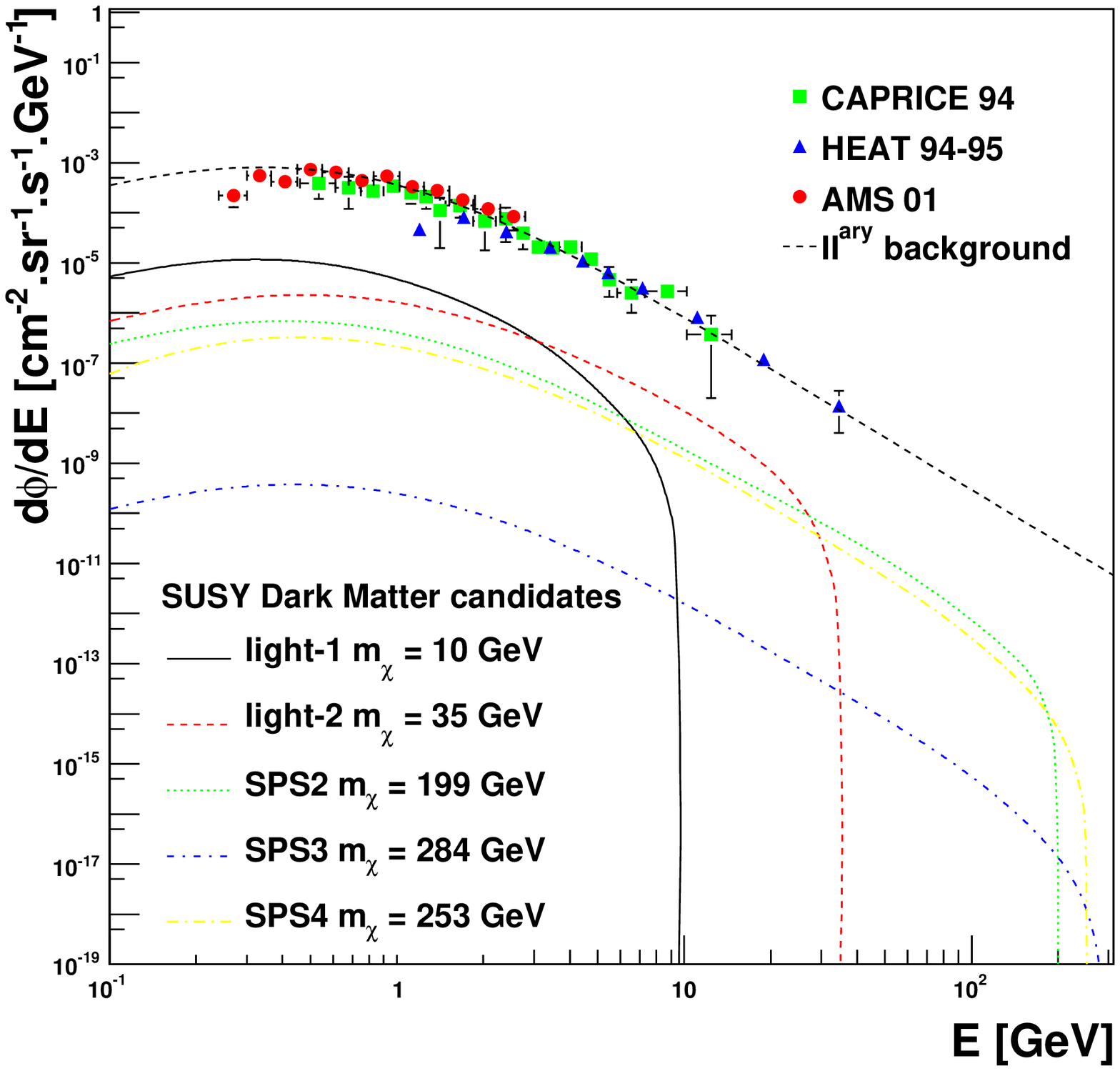}
\includegraphics[width=\columnwidth, clip]{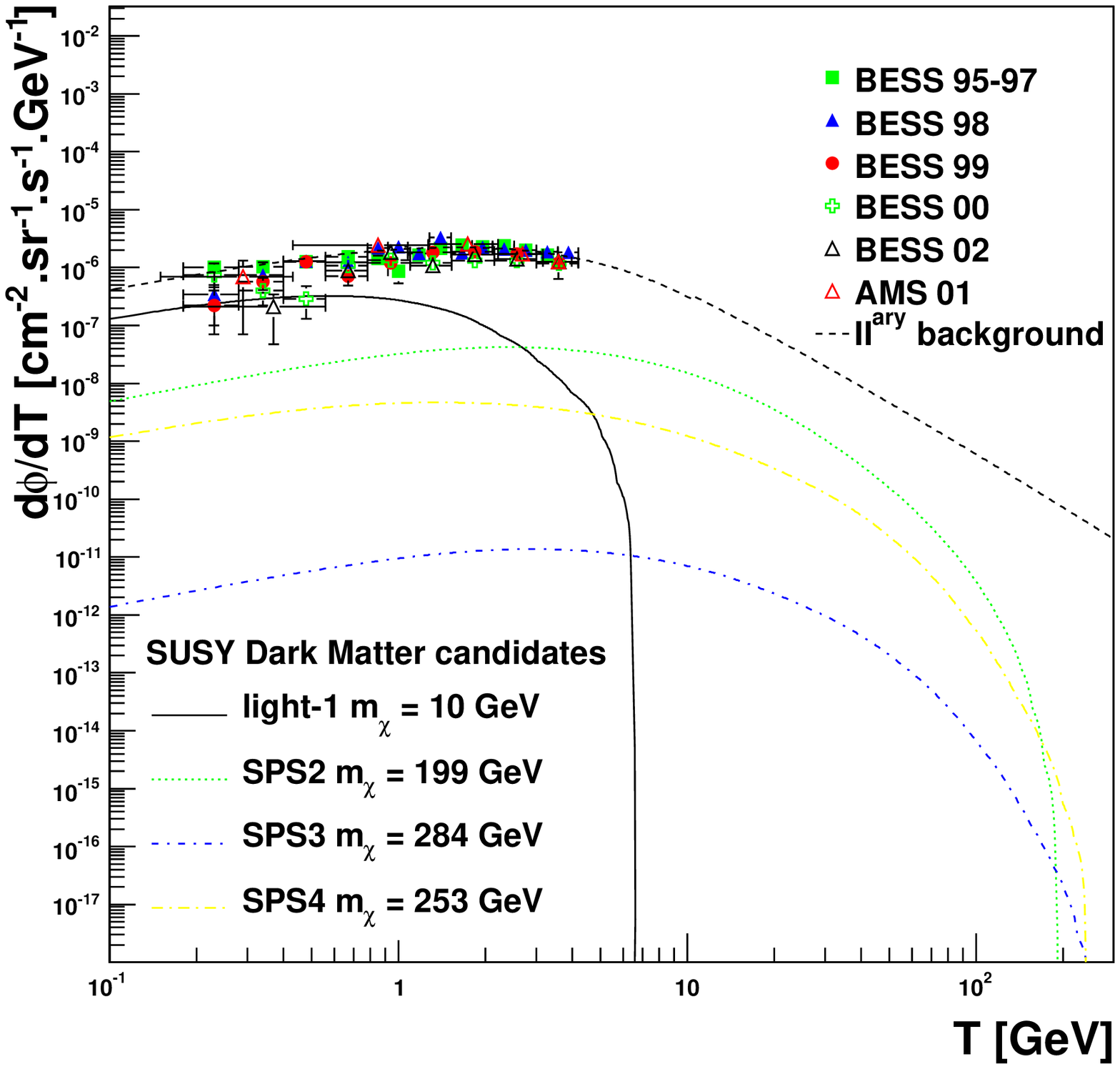}
\includegraphics[width=\columnwidth, clip]{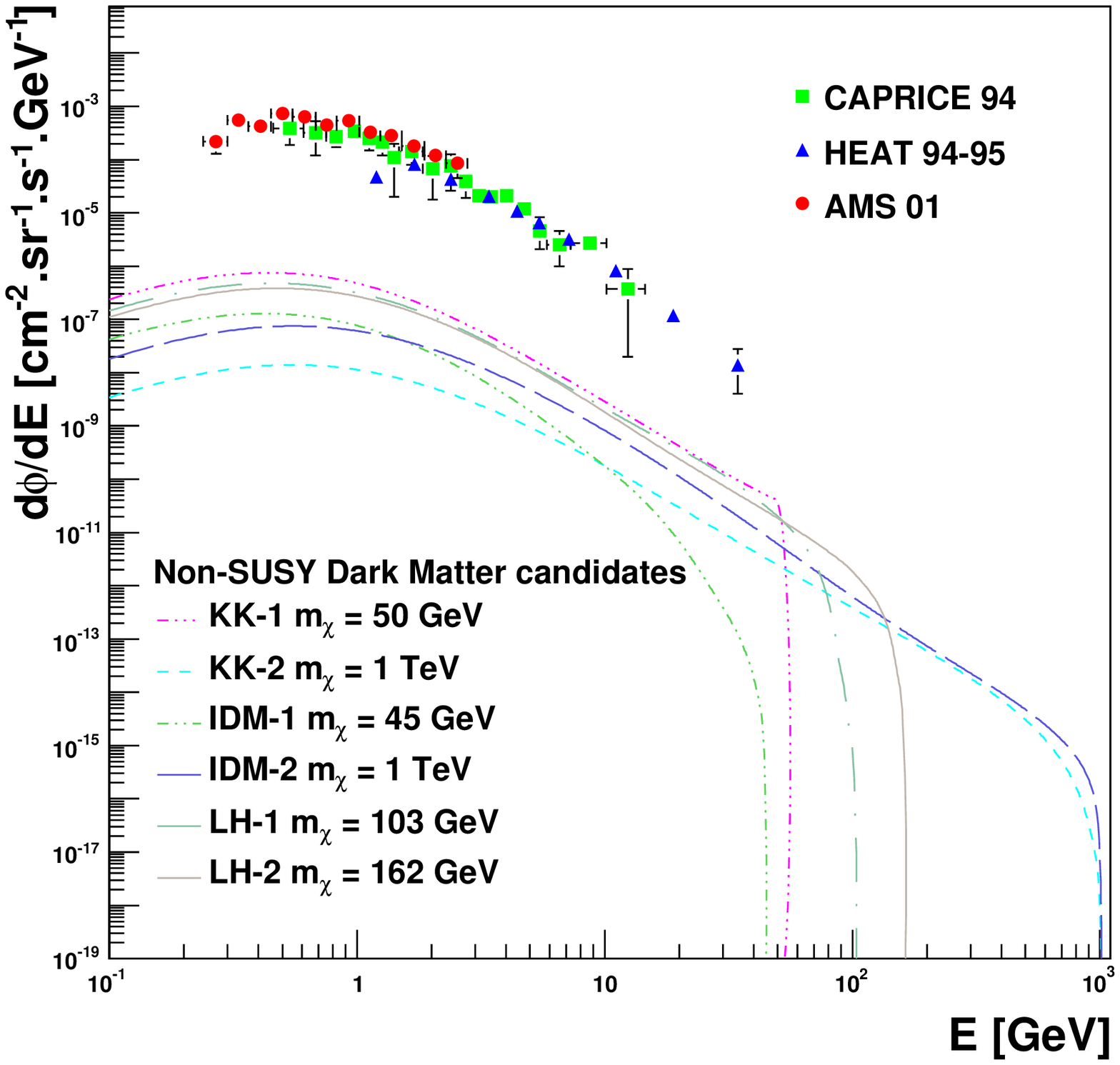}
\includegraphics[width=\columnwidth, clip]{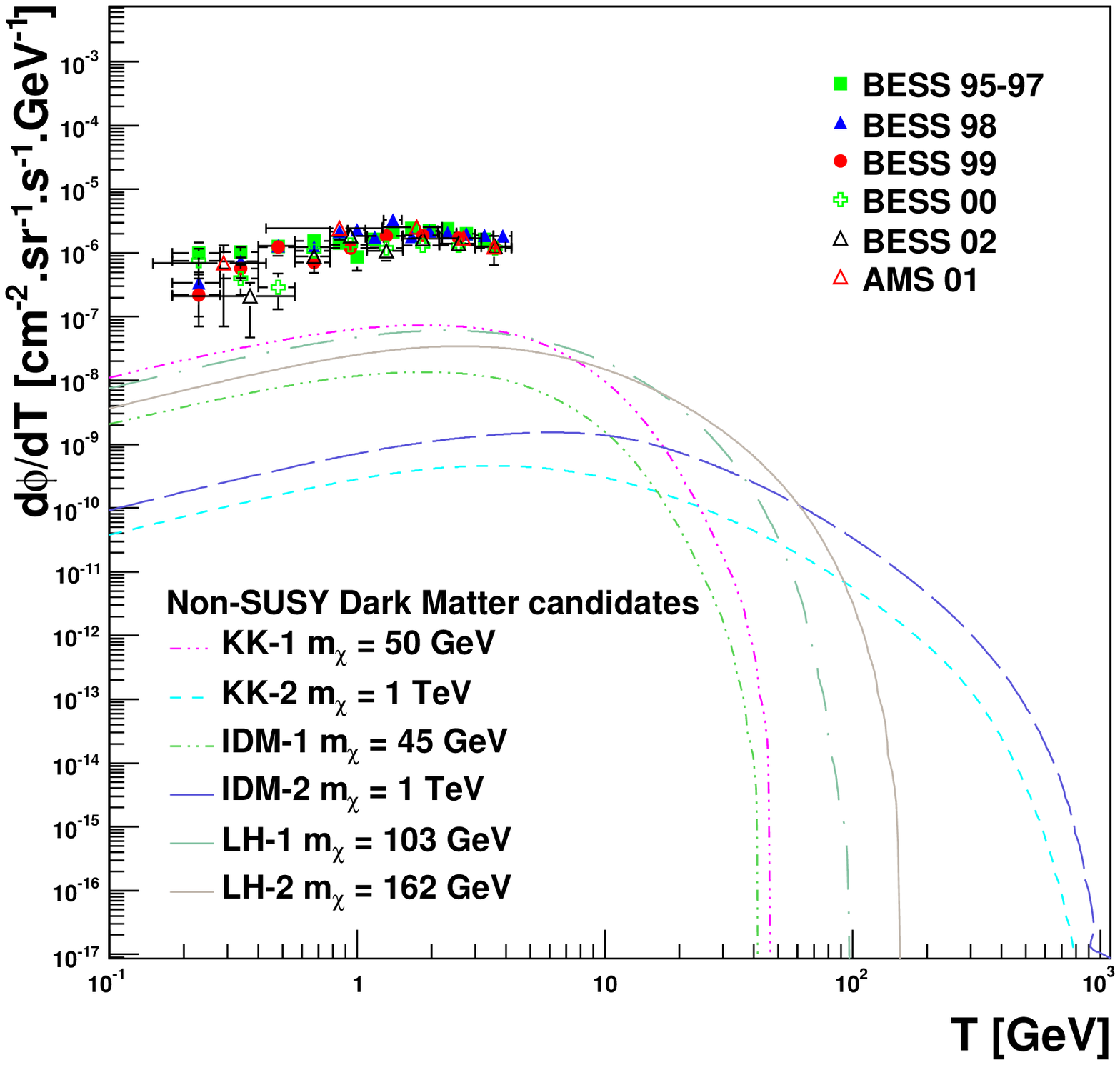}
\caption{\small Left: positron fluxes for SUSY dark matter (top) and non-SUSY 
  dark matter (bottom). Right: same as left panels but for antiprotons. For 
comparison, the secondary background for positrons (antiprotons, respectively) 
as calculated by~\cite{2008arXiv0809.5268D} (\cite{2001ApJ...563..172D}) is 
reported on the top left (right) panel.}
\label{fig:all_models}
\end{center}
\end{figure*}

We have derived predictions for all dark matter candidates presented in 
Sect.~\ref{sec:wimp}, of which the features are summarized in 
Tab.~\ref{tab:models} -- some of them are detectable at the LHC, with well 
defined related signatures. The positron (left) and antiproton (right) fluxes 
are reported in Fig.~\ref{fig:all_models} for SUSY (top) and non-SUSY 
particles (bottom panels, respectively). On the same panels are pictured the 
existing data (from~\cite{2000ApJ...532..653B,2001ApJ...559..296D,
  2002PhR...366..331A} for positrons, and from~\cite{2000PhRvL..84.1078O,
  2001APh....16..121M,2002PhRvL..88e1101A,2005ICRC....3...13H,
  2002PhR...366..331A} for antiprotons). In the top panels, we also 
report the secondary backgrounds computed by~\cite{2008arXiv0809.5268D} for 
positrons (top left panel), and by~\cite{2001ApJ...563..172D} for antiprotons 
(top right panel), both using the same median set of propagation parameters 
employed here for primaries. From this figure, we clearly see that only the 
lightest SUSY candidate is about to imprint the antiproton spectrum, but at 
very low energies ($\lesssim 1$ GeV), a spectral region which very much 
suffers from solar modulation effects (the scatter in the low energy BESS data 
points is mainly due to different solar activities during the observations). 
All other candidates remain far lower than the existing data, and are 
therefore poorly constrained. Even a one or two order of magnitude 
enhancement, due to the various effects considered above, would not be enough 
to detect nor constrain the studied dark matter candidates with the current 
data. We finally stress that even primary signals high enough to be 
detectable would be constrained to not exceed too much the secondary 
contributions, which are already in rather good agreement with the data.

\begin{figure*}[t]
\begin{center}
\includegraphics[width=\columnwidth, clip]{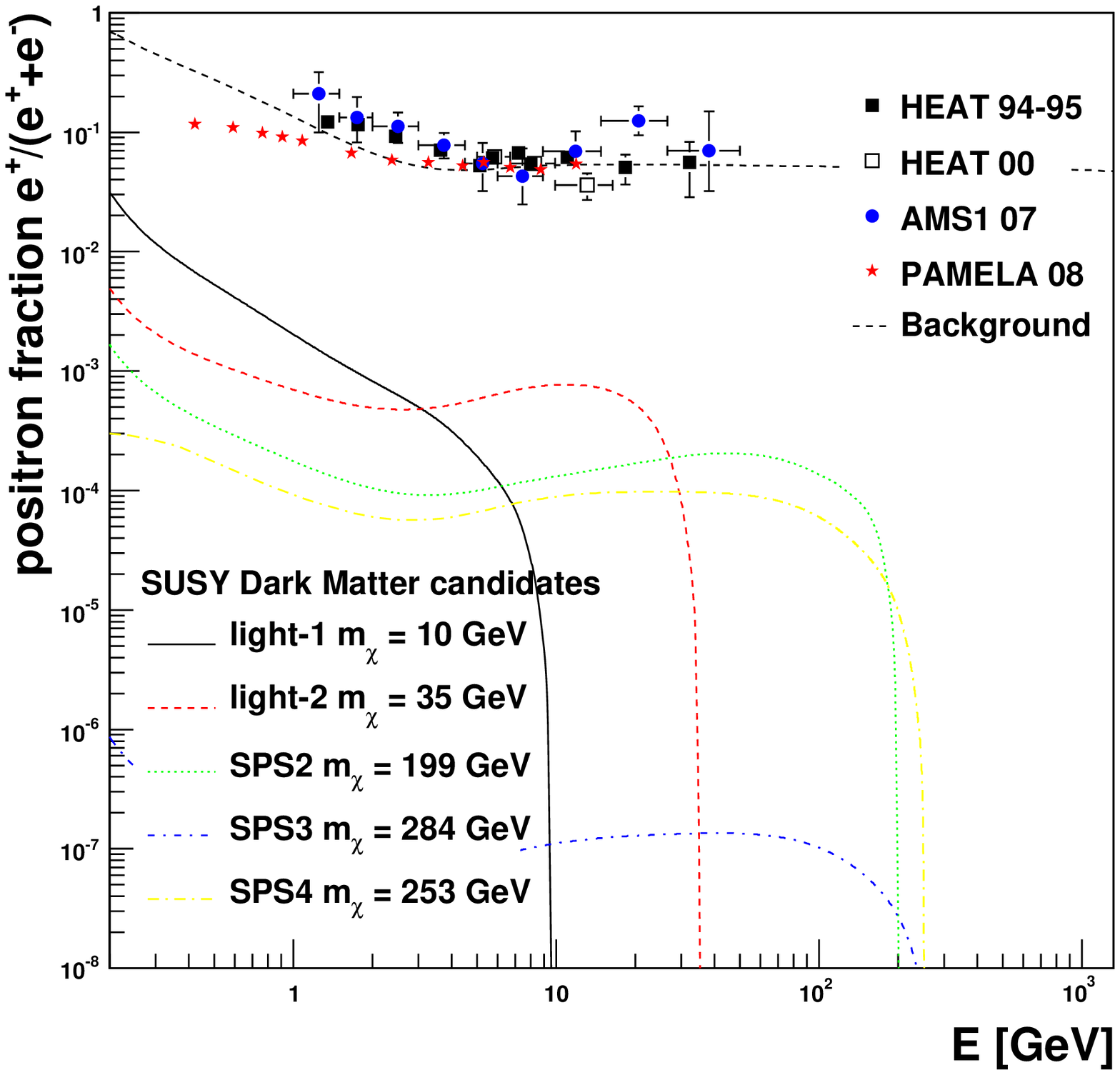}
\includegraphics[width=\columnwidth, clip]{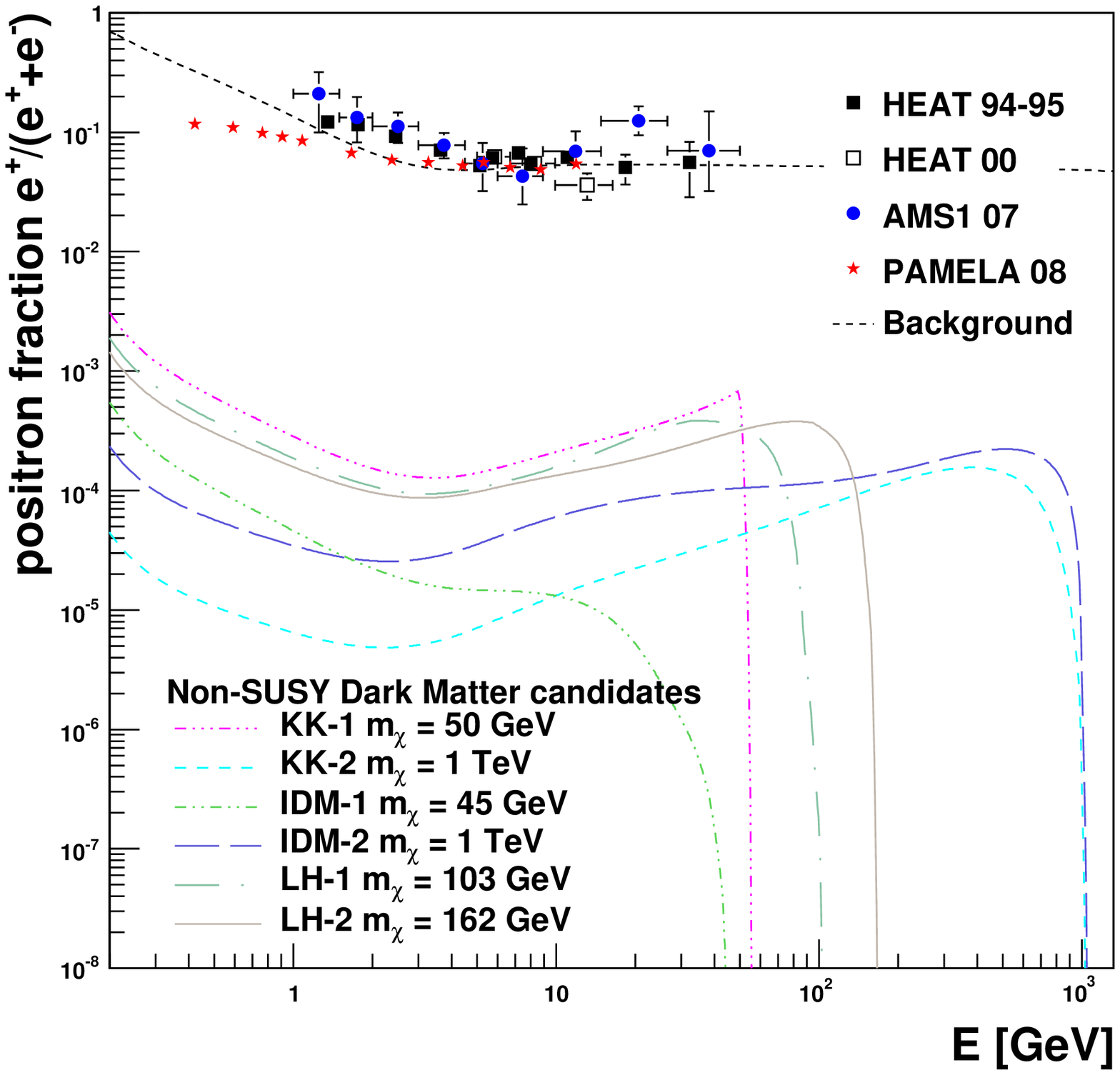}
\caption{\small Positron fraction for SUSY dark matter (left panel) and 
  non-SUSY dark matter (right panel). The secondary positron background is 
  that calculated in~\cite{2008arXiv0809.5268D} with the median set of 
  propagation parameters.}
\label{fig:pos_frac}
\end{center}
\end{figure*}

Another trendy quantity which has recently strongly drawn 
attention is the positron fraction, that is the ratio 
$\phi_{e^+}/(\phi_{e^-}+\phi_{e^+})$ --- where $\phi_{e^-}$ is the electron 
flux. The focus has hold since August (2008), when the PAMELA collaboration 
has presented some preliminary measurements up 
to 270 GeV at the IDM08 conference in Stockholm, which indicate that the 
positron fraction significantly increases with energy contrarily to what is 
expected e.g. from the predictions of~\cite{1998ApJ...493..694M} of both the 
astrophysical electron and secondary positron cosmic ray fluxes. These 
preliminary results confirm the trend previously observed e.g. in the HEAT and 
AMS experiments~\cite{1997ApJ...482L.191B,2004PhRvL..93x1102B,
  2007PhLB..646..145A}. Furthermore, 
newer predictions of the secondary positron flux have just been published 
in~\cite{2008arXiv0809.5268D}, which have been derived in the same slab 
diffusion model as we use in this paper. The authors have also calculated the 
theoretical uncertainties coming from various origins, and found a rather 
good agreement between their predictions of the secondary positron flux and 
the existing data on the positron flux (see Fig.~\ref{fig:all_models}, top 
left panel). In their discussion on the positron fraction, they have 
emphasized the lack of knowledge of the electron cosmic ray flux above a few 
tens of GeV, where it is not well constrained. They have nicely shown how, 
when describing the high energy part of the electron flux with a single 
power law, very slight modifications of the spectral index can lead to quite 
different interpretations of the positron fraction data, spanning from a large 
excess to no excess at all with respect to the predictions. Though we have 
already shown in Fig.~\ref{fig:all_models} that our predictions for the 
primary positrons lie well below the data and the expected background, we also 
plot in Fig.~\ref{fig:pos_frac} the corresponding positron fraction for the 
sake of completeness. To this aim, we have modeled the electron flux like 
in~\cite{2008arXiv0809.5268D}, with a fit on the AMS data and constrain the 
spectral part above a few GeV to be a single power 
law with index -3.44 corresponding to the best fit value obtained 
in~\cite{2004ApJ...612..262C}, in which data up to $\sim 1$ TeV were utilized. 
Note that this electron spectrum differs from that 
of~\cite{2008arXiv0809.5268D}, since it does not correspond to their 
\emph{hard}, neither \emph{soft} indices, but to a median one instead.
In Fig.~\ref{fig:pos_frac}, we also report the positron fraction data released 
in~\cite{1997ApJ...482L.191B,2004PhRvL..93x1102B,2007PhLB..646..145A}, and we 
also show the PAMELA data but up to 10 GeV only~\cite{picozza_blois_08}, since 
higher energy points are not yet officially released. The positron fractions 
associated with the SUSY dark matter models are plotted in the left panel of 
Fig.~\ref{fig:pos_frac}; those corresponding to the non-SUSY models are 
featured in the right panel. We see again that all primary signals are orders 
of magnitude far below not only from the data but also from the background 
expectation. Modifying the electron spectrum would not change our conclusions 
here. If one still tries to accomodate the positron fraction flux by tuning 
the dark matter properties, the counter part in antiprotons will be a strong 
limit, unless the antiproton production is suppressed, which is the case for 
our light neutralino model of 35 GeV for which only the leptonic couplings 
are relevant (see Tab.~\ref{tab:models}). Such a conclusion has also been 
stressed in~\cite{2008arXiv0809.2409C}.

All statements in this subsection rely of course on the fact that the 
propagation model we used is defined by the median set of propagation 
parameters of Tab.~\ref{tab:prop}. A more favorable set, which for instance 
would involve a larger diffusion coefficient $K$ and a larger vertical 
extension $L$ of the diffusion zone (such that $K/L$, the main constrained 
quantity, remains constant), could lead up to an additional order of magnitude 
in terms of fluxes (see e.g.~\cite{2004PhRvD..69f3501D,
  2008A&A...479..427L,2008PhRvD..77f3527D}), and could therefore make 
of the current data very strong limits. This points towards the necessity to 
have more precise and over larger energy ranges data of secondary/primary 
cosmic rays to much better bound the propagation allowed configurations. 
Furthermore, we see on Fig.~\ref{fig:all_models} that there is an overwhelming 
lack of high energy data above $\sim 10$ GeV, precisely in spectral regions 
where the astrophysical background coming from secondaries is expected to be 
power law suppressed. Higher energy data could still exhibit spectral 
transitions due to some exotic components, provided the dark matter candidate 
is massive enough. However, this would require dark matter particles and/or 
spatial distributions with quite different properties from those used in this 
paper, since we have have shown them tho have contribution up to several 
orders of magnitude below the existing data. The PAMELA satellite is currently 
improving the measurements on antiprotons and positrons in a slightly higher 
energy range than before~\cite{2007arXiv0708.1808C}, and the GLAST satellite 
will also have the potential to detect high energy electrons and 
positrons~\cite{2007arXiv0706.0882M}.

\subsection{Limits of the N-body framework}
\label{subsec:dm_limits}
N-body simulations are the only way to scrutinize structures that have 
virialized in the non linear regime, and especially when one is interested 
to check some peculiar effects, like those coming from inhomogeneities 
or asphericity. Nevertheless, such a framework still suffers from many 
drawbacks, due to their incompleteness. First of all, the limited spatial 
resolution prevents the study of the smallest scales that substructures 
can have, and which are fixed by the free streaming length of dark matter 
particles in the early universe. Some authors have tried to tackle this issue, 
by performing simulations resolving $\sim 10^{-6}\Msun$ objects, but their 
results hold only at a redshift $z=26$, when galactic halos are not yet 
formed~\cite{2005Natur.433..389D}. Even by taking the fastest available
supercomputers for cosmological simulations, the authors 
of~\cite{2008arXiv0805.1244D} only reached a resolution allowing the 
identification of $\sim 10^5\Msun$ objects, that is more than 10 orders of 
magnitude beyond what is expected from WIMP cold dark matter.

Moreover, it is noteworthy to recall that baryons are likely to play an 
important role in the dynamics of galaxies, and they are not yet fully 
incorporated in cosmological simulations down to the galactic scale. For 
instance, galactic bars observed in the centers of spirals have long been 
shown to exchange momentum with the dark matter halo, and are therefore 
expected to modify the dark matter distribution (see e.g. a recent analysis 
in~\cite{bar_lia_07}, and references therein). Moreover, it is well known, but 
scarcely mentioned in the present context, that the available data 
on the dynamics of the Milky Way, the target of interest in this paper, 
strongly disfavor cuspy halo models (see more detailed arguments in 
e.g.~\cite{2004MNRAS.351..903G}).

\begin{figure*}[t]
\begin{center}
\includegraphics[width=\columnwidth, clip]{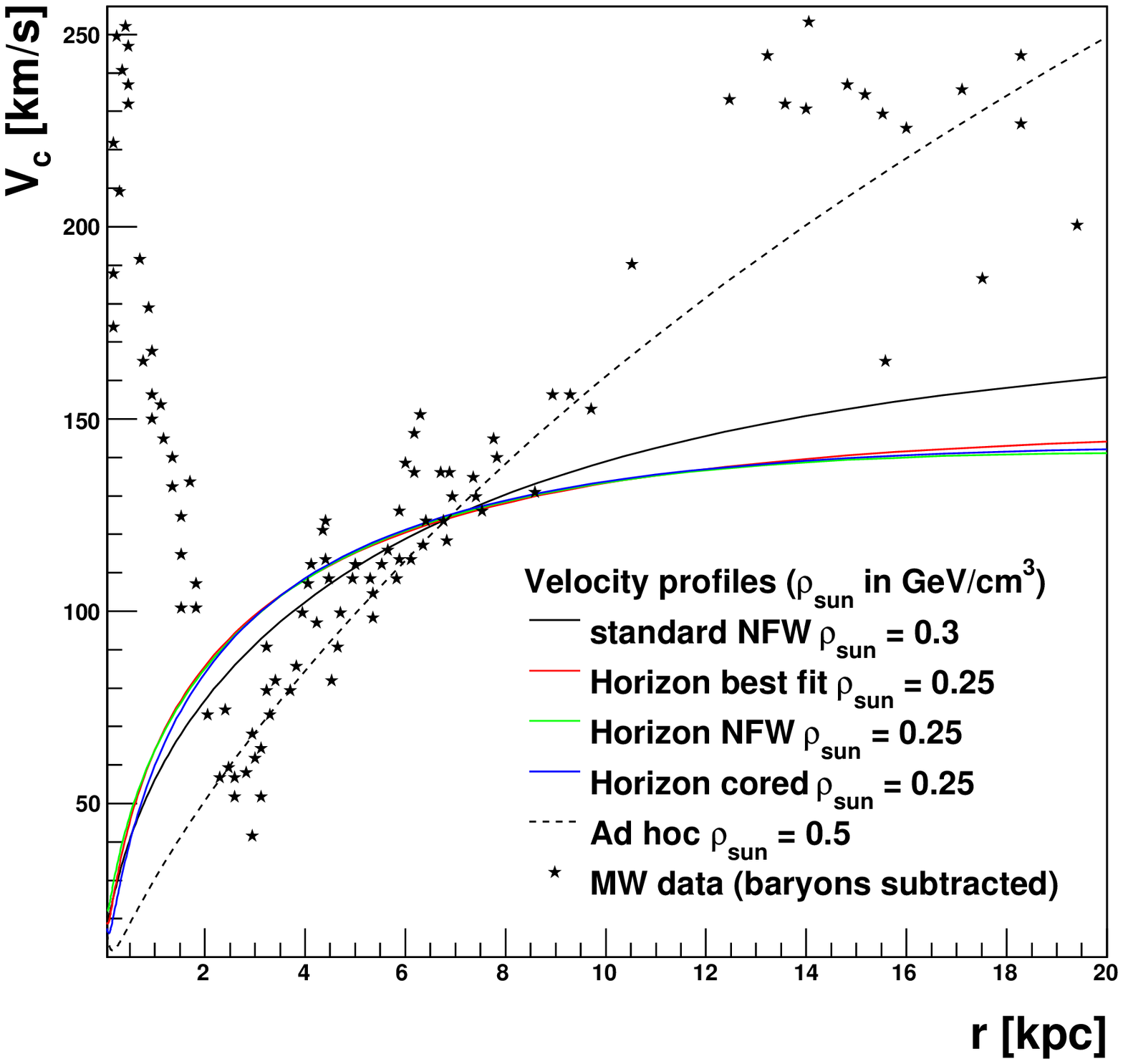}
\includegraphics[width=\columnwidth, clip]{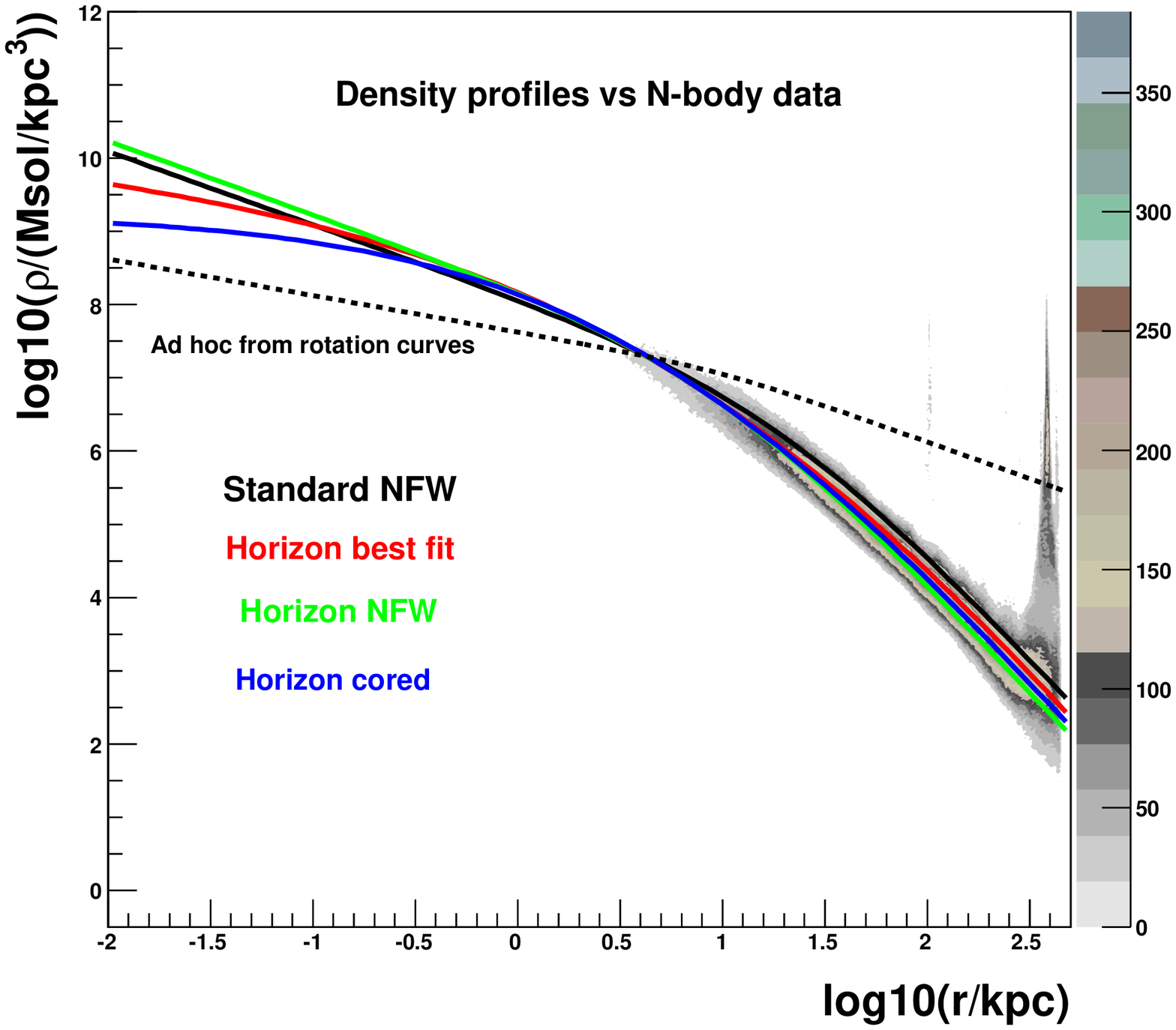}
\caption{\small Left: Radial velocities inferred from the spherical dark matter 
halo models discussed in Sect.~\ref{sec:horizon}, plus an \emph{ad hoc} 
profile with $\alpha = 2,\,\beta=1,\,\gamma=0.5,\, 
r_s = 10\,{\rm kpc},\, \rhosun = 0.5\,{\rm GeV.cm^{-3}}$. The black stars 
represent the kinematic data from which the baryonic contribution has been 
subtracted; cf. Fig. 3 of~\cite{2006CeMDA..94..369E}. Right: same density 
profiles reported on the N-body density data, for comparison.}
\label{fig:rotation_curves}
\end{center}
\end{figure*}

In order to illustrate this latter statement, we have used the kinematic data 
provided by~\cite{2006CeMDA..94..369E}, for which the authors subtracted the 
contribution of the baryons as described by the bulge-disc-bar model 
of~\cite{2002MNRAS.330..591B}. In Fig.~\ref{fig:rotation_curves} (left panel), 
we have reported the circular velocities against the galactocentric radius, 
inferred from the smooth dark matter profiles fitted on our N-body data and 
presented in Sect.~\ref{sec:horizon}, from the standard NFW profile, and also 
from a \emph{ad hoc} profile with parameters $\alpha = 2,\,\beta=1,\,
\gamma=0.5,\, r_s = 10\,{\rm kpc},\, \rhosun = 0.5\,{\rm GeV.cm^{-3}}$ 
(cf.~Eq.~\ref{eq:profile}). The latter has no physical justification at all, 
and has only been tuned to correctly adjust the velocity data. For comparison, 
we have plotted the same profiles against the density data extracted from our 
N-body simulation (right panel).

Disregarding the \emph{ad hoc} profile for the moment, it is interesting to 
remark that while the standard NFW hardly better fits the velocity curves than 
the others at large radii ($r \gtrsim 6$ kpc), all the cuspy profiles fail to 
reproduce the dark matter rotation curve in the inner 4 kpc --- the 
central parts of our Galaxy are influenced by the bar, which perturbs the
kinematics of the gas; thus the points inside about 3 kpc are not to be 
considered. These cusps 
systematically lead to an overestimate of the dark matter mass. Between 4 and 
8 kpc the fits are also bad, but take advantage of the large dispersion of the 
observational points to remain within the cloud of points. However, they 
adjust the highest values at the inner radii and the lowest values at the 
largest radii, i.e. are not acceptable in the $\chi^2$ sense. Beyond 8 kpc the 
values predicted by the cuspy profiles are by far too low compared to the 
observations, even when one takes into account the considerable uncertainties 
of the observations outside the solar radius. The \emph{ad hoc} profile, which 
is close to a cored isothermal profile, provides a good fit to the 
observations though it may overestimate the velocities at 
large radius (see Fig.~\ref{fig:rotation_curves}, left). This is not 
surprising since it was tuned for this purpose. Nevertheless, interestingly, 
this profile does not fit the N-body data (see Fig.~\ref{fig:rotation_curves}, 
right). This underlines one of the well known and much discussed discrepancies 
between observations and the cosmological N-body simulations in 
the scale of galaxies (see e.g.~\cite{2004IAUS..220...39B} 
and references therein). Such discrepancies are believed to be due to the fact 
that cosmological N-body simulations do not include yet the baryonic 
component, of which the presence could lower the cusp into a core.

Hence, we stress that the predictions we provided in the previous section 
should be better considered as indicative, because they do certainly not 
contain the full theoretical uncertainties, especially those coming from our 
ignorance of the baryon influence. Nevertheless, our calculations, as 
performed in an N-body context, still allow a better understanding of the 
uncertainties which come from the possible fluctuations of our local dark 
environment. Computing the fluxes by using the previously discussed 
\emph{ad hoc} profile would give exactly the same trend as those reported in 
Fig.~\ref{fig:comp_smooth}, with a slightly increased asymptotic value at 
short cosmic ray propagation scales, by a factor of 
$(\rho_0^{\rm ad\,hoc}/\rho_0^{\rm horizon})^2 = 4$. In order to not 
further complexify the reading, we do not plot the corresponding predictions, 
since this profile does not rely on any theoretical motivation. Eventually, 
note that despite the already mentionned drawbacks, our analysis sketches the 
bases for further studies using more complicated frameworks (including baryons 
for instance).

\section{Conclusion}
\label{sec:concl}

In this paper, we have studied the theoretical uncertainties associated with 
dark matter density fluctuations affecting the primary antimatter cosmic ray 
fluxes (positrons and antiprotons), possibly produced by dark matter 
annihilation in the Galaxy, by using an N-body simulation together with 
analytical descriptions of a galaxy-like dark matter halo. This is the first 
attempt in this research field to directly connect the source term of the 
cosmic ray propagation equation to a 3D density map coming from a cosmological 
N-body framework, while this has already often been used for predictions 
of gamma-ray fluxes. Indeed, the latter situation does not involve cosmic 
ray propagation, which seriously complicates the calculations, and makes them 
much more time consuming. The main purposes of this work was to 
quantify those uncertainties, and in some cases to try to understand them by 
analytical means. While this was achieved in a WIMP model-independent setting, 
we also aimed at reviewing the status of predictions 
for many different well motivated dark matter candidates in particle physics, 
supersymmetric or not.

The N-body framework led us to study several effects, such as non-spherical 
dark matter distributions (Sect.~\ref{subsubsec:ellipt}) and density 
fluctuations (Sect.~\ref{subsubsec:inhom}), which are not accounted for 
when using the traditional spherical smooth halo models. We have shown that, 
as expected, fluxes are mainly affected at cosmic ray energies corresponding 
to short propagation characteristic lengths (low/high for 
antiprotons/positrons) and to local contributions, while when looking at 
energies of large propagation lengths, the smoothing due to diffusion erases 
the peculiarities in the spectra, and is well reproduced by a spherical halo 
model. The main uncertainties on the predicted fluxes come therefore from 
those on the local dark matter environment, which we have shown to fluctuate 
a lot in the N-body framework. This can lead to $\pm$ 1 order of magnitude in 
terms of flux, at high energy for positrons, and low energy for antiprotons. 
This could be more constrained with much better limits on the local dark 
matter density.

Moreover, though our simulation has a too poor spatial resolution to resolve 
substructures inside the galactic halo, we have extrapolated our results in 
order to include the potential effects of the presence of sub-halos by 
using the analytical method detailed in~\cite{2008A&A...479..427L}. 
We have considered two situations: one \emph{maximal}, analytically extending 
the clump mass spectrum down to $10^{-6}\Msun$, led to a flux enhancement by 
an energy-dependent factor reaching $\sim 4$ at maximum; a second, 
\emph{Via-Lactea-like}, involved clumps down to $10^6\Msun$ only and did not 
result in any flux enhancement.

Consequently to the previous points, the smooth and spherical modeling 
of the dark matter halo leads to a very good approximation of a more complete 
and complex situation, as derived from an N-body simulation, at large 
characteristic propagation lengths for antimatter cosmic rays (low/high 
energies for positrons/antiprotons). Nevertheless, though predictions in the 
spectral regions associated with those large characteristic scales are also 
the less affected by the uncertainties coming from dark matter density 
fluctuations, we remind that they are still very sensitive to the used 
propagation model.

Beside this trial for directly \emph{measuring} the theoretical uncertainties 
with an N-body experiment, we reviewed the predictions of the positron and 
antiproton primary contribution associated with some specific particle 
physics dark matter candidates (Sect.~\ref{subsec:dm_models}). We have not 
only included some popular 
supersymmetric and extra-dimensional models, but also some more specific 
ones based on minimality arguments, such as the little Higgs model, or the 
inert doublet model. We have shown that predictions are well lower than the 
existing data, except perhaps for the antiproton flux associated with the 
lightest neutralino model, just because of the more favorable $1/\mchi^2$ 
factor appearing in the flux expression. Nevertheless, even if the latter 
appears in tension with the current data, its contribution to the antiproton 
flux occurs at energies where solar modulation effects are expected to be 
important, and it would be hard to find a clear signature in this spectral 
region. The other models are well below the experimental measures, but higher 
energy measurements could perhaps unveil some spectral transitions followed 
by an energy cut-off which could possibly be attributed to dark matter. In 
this case, uncertainties associated with the dark matter density fluctuations 
would be much less stringent for the antiproton signal than for the positron 
because of propagation scale arguments. On the contrary, a nearby clump would 
favor a detection with positrons at high energy, much more than with 
antiprotons. Anyway, higher energy data are necessary in this field, for 
these antimatter species as well as for standard nuclei species which will 
yield much better constraints to the propagation modeling. Obviously, 
indications of a WIMP candidate mass, as expected from the LHC, would provide 
the relevant spectral region where to concentrate the astrophysical searches, 
and would allow to concentrate on much more subtle effects.

Finally, we made a self-criticism exercise -- Sect.~\ref{subsec:dm_limits} -- 
by comparing our different 
halo models to the star kinematic data available for our Milky-Way Galaxy, 
after subtraction of the baryonic contribution as modeled 
by~\cite{2006CeMDA..94..369E}. We show that the dark matter contribution 
to the velocity field as inferred from our N-body simulation, as well as 
more general ones, are far to reproduce the kinematic data, systematically 
leading to a mass excess in the central regions ($\lesssim 4$ kpc) of the 
Galaxy. This clearly calls for restraint in our potential claims, and clearly 
points towards some large amount of ignorance in the intimate (and 
gravitational) relations between dark matter and baryons.

\section*{Acknowledgements}
We would like to thank A.~Bosma, N.~Fornengo and P.~Salati for interesting 
discussions during this study. This work took advantage of the 
favorable environment encountered during some of the French GDR SUSY meetings; 
we are grateful to its current headmaster J.~Orloff and to its Dark Matter 
subgroup coordinators C.~Goy and G.~Moultaka for having supported this 
collaboration. F.-S. Ling has funding from the Belgian FNRS. This work 
was partly supported by grant ANR-06-BLAN-0172.

\bibliography{lavalle_bib}

\begin{thebibliography}{95}
\expandafter\ifx\csname natexlab\endcsname\relax\def\natexlab#1{#1}\fi
\expandafter\ifx\csname bibnamefont\endcsname\relax
  \def\bibnamefont#1{#1}\fi
\expandafter\ifx\csname bibfnamefont\endcsname\relax
  \def\bibfnamefont#1{#1}\fi
\expandafter\ifx\csname citenamefont\endcsname\relax
  \def\citenamefont#1{#1}\fi
\expandafter\ifx\csname url\endcsname\relax
  \def\url#1{\texttt{#1}}\fi
\expandafter\ifx\csname urlprefix\endcsname\relax\def\urlprefix{URL }\fi
\providecommand{\bibinfo}[2]{#2}
\providecommand{\eprint}[2][]{\url{#2}}

\bibitem[{\citenamefont{{Murayama}}(2007)}]{review_dm_murayama_07}
\bibinfo{author}{\bibfnamefont{H.}~\bibnamefont{{Murayama}}}
  (\bibinfo{year}{2007}), \eprint{arXiv:0704.2276}.

\bibitem[{\citenamefont{{Primack}}(2007)}]{2007NuPhS.173....1P}
\bibinfo{author}{\bibfnamefont{J.~R.} \bibnamefont{{Primack}}},
  \bibinfo{journal}{Nuclear Physics B Proceedings Supplements}
  \textbf{\bibinfo{volume}{173}}, \bibinfo{pages}{1} (\bibinfo{year}{2007}),
  \eprint{arXiv:astro-ph/0609541}.

\bibitem[{\citenamefont{{Jungman} et~al.}(1996)\citenamefont{{Jungman},
  {Kamionkowski}, and {Griest}}}]{susy_dm_jungman_etal_96}
\bibinfo{author}{\bibfnamefont{G.}~\bibnamefont{{Jungman}}},
  \bibinfo{author}{\bibfnamefont{M.}~\bibnamefont{{Kamionkowski}}},
  \bibnamefont{and} \bibinfo{author}{\bibfnamefont{K.}~\bibnamefont{{Griest}}},
  \bibinfo{journal}{Phys. Rept.} \textbf{\bibinfo{volume}{267}},
  \bibinfo{pages}{195} (\bibinfo{year}{1996}), \eprint{arXiv:hep-ph/9506380}.

\bibitem[{\citenamefont{{Bergstr\"om}}(2000)}]{review_dm_bergstrom_00}
\bibinfo{author}{\bibfnamefont{L.}~\bibnamefont{{Bergstr\"om}}},
  \bibinfo{journal}{Reports of Progress in Physics}
  \textbf{\bibinfo{volume}{63}}, \bibinfo{pages}{793} (\bibinfo{year}{2000}),
  \eprint{arXiv:hep-ph/0002126}.

\bibitem[{\citenamefont{{Bertone} et~al.}(2005)\citenamefont{{Bertone},
  {Hooper}, and {Silk}}}]{review_dm_bertone_etal_05}
\bibinfo{author}{\bibfnamefont{G.}~\bibnamefont{{Bertone}}},
  \bibinfo{author}{\bibfnamefont{D.}~\bibnamefont{{Hooper}}}, \bibnamefont{and}
  \bibinfo{author}{\bibfnamefont{J.}~\bibnamefont{{Silk}}},
  \bibinfo{journal}{Phys. Rept.} \textbf{\bibinfo{volume}{405}},
  \bibinfo{pages}{279} (\bibinfo{year}{2005}), \eprint{arXiv:hep-ph/0404175}.

\bibitem[{\citenamefont{{Carr} et~al.}(2006)\citenamefont{{Carr}, {Lamanna},
  and {Lavalle}}}]{review_dm_carr_etal_06}
\bibinfo{author}{\bibfnamefont{J.}~\bibnamefont{{Carr}}},
  \bibinfo{author}{\bibfnamefont{G.}~\bibnamefont{{Lamanna}}},
  \bibnamefont{and}
  \bibinfo{author}{\bibfnamefont{J.}~\bibnamefont{{Lavalle}}},
  \bibinfo{journal}{Reports of Progress in Physics}
  \textbf{\bibinfo{volume}{69}}, \bibinfo{pages}{2475} (\bibinfo{year}{2006}).

\bibitem[{\citenamefont{{Mambrini} et~al.}(2006)\citenamefont{{Mambrini},
  {Mu{\~n}oz}, and {Nezri}}}]{2006JCAP...12..003M}
\bibinfo{author}{\bibfnamefont{Y.}~\bibnamefont{{Mambrini}}},
  \bibinfo{author}{\bibfnamefont{C.}~\bibnamefont{{Mu{\~n}oz}}},
  \bibnamefont{and} \bibinfo{author}{\bibfnamefont{E.}~\bibnamefont{{Nezri}}},
  \bibinfo{journal}{Journal of Cosmology and Astro-Particle Physics}
  \textbf{\bibinfo{volume}{12}}, \bibinfo{pages}{3} (\bibinfo{year}{2006}),
  \eprint{arXiv:hep-ph/0607266}.

\bibitem[{\citenamefont{{Gunn} et~al.}(1978)\citenamefont{{Gunn}, {Lee},
  {Lerche}, {Schramm}, and {Steigman}}}]{1978ApJ...223.1015G}
\bibinfo{author}{\bibfnamefont{J.~E.} \bibnamefont{{Gunn}}},
  \bibinfo{author}{\bibfnamefont{B.~W.} \bibnamefont{{Lee}}},
  \bibinfo{author}{\bibfnamefont{I.}~\bibnamefont{{Lerche}}},
  \bibinfo{author}{\bibfnamefont{D.~N.} \bibnamefont{{Schramm}}},
  \bibnamefont{and}
  \bibinfo{author}{\bibfnamefont{G.}~\bibnamefont{{Steigman}}},
  \bibinfo{journal}{\apj} \textbf{\bibinfo{volume}{223}}, \bibinfo{pages}{1015}
  (\bibinfo{year}{1978}).

\bibitem[{\citenamefont{{Silk} and {Srednicki}}(1984)}]{1984PhRvL..53..624S}
\bibinfo{author}{\bibfnamefont{J.}~\bibnamefont{{Silk}}} \bibnamefont{and}
  \bibinfo{author}{\bibfnamefont{M.}~\bibnamefont{{Srednicki}}},
  \bibinfo{journal}{Physical Review Letters} \textbf{\bibinfo{volume}{53}},
  \bibinfo{pages}{624} (\bibinfo{year}{1984}).

\bibitem[{\citenamefont{{Stoehr} et~al.}(2003)\citenamefont{{Stoehr}, {White},
  {Springel}, {Tormen}, and {Yoshida}}}]{2003MNRAS.345.1313S}
\bibinfo{author}{\bibfnamefont{F.}~\bibnamefont{{Stoehr}}},
  \bibinfo{author}{\bibfnamefont{S.~D.~M.} \bibnamefont{{White}}},
  \bibinfo{author}{\bibfnamefont{V.}~\bibnamefont{{Springel}}},
  \bibinfo{author}{\bibfnamefont{G.}~\bibnamefont{{Tormen}}}, \bibnamefont{and}
  \bibinfo{author}{\bibfnamefont{N.}~\bibnamefont{{Yoshida}}},
  \bibinfo{journal}{Mon. Not. R. Astron. Soc.} \textbf{\bibinfo{volume}{345}},
  \bibinfo{pages}{1313} (\bibinfo{year}{2003}),
  \eprint{arXiv:astro-ph/0307026}.

\bibitem[{\citenamefont{{Diemand}
  et~al.}(2007{\natexlab{a}})\citenamefont{{Diemand}, {Kuhlen}, and
  {Madau}}}]{2007ApJ...657..262D}
\bibinfo{author}{\bibfnamefont{J.}~\bibnamefont{{Diemand}}},
  \bibinfo{author}{\bibfnamefont{M.}~\bibnamefont{{Kuhlen}}}, \bibnamefont{and}
  \bibinfo{author}{\bibfnamefont{P.}~\bibnamefont{{Madau}}},
  \bibinfo{journal}{\apj} \textbf{\bibinfo{volume}{657}}, \bibinfo{pages}{262}
  (\bibinfo{year}{2007}{\natexlab{a}}).

\bibitem[{\citenamefont{{Athanassoula}
  et~al.}(2008)\citenamefont{{Athanassoula}, {Ling}, {Nezri}, and
  {Teyssier}}}]{2008arXiv0801.4673A}
\bibinfo{author}{\bibfnamefont{E.}~\bibnamefont{{Athanassoula}}},
  \bibinfo{author}{\bibfnamefont{F.~.} \bibnamefont{{Ling}}},
  \bibinfo{author}{\bibfnamefont{E.}~\bibnamefont{{Nezri}}}, \bibnamefont{and}
  \bibinfo{author}{\bibfnamefont{R.}~\bibnamefont{{Teyssier}}},
  \bibinfo{journal}{ArXiv e-prints} \textbf{\bibinfo{volume}{801}}
  (\bibinfo{year}{2008}), \eprint{0801.4673}.

\bibitem[{\citenamefont{{Kuhlen} et~al.}(2008)\citenamefont{{Kuhlen},
  {Diemand}, and {Madau}}}]{2008arXiv0805.4416K}
\bibinfo{author}{\bibfnamefont{M.}~\bibnamefont{{Kuhlen}}},
  \bibinfo{author}{\bibfnamefont{J.}~\bibnamefont{{Diemand}}},
  \bibnamefont{and} \bibinfo{author}{\bibfnamefont{P.}~\bibnamefont{{Madau}}},
  \bibinfo{journal}{ArXiv e-prints} \textbf{\bibinfo{volume}{805}}
  (\bibinfo{year}{2008}), \eprint{0805.4416}.

\bibitem[{\citenamefont{{Bergstr{\"o}m}
  et~al.}(1999{\natexlab{a}})\citenamefont{{Bergstr{\"o}m}, {Edsj{\"o}},
  {Gondolo}, and {Ullio}}}]{1999PhRvD..59d3506B}
\bibinfo{author}{\bibfnamefont{L.}~\bibnamefont{{Bergstr{\"o}m}}},
  \bibinfo{author}{\bibfnamefont{J.}~\bibnamefont{{Edsj{\"o}}}},
  \bibinfo{author}{\bibfnamefont{P.}~\bibnamefont{{Gondolo}}},
  \bibnamefont{and} \bibinfo{author}{\bibfnamefont{P.}~\bibnamefont{{Ullio}}},
  \bibinfo{journal}{\prd} \textbf{\bibinfo{volume}{59}},
  \bibinfo{pages}{043506} (\bibinfo{year}{1999}{\natexlab{a}}),
  \eprint{arXiv:astro-ph/9806072}.

\bibitem[{\citenamefont{{Ullio} et~al.}(2002)\citenamefont{{Ullio},
  {Bergstr{\"o}m}, {Edsj{\"o}}, and {Lacey}}}]{2002PhRvD..66l3502U}
\bibinfo{author}{\bibfnamefont{P.}~\bibnamefont{{Ullio}}},
  \bibinfo{author}{\bibfnamefont{L.}~\bibnamefont{{Bergstr{\"o}m}}},
  \bibinfo{author}{\bibfnamefont{J.}~\bibnamefont{{Edsj{\"o}}}},
  \bibnamefont{and} \bibinfo{author}{\bibfnamefont{C.}~\bibnamefont{{Lacey}}},
  \bibinfo{journal}{\prd} \textbf{\bibinfo{volume}{66}},
  \bibinfo{pages}{123502} (\bibinfo{year}{2002}),
  \eprint{arXiv:astro-ph/0207125}.

\bibitem[{\citenamefont{{Berezinsky} et~al.}(2003)\citenamefont{{Berezinsky},
  {Dokuchaev}, and {Eroshenko}}}]{berezinsky_etal_03}
\bibinfo{author}{\bibfnamefont{V.}~\bibnamefont{{Berezinsky}}},
  \bibinfo{author}{\bibfnamefont{V.}~\bibnamefont{{Dokuchaev}}},
  \bibnamefont{and}
  \bibinfo{author}{\bibfnamefont{Y.}~\bibnamefont{{Eroshenko}}},
  \bibinfo{journal}{\prd} \textbf{\bibinfo{volume}{68}},
  \bibinfo{pages}{103003} (\bibinfo{year}{2003}),
  \eprint{arXiv:astro-ph/0301551}.

\bibitem[{\citenamefont{{Berezinsky} et~al.}(2006)\citenamefont{{Berezinsky},
  {Dokuchaev}, and {Eroshenko}}}]{berezinsky_etal_06}
\bibinfo{author}{\bibfnamefont{V.}~\bibnamefont{{Berezinsky}}},
  \bibinfo{author}{\bibfnamefont{V.}~\bibnamefont{{Dokuchaev}}},
  \bibnamefont{and}
  \bibinfo{author}{\bibfnamefont{Y.}~\bibnamefont{{Eroshenko}}},
  \bibinfo{journal}{\prd} \textbf{\bibinfo{volume}{73}},
  \bibinfo{pages}{063504} (\bibinfo{year}{2006}),
  \eprint{arXiv:astro-ph/0511494}.

\bibitem[{\citenamefont{{Pieri} et~al.}(2008)\citenamefont{{Pieri}, {Bertone},
  and {Branchini}}}]{2008MNRAS.384.1627P}
\bibinfo{author}{\bibfnamefont{L.}~\bibnamefont{{Pieri}}},
  \bibinfo{author}{\bibfnamefont{G.}~\bibnamefont{{Bertone}}},
  \bibnamefont{and}
  \bibinfo{author}{\bibfnamefont{E.}~\bibnamefont{{Branchini}}},
  \bibinfo{journal}{Mon. Not. R. Astron. Soc.} \textbf{\bibinfo{volume}{384}},
  \bibinfo{pages}{1627} (\bibinfo{year}{2008}), \eprint{arXiv:0706.2101}.

\bibitem[{\citenamefont{{Lavalle} et~al.}(2007)\citenamefont{{Lavalle},
  {Pochon}, {Salati}, and {Taillet}}}]{2007A&A...462..827L}
\bibinfo{author}{\bibfnamefont{J.}~\bibnamefont{{Lavalle}}},
  \bibinfo{author}{\bibfnamefont{J.}~\bibnamefont{{Pochon}}},
  \bibinfo{author}{\bibfnamefont{P.}~\bibnamefont{{Salati}}}, \bibnamefont{and}
  \bibinfo{author}{\bibfnamefont{R.}~\bibnamefont{{Taillet}}},
  \bibinfo{journal}{A\&A} \textbf{\bibinfo{volume}{462}}, \bibinfo{pages}{827}
  (\bibinfo{year}{2007}).

\bibitem[{\citenamefont{{Lavalle} et~al.}(2008)\citenamefont{{Lavalle}, {Yuan},
  {Maurin}, and {Bi}}}]{2008A&A...479..427L}
\bibinfo{author}{\bibfnamefont{J.}~\bibnamefont{{Lavalle}}},
  \bibinfo{author}{\bibfnamefont{Q.}~\bibnamefont{{Yuan}}},
  \bibinfo{author}{\bibfnamefont{D.}~\bibnamefont{{Maurin}}}, \bibnamefont{and}
  \bibinfo{author}{\bibfnamefont{X.-J.} \bibnamefont{{Bi}}},
  \bibinfo{journal}{A\&A} \textbf{\bibinfo{volume}{479}}, \bibinfo{pages}{427}
  (\bibinfo{year}{2008}), \eprint{arXiv:0709.3634}.

\bibitem[{\citenamefont{{Donato} et~al.}(2004)\citenamefont{{Donato},
  {Fornengo}, {Maurin}, {Salati}, and {Taillet}}}]{2004PhRvD..69f3501D}
\bibinfo{author}{\bibfnamefont{F.}~\bibnamefont{{Donato}}},
  \bibinfo{author}{\bibfnamefont{N.}~\bibnamefont{{Fornengo}}},
  \bibinfo{author}{\bibfnamefont{D.}~\bibnamefont{{Maurin}}},
  \bibinfo{author}{\bibfnamefont{P.}~\bibnamefont{{Salati}}}, \bibnamefont{and}
  \bibinfo{author}{\bibfnamefont{R.}~\bibnamefont{{Taillet}}},
  \bibinfo{journal}{\prd} \textbf{\bibinfo{volume}{69}},
  \bibinfo{pages}{063501} (\bibinfo{year}{2004}).

\bibitem[{\citenamefont{{Lionetto} et~al.}(2005)\citenamefont{{Lionetto},
  {Morselli}, and {Zdravkovic}}}]{2005JCAP...09..010L}
\bibinfo{author}{\bibfnamefont{A.~M.} \bibnamefont{{Lionetto}}},
  \bibinfo{author}{\bibfnamefont{A.}~\bibnamefont{{Morselli}}},
  \bibnamefont{and}
  \bibinfo{author}{\bibfnamefont{V.}~\bibnamefont{{Zdravkovic}}},
  \bibinfo{journal}{Journal of Cosmology and Astro-Particle Physics}
  \textbf{\bibinfo{volume}{9}}, \bibinfo{pages}{10} (\bibinfo{year}{2005}).

\bibitem[{\citenamefont{{Delahaye}
  et~al.}(2008{\natexlab{a}})\citenamefont{{Delahaye}, {Lineros}, {Donato},
  {Fornengo}, and {Salati}}}]{2008PhRvD..77f3527D}
\bibinfo{author}{\bibfnamefont{T.}~\bibnamefont{{Delahaye}}},
  \bibinfo{author}{\bibfnamefont{R.}~\bibnamefont{{Lineros}}},
  \bibinfo{author}{\bibfnamefont{F.}~\bibnamefont{{Donato}}},
  \bibinfo{author}{\bibfnamefont{N.}~\bibnamefont{{Fornengo}}},
  \bibnamefont{and} \bibinfo{author}{\bibfnamefont{P.}~\bibnamefont{{Salati}}},
  \bibinfo{journal}{\prd} \textbf{\bibinfo{volume}{77}},
  \bibinfo{pages}{063527} (\bibinfo{year}{2008}{\natexlab{a}}),
  \eprint{0712.2312}.

\bibitem[{\citenamefont{{Diemand} et~al.}(2008)\citenamefont{{Diemand},
  {Kuhlen}, {Madau}, {Zemp}, {Moore}, {Potter}, and
  {Stadel}}}]{2008arXiv0805.1244D}
\bibinfo{author}{\bibfnamefont{J.}~\bibnamefont{{Diemand}}},
  \bibinfo{author}{\bibfnamefont{M.}~\bibnamefont{{Kuhlen}}},
  \bibinfo{author}{\bibfnamefont{P.}~\bibnamefont{{Madau}}},
  \bibinfo{author}{\bibfnamefont{M.}~\bibnamefont{{Zemp}}},
  \bibinfo{author}{\bibfnamefont{B.}~\bibnamefont{{Moore}}},
  \bibinfo{author}{\bibfnamefont{D.}~\bibnamefont{{Potter}}}, \bibnamefont{and}
  \bibinfo{author}{\bibfnamefont{J.}~\bibnamefont{{Stadel}}},
  \bibinfo{journal}{ArXiv e-prints} \textbf{\bibinfo{volume}{805}}
  (\bibinfo{year}{2008}), \eprint{0805.1244}.

\bibitem[{\citenamefont{{Teyssier}}(2002)}]{2002A&A...385..337T}
\bibinfo{author}{\bibfnamefont{R.}~\bibnamefont{{Teyssier}}},
  \bibinfo{journal}{A\&A} \textbf{\bibinfo{volume}{385}}, \bibinfo{pages}{337}
  (\bibinfo{year}{2002}), \eprint{arXiv:astro-ph/0111367}.

\bibitem[{\citenamefont{{Casertano} and {Hut}}(1985)}]{1985ApJ...298...80C}
\bibinfo{author}{\bibfnamefont{S.}~\bibnamefont{{Casertano}}} \bibnamefont{and}
  \bibinfo{author}{\bibfnamefont{P.}~\bibnamefont{{Hut}}},
  \bibinfo{journal}{\apj} \textbf{\bibinfo{volume}{298}}, \bibinfo{pages}{80}
  (\bibinfo{year}{1985}).

\bibitem[{\citenamefont{{Delahaye}
  et~al.}(2008{\natexlab{b}})\citenamefont{{Delahaye}, {Donato}, {Fornengo},
  {Lavalle}, {Lineros}, {Salati}, {Taillet}, and {.}}}]{2008arXiv0809.5268D}
\bibinfo{author}{\bibfnamefont{T.}~\bibnamefont{{Delahaye}}},
  \bibinfo{author}{\bibfnamefont{F.}~\bibnamefont{{Donato}}},
  \bibinfo{author}{\bibfnamefont{N.}~\bibnamefont{{Fornengo}}},
  \bibinfo{author}{\bibfnamefont{J.}~\bibnamefont{{Lavalle}}},
  \bibinfo{author}{\bibfnamefont{R.}~\bibnamefont{{Lineros}}},
  \bibinfo{author}{\bibfnamefont{P.}~\bibnamefont{{Salati}}},
  \bibinfo{author}{\bibfnamefont{R.}~\bibnamefont{{Taillet}}},
  \bibnamefont{and} \bibinfo{author}{\bibnamefont{{.}}},
  \bibinfo{journal}{ArXiv e-prints}  (\bibinfo{year}{2008}{\natexlab{b}}),
  \eprint{arXiv:0809.5268}.

\bibitem[{\citenamefont{{Donato} et~al.}(2001)\citenamefont{{Donato}, {Maurin},
  {Salati}, {Barrau}, {Boudoul}, and {Taillet}}}]{2001ApJ...563..172D}
\bibinfo{author}{\bibfnamefont{F.}~\bibnamefont{{Donato}}},
  \bibinfo{author}{\bibfnamefont{D.}~\bibnamefont{{Maurin}}},
  \bibinfo{author}{\bibfnamefont{P.}~\bibnamefont{{Salati}}},
  \bibinfo{author}{\bibfnamefont{A.}~\bibnamefont{{Barrau}}},
  \bibinfo{author}{\bibfnamefont{G.}~\bibnamefont{{Boudoul}}},
  \bibnamefont{and}
  \bibinfo{author}{\bibfnamefont{R.}~\bibnamefont{{Taillet}}},
  \bibinfo{journal}{\apj} \textbf{\bibinfo{volume}{563}}, \bibinfo{pages}{172}
  (\bibinfo{year}{2001}).

\bibitem[{\citenamefont{{Strong} and {Moskalenko}}(1998)}]{1998ApJ...509..212S}
\bibinfo{author}{\bibfnamefont{A.~W.} \bibnamefont{{Strong}}} \bibnamefont{and}
  \bibinfo{author}{\bibfnamefont{I.~V.} \bibnamefont{{Moskalenko}}},
  \bibinfo{journal}{\apj} \textbf{\bibinfo{volume}{509}}, \bibinfo{pages}{212}
  (\bibinfo{year}{1998}), \eprint{arXiv:astro-ph/9807150}.

\bibitem[{\citenamefont{{Berezinskii} et~al.}(1990)\citenamefont{{Berezinskii},
  {Bulanov}, {Dogiel}, and {Ptuskin}}}]{berezinsky_book_90}
\bibinfo{author}{\bibfnamefont{V.~S.} \bibnamefont{{Berezinskii}}},
  \bibinfo{author}{\bibfnamefont{S.~V.} \bibnamefont{{Bulanov}}},
  \bibinfo{author}{\bibfnamefont{V.~A.} \bibnamefont{{Dogiel}}},
  \bibnamefont{and} \bibinfo{author}{\bibfnamefont{V.~S.}
  \bibnamefont{{Ptuskin}}}, \emph{\bibinfo{title}{{Astrophysics of cosmic
  rays}}} (\bibinfo{publisher}{Amsterdam: North-Holland, 1990, edited by
  Ginzburg, V.L.}, \bibinfo{year}{1990}).

\bibitem[{\citenamefont{{Maurin} et~al.}(2001)\citenamefont{{Maurin}, {Donato},
  {Taillet}, and {Salati}}}]{maurin_etal_01}
\bibinfo{author}{\bibfnamefont{D.}~\bibnamefont{{Maurin}}},
  \bibinfo{author}{\bibfnamefont{F.}~\bibnamefont{{Donato}}},
  \bibinfo{author}{\bibfnamefont{R.}~\bibnamefont{{Taillet}}},
  \bibnamefont{and} \bibinfo{author}{\bibfnamefont{P.}~\bibnamefont{{Salati}}},
  \bibinfo{journal}{\apj} \textbf{\bibinfo{volume}{555}}, \bibinfo{pages}{585}
  (\bibinfo{year}{2001}).

\bibitem[{\citenamefont{{Maurin} et~al.}(2002)\citenamefont{{Maurin},
  {Taillet}, and {Donato}}}]{2002A&A...394.1039M}
\bibinfo{author}{\bibfnamefont{D.}~\bibnamefont{{Maurin}}},
  \bibinfo{author}{\bibfnamefont{R.}~\bibnamefont{{Taillet}}},
  \bibnamefont{and} \bibinfo{author}{\bibfnamefont{F.}~\bibnamefont{{Donato}}},
  \bibinfo{journal}{Astronomy and Astrophys.} \textbf{\bibinfo{volume}{394}},
  \bibinfo{pages}{1039} (\bibinfo{year}{2002}).

\bibitem[{\citenamefont{{Donato} et~al.}(2002)\citenamefont{{Donato}, {Maurin},
  and {Taillet}}}]{2002A&A...381..539D}
\bibinfo{author}{\bibfnamefont{F.}~\bibnamefont{{Donato}}},
  \bibinfo{author}{\bibfnamefont{D.}~\bibnamefont{{Maurin}}}, \bibnamefont{and}
  \bibinfo{author}{\bibfnamefont{R.}~\bibnamefont{{Taillet}}},
  \bibinfo{journal}{Astronomy and Astrophys.} \textbf{\bibinfo{volume}{381}},
  \bibinfo{pages}{539} (\bibinfo{year}{2002}).

\bibitem[{\citenamefont{{Perko}}(1987)}]{1987A&A...184..119P}
\bibinfo{author}{\bibfnamefont{J.~S.} \bibnamefont{{Perko}}},
  \bibinfo{journal}{A\&A} \textbf{\bibinfo{volume}{184}}, \bibinfo{pages}{119}
  (\bibinfo{year}{1987}).

\bibitem[{\citenamefont{{Maurin} et~al.}(2004)\citenamefont{{Maurin},
  {Taillet}, {Donato}, {Salati}, {Barrau}, and
  {Boudoul}}}]{2002astro.ph.12111M}
\bibinfo{author}{\bibfnamefont{D.}~\bibnamefont{{Maurin}}},
  \bibinfo{author}{\bibfnamefont{R.}~\bibnamefont{{Taillet}}},
  \bibinfo{author}{\bibfnamefont{F.}~\bibnamefont{{Donato}}},
  \bibinfo{author}{\bibfnamefont{P.}~\bibnamefont{{Salati}}},
  \bibinfo{author}{\bibfnamefont{A.}~\bibnamefont{{Barrau}}}, \bibnamefont{and}
  \bibinfo{author}{\bibfnamefont{G.}~\bibnamefont{{Boudoul}}},
  \bibinfo{journal}{Research Signposts, Recent Research Developments in
  Astronomy and Astrophys.} \textbf{\bibinfo{volume}{2}}, \bibinfo{pages}{193}
  (\bibinfo{year}{2004}), \eprint{arXiv:astro-ph/0212111}.

\bibitem[{\citenamefont{{Barrau} et~al.}(2002)\citenamefont{{Barrau},
  {Boudoul}, {Donato}, {Maurin}, {Salati}, and
  {Taillet}}}]{2002A&A...388..676B}
\bibinfo{author}{\bibfnamefont{A.}~\bibnamefont{{Barrau}}},
  \bibinfo{author}{\bibfnamefont{G.}~\bibnamefont{{Boudoul}}},
  \bibinfo{author}{\bibfnamefont{F.}~\bibnamefont{{Donato}}},
  \bibinfo{author}{\bibfnamefont{D.}~\bibnamefont{{Maurin}}},
  \bibinfo{author}{\bibfnamefont{P.}~\bibnamefont{{Salati}}}, \bibnamefont{and}
  \bibinfo{author}{\bibfnamefont{R.}~\bibnamefont{{Taillet}}},
  \bibinfo{journal}{A\&A} \textbf{\bibinfo{volume}{388}}, \bibinfo{pages}{676}
  (\bibinfo{year}{2002}), \eprint{arXiv:astro-ph/0112486}.

\bibitem[{\citenamefont{{Barrau} et~al.}(2005)\citenamefont{{Barrau}, {Salati},
  {Servant}, {Donato}, {Grain}, {Maurin}, and {Taillet}}}]{2005PhRvD..72f3507B}
\bibinfo{author}{\bibfnamefont{A.}~\bibnamefont{{Barrau}}},
  \bibinfo{author}{\bibfnamefont{P.}~\bibnamefont{{Salati}}},
  \bibinfo{author}{\bibfnamefont{G.}~\bibnamefont{{Servant}}},
  \bibinfo{author}{\bibfnamefont{F.}~\bibnamefont{{Donato}}},
  \bibinfo{author}{\bibfnamefont{J.}~\bibnamefont{{Grain}}},
  \bibinfo{author}{\bibfnamefont{D.}~\bibnamefont{{Maurin}}}, \bibnamefont{and}
  \bibinfo{author}{\bibfnamefont{R.}~\bibnamefont{{Taillet}}},
  \bibinfo{journal}{\prd} \textbf{\bibinfo{volume}{72}},
  \bibinfo{pages}{063507} (\bibinfo{year}{2005}).

\bibitem[{\citenamefont{{Bringmann} and {Salati}}(2007)}]{2007PhRvD..75h3006B}
\bibinfo{author}{\bibfnamefont{T.}~\bibnamefont{{Bringmann}}} \bibnamefont{and}
  \bibinfo{author}{\bibfnamefont{P.}~\bibnamefont{{Salati}}},
  \bibinfo{journal}{\prd} \textbf{\bibinfo{volume}{75}},
  \bibinfo{pages}{083006} (\bibinfo{year}{2007}).

\bibitem[{\citenamefont{{Brun} et~al.}(2007)\citenamefont{{Brun}, {Bertone},
  {Lavalle}, {Salati}, and {Taillet}}}]{brun_etal_07}
\bibinfo{author}{\bibfnamefont{P.}~\bibnamefont{{Brun}}},
  \bibinfo{author}{\bibfnamefont{G.}~\bibnamefont{{Bertone}}},
  \bibinfo{author}{\bibfnamefont{J.}~\bibnamefont{{Lavalle}}},
  \bibinfo{author}{\bibfnamefont{P.}~\bibnamefont{{Salati}}}, \bibnamefont{and}
  \bibinfo{author}{\bibfnamefont{R.}~\bibnamefont{{Taillet}}},
  \bibinfo{journal}{\prd} \textbf{\bibinfo{volume}{76}},
  \bibinfo{pages}{083506} (\bibinfo{year}{2007}), \eprint{arXiv:0704.2543}.

\bibitem[{\citenamefont{{Bulanov} and {Dogel}}(1974)}]{1974Ap&SS..29..305B}
\bibinfo{author}{\bibfnamefont{S.~V.} \bibnamefont{{Bulanov}}}
  \bibnamefont{and} \bibinfo{author}{\bibfnamefont{V.~A.}
  \bibnamefont{{Dogel}}}, \bibinfo{journal}{Astrophysics and Space Science}
  \textbf{\bibinfo{volume}{29}}, \bibinfo{pages}{305} (\bibinfo{year}{1974}).

\bibitem[{\citenamefont{{Baltz} and {Edsj{\"o}}}(1999)}]{1999PhRvD..59b3511B}
\bibinfo{author}{\bibfnamefont{E.~A.} \bibnamefont{{Baltz}}} \bibnamefont{and}
  \bibinfo{author}{\bibfnamefont{J.}~\bibnamefont{{Edsj{\"o}}}},
  \bibinfo{journal}{\prd} \textbf{\bibinfo{volume}{59}},
  \bibinfo{pages}{023511} (\bibinfo{year}{1999}), \eprint{astro-ph/9808243}.

\bibitem[{\citenamefont{{Taillet} and {Maurin}}(2003)}]{2003A&A...402..971T}
\bibinfo{author}{\bibfnamefont{R.}~\bibnamefont{{Taillet}}} \bibnamefont{and}
  \bibinfo{author}{\bibfnamefont{D.}~\bibnamefont{{Maurin}}},
  \bibinfo{journal}{A\&A} \textbf{\bibinfo{volume}{402}}, \bibinfo{pages}{971}
  (\bibinfo{year}{2003}), \eprint{astro-ph/0212112}.

\bibitem[{\citenamefont{{Sj{\"o}strand}
  et~al.}(2006)\citenamefont{{Sj{\"o}strand}, {Mrenna}, and
  {Skands}}}]{2006JHEP...05..026S}
\bibinfo{author}{\bibfnamefont{T.}~\bibnamefont{{Sj{\"o}strand}}},
  \bibinfo{author}{\bibfnamefont{S.}~\bibnamefont{{Mrenna}}}, \bibnamefont{and}
  \bibinfo{author}{\bibfnamefont{P.}~\bibnamefont{{Skands}}},
  \bibinfo{journal}{Journal of High Energy Physics}
  \textbf{\bibinfo{volume}{5}}, \bibinfo{pages}{26} (\bibinfo{year}{2006}),
  \eprint{arXiv:hep-ph/0603175}.

\bibitem[{\citenamefont{{B{\'e}langer}
  et~al.}(2006)\citenamefont{{B{\'e}langer}, {Boudjema}, {Pukhov}, and
  {Semenov}}}]{2006CoPhC.174..577B}
\bibinfo{author}{\bibfnamefont{G.}~\bibnamefont{{B{\'e}langer}}},
  \bibinfo{author}{\bibfnamefont{F.}~\bibnamefont{{Boudjema}}},
  \bibinfo{author}{\bibfnamefont{A.}~\bibnamefont{{Pukhov}}}, \bibnamefont{and}
  \bibinfo{author}{\bibfnamefont{A.}~\bibnamefont{{Semenov}}},
  \bibinfo{journal}{Computer Physics Communications}
  \textbf{\bibinfo{volume}{174}}, \bibinfo{pages}{577} (\bibinfo{year}{2006}),
  \eprint{arXiv:hep-ph/0405253}.

\bibitem[{\citenamefont{Fayet and Ferrara}(1977)}]{Fayet:1976cr}
\bibinfo{author}{\bibfnamefont{P.}~\bibnamefont{Fayet}} \bibnamefont{and}
  \bibinfo{author}{\bibfnamefont{S.}~\bibnamefont{Ferrara}},
  \bibinfo{journal}{Phys. Rept.} \textbf{\bibinfo{volume}{32}},
  \bibinfo{pages}{249} (\bibinfo{year}{1977}).

\bibitem[{\citenamefont{Haber and Kane}(1985)}]{Haber:1984rc}
\bibinfo{author}{\bibfnamefont{H.~E.} \bibnamefont{Haber}} \bibnamefont{and}
  \bibinfo{author}{\bibfnamefont{G.~L.} \bibnamefont{Kane}},
  \bibinfo{journal}{Phys. Rept.} \textbf{\bibinfo{volume}{117}},
  \bibinfo{pages}{75} (\bibinfo{year}{1985}).

\bibitem[{\citenamefont{{Ellis} et~al.}(1983)\citenamefont{{Ellis},
  {Nanopoulos}, and {Tamvakis}}}]{1983PhLB..121..123E}
\bibinfo{author}{\bibfnamefont{J.}~\bibnamefont{{Ellis}}},
  \bibinfo{author}{\bibfnamefont{D.~V.} \bibnamefont{{Nanopoulos}}},
  \bibnamefont{and}
  \bibinfo{author}{\bibfnamefont{K.}~\bibnamefont{{Tamvakis}}},
  \bibinfo{journal}{Physics Letters B} \textbf{\bibinfo{volume}{121}},
  \bibinfo{pages}{123} (\bibinfo{year}{1983}).

\bibitem[{\citenamefont{{Kane} et~al.}(1994)\citenamefont{{Kane}, {Kolda},
  {Roszkowski}, and {Wells}}}]{1994PhRvD..49.6173K}
\bibinfo{author}{\bibfnamefont{G.~L.} \bibnamefont{{Kane}}},
  \bibinfo{author}{\bibfnamefont{C.}~\bibnamefont{{Kolda}}},
  \bibinfo{author}{\bibfnamefont{L.}~\bibnamefont{{Roszkowski}}},
  \bibnamefont{and} \bibinfo{author}{\bibfnamefont{J.~D.}
  \bibnamefont{{Wells}}}, \bibinfo{journal}{\prd}
  \textbf{\bibinfo{volume}{49}}, \bibinfo{pages}{6173} (\bibinfo{year}{1994}),
  \eprint{arXiv:hep-ph/9312272}.

\bibitem[{\citenamefont{{Bottino} et~al.}(1998)\citenamefont{{Bottino},
  {Donato}, {Fornengo}, and {Salati}}}]{1998PhRvD..58l3503B}
\bibinfo{author}{\bibfnamefont{A.}~\bibnamefont{{Bottino}}},
  \bibinfo{author}{\bibfnamefont{F.}~\bibnamefont{{Donato}}},
  \bibinfo{author}{\bibfnamefont{N.}~\bibnamefont{{Fornengo}}},
  \bibnamefont{and} \bibinfo{author}{\bibfnamefont{P.}~\bibnamefont{{Salati}}},
  \bibinfo{journal}{\prd} \textbf{\bibinfo{volume}{58}},
  \bibinfo{pages}{123503} (\bibinfo{year}{1998}),
  \eprint{arXiv:astro-ph/9804137}.

\bibitem[{\citenamefont{{Bergstr{\"o}m}
  et~al.}(1999{\natexlab{b}})\citenamefont{{Bergstr{\"o}m}, {Edsj{\"o}}, and
  {Ullio}}}]{1999ApJ...526..215B}
\bibinfo{author}{\bibfnamefont{L.}~\bibnamefont{{Bergstr{\"o}m}}},
  \bibinfo{author}{\bibfnamefont{J.}~\bibnamefont{{Edsj{\"o}}}},
  \bibnamefont{and} \bibinfo{author}{\bibfnamefont{P.}~\bibnamefont{{Ullio}}},
  \bibinfo{journal}{\apj} \textbf{\bibinfo{volume}{526}}, \bibinfo{pages}{215}
  (\bibinfo{year}{1999}{\natexlab{b}}), \eprint{arXiv:astro-ph/9902012}.

\bibitem[{\citenamefont{{Allanach} et~al.}(2002)\citenamefont{{Allanach},
  {Battaglia}, {Blair}, {Carena}, {de Roeck}, {Dedes}, {Djouadi}, {Gerdes},
  {Ghodbane}, {Gunion} et~al.}}]{2002EPJC...25..113A}
\bibinfo{author}{\bibfnamefont{B.~C.} \bibnamefont{{Allanach}}},
  \bibinfo{author}{\bibfnamefont{M.}~\bibnamefont{{Battaglia}}},
  \bibinfo{author}{\bibfnamefont{G.~A.} \bibnamefont{{Blair}}},
  \bibinfo{author}{\bibfnamefont{M.}~\bibnamefont{{Carena}}},
  \bibinfo{author}{\bibfnamefont{A.}~\bibnamefont{{de Roeck}}},
  \bibinfo{author}{\bibfnamefont{A.}~\bibnamefont{{Dedes}}},
  \bibinfo{author}{\bibfnamefont{A.}~\bibnamefont{{Djouadi}}},
  \bibinfo{author}{\bibfnamefont{D.}~\bibnamefont{{Gerdes}}},
  \bibinfo{author}{\bibfnamefont{N.}~\bibnamefont{{Ghodbane}}},
  \bibinfo{author}{\bibfnamefont{J.}~\bibnamefont{{Gunion}}},
  \bibnamefont{et~al.}, \bibinfo{journal}{European Physical Journal C}
  \textbf{\bibinfo{volume}{25}}, \bibinfo{pages}{113} (\bibinfo{year}{2002}),
  \eprint{arXiv:hep-ph/0202233}.

\bibitem[{\citenamefont{{Bottino} et~al.}(2005)\citenamefont{{Bottino},
  {Donato}, {Fornengo}, and {Salati}}}]{2005PhRvD..72h3518B}
\bibinfo{author}{\bibfnamefont{A.}~\bibnamefont{{Bottino}}},
  \bibinfo{author}{\bibfnamefont{F.}~\bibnamefont{{Donato}}},
  \bibinfo{author}{\bibfnamefont{N.}~\bibnamefont{{Fornengo}}},
  \bibnamefont{and} \bibinfo{author}{\bibfnamefont{P.}~\bibnamefont{{Salati}}},
  \bibinfo{journal}{\prd} \textbf{\bibinfo{volume}{72}},
  \bibinfo{pages}{083518} (\bibinfo{year}{2005}),
  \eprint{arXiv:hep-ph/0507086}.

\bibitem[{\citenamefont{{Bottino} et~al.}(2008)\citenamefont{{Bottino},
  {Fornengo}, {Polesello}, and {Scopel}}}]{2008PhRvD..77k5026B}
\bibinfo{author}{\bibfnamefont{A.}~\bibnamefont{{Bottino}}},
  \bibinfo{author}{\bibfnamefont{N.}~\bibnamefont{{Fornengo}}},
  \bibinfo{author}{\bibfnamefont{G.}~\bibnamefont{{Polesello}}},
  \bibnamefont{and} \bibinfo{author}{\bibfnamefont{S.}~\bibnamefont{{Scopel}}},
  \bibinfo{journal}{\prd} \textbf{\bibinfo{volume}{77}},
  \bibinfo{pages}{115026} (\bibinfo{year}{2008}), \eprint{arXiv:0801.3334}.

\bibitem[{\citenamefont{{Appelquist} et~al.}(2001)\citenamefont{{Appelquist},
  {Cheng}, and {Dobrescu}}}]{2001PhRvD..64c5002A}
\bibinfo{author}{\bibfnamefont{T.}~\bibnamefont{{Appelquist}}},
  \bibinfo{author}{\bibfnamefont{H.-C.} \bibnamefont{{Cheng}}},
  \bibnamefont{and} \bibinfo{author}{\bibfnamefont{B.~A.}
  \bibnamefont{{Dobrescu}}}, \bibinfo{journal}{\prd}
  \textbf{\bibinfo{volume}{64}}, \bibinfo{pages}{035002}
  (\bibinfo{year}{2001}), \eprint{arXiv:hep-ph/0012100}.

\bibitem[{\citenamefont{{Randall} and
  {Sundrum}}(1999{\natexlab{a}})}]{1999PhRvL..83.4690R}
\bibinfo{author}{\bibfnamefont{L.}~\bibnamefont{{Randall}}} \bibnamefont{and}
  \bibinfo{author}{\bibfnamefont{R.}~\bibnamefont{{Sundrum}}},
  \bibinfo{journal}{Physical Review Letters} \textbf{\bibinfo{volume}{83}},
  \bibinfo{pages}{4690} (\bibinfo{year}{1999}{\natexlab{a}}),
  \eprint{arXiv:hep-th/9906064}.

\bibitem[{\citenamefont{{Randall} and
  {Sundrum}}(1999{\natexlab{b}})}]{1999PhRvL..83.3370R}
\bibinfo{author}{\bibfnamefont{L.}~\bibnamefont{{Randall}}} \bibnamefont{and}
  \bibinfo{author}{\bibfnamefont{R.}~\bibnamefont{{Sundrum}}},
  \bibinfo{journal}{Physical Review Letters} \textbf{\bibinfo{volume}{83}},
  \bibinfo{pages}{3370} (\bibinfo{year}{1999}{\natexlab{b}}),
  \eprint{arXiv:hep-ph/9905221}.

\bibitem[{\citenamefont{{Servant} and {Tait}}(2003)}]{2003NuPhB.650..391S}
\bibinfo{author}{\bibfnamefont{G.}~\bibnamefont{{Servant}}} \bibnamefont{and}
  \bibinfo{author}{\bibfnamefont{T.~M.~P.} \bibnamefont{{Tait}}},
  \bibinfo{journal}{Nuclear Physics B} \textbf{\bibinfo{volume}{650}},
  \bibinfo{pages}{391} (\bibinfo{year}{2003}), \eprint{arXiv:hep-ph/0206071}.

\bibitem[{\citenamefont{{Bertone} et~al.}(2003)\citenamefont{{Bertone},
  {Servant}, and {Sigl}}}]{2003PhRvD..68d4008B}
\bibinfo{author}{\bibfnamefont{G.}~\bibnamefont{{Bertone}}},
  \bibinfo{author}{\bibfnamefont{G.}~\bibnamefont{{Servant}}},
  \bibnamefont{and} \bibinfo{author}{\bibfnamefont{G.}~\bibnamefont{{Sigl}}},
  \bibinfo{journal}{\prd} \textbf{\bibinfo{volume}{68}},
  \bibinfo{pages}{044008} (\bibinfo{year}{2003}),
  \eprint{arXiv:hep-ph/0211342}.

\bibitem[{\citenamefont{{Agashe} and {Servant}}(2004)}]{2004PhRvL..93w1805A}
\bibinfo{author}{\bibfnamefont{K.}~\bibnamefont{{Agashe}}} \bibnamefont{and}
  \bibinfo{author}{\bibfnamefont{G.}~\bibnamefont{{Servant}}},
  \bibinfo{journal}{Physical Review Letters} \textbf{\bibinfo{volume}{93}},
  \bibinfo{pages}{231805} (\bibinfo{year}{2004}),
  \eprint{arXiv:hep-ph/0403143}.

\bibitem[{\citenamefont{{Agashe} and {Servant}}(2005)}]{2005JCAP...02..002A}
\bibinfo{author}{\bibfnamefont{K.}~\bibnamefont{{Agashe}}} \bibnamefont{and}
  \bibinfo{author}{\bibfnamefont{G.}~\bibnamefont{{Servant}}},
  \bibinfo{journal}{Journal of Cosmology and Astro-Particle Physics}
  \textbf{\bibinfo{volume}{2}}, \bibinfo{pages}{2} (\bibinfo{year}{2005}),
  \eprint{arXiv:hep-ph/0411254}.

\bibitem[{\citenamefont{{Hooper} and {Servant}}(2005)}]{lzp_hooper_servant_05}
\bibinfo{author}{\bibfnamefont{D.}~\bibnamefont{{Hooper}}} \bibnamefont{and}
  \bibinfo{author}{\bibfnamefont{G.}~\bibnamefont{{Servant}}},
  \bibinfo{journal}{Astroparticle Physics} \textbf{\bibinfo{volume}{24}},
  \bibinfo{pages}{231} (\bibinfo{year}{2005}), \eprint{arXiv:hep-ph/0502247}.

\bibitem[{\citenamefont{Barbieri et~al.}(2006)\citenamefont{Barbieri, Hall, and
  Rychkov}}]{Barbieri:2006dq}
\bibinfo{author}{\bibfnamefont{R.}~\bibnamefont{Barbieri}},
  \bibinfo{author}{\bibfnamefont{L.~J.} \bibnamefont{Hall}}, \bibnamefont{and}
  \bibinfo{author}{\bibfnamefont{V.~S.} \bibnamefont{Rychkov}},
  \bibinfo{journal}{Phys. Rev.} \textbf{\bibinfo{volume}{D74}},
  \bibinfo{pages}{015007} (\bibinfo{year}{2006}), \eprint{hep-ph/0603188}.

\bibitem[{\citenamefont{Lopez~Honorez et~al.}(2007)\citenamefont{Lopez~Honorez,
  Nezri, Oliver, and Tytgat}}]{LopezHonorez:2006gr}
\bibinfo{author}{\bibfnamefont{L.}~\bibnamefont{Lopez~Honorez}},
  \bibinfo{author}{\bibfnamefont{E.}~\bibnamefont{Nezri}},
  \bibinfo{author}{\bibfnamefont{J.~F.} \bibnamefont{Oliver}},
  \bibnamefont{and} \bibinfo{author}{\bibfnamefont{M.~H.~G.}
  \bibnamefont{Tytgat}}, \bibinfo{journal}{JCAP}
  \textbf{\bibinfo{volume}{0702}}, \bibinfo{pages}{028} (\bibinfo{year}{2007}),
  \eprint{hep-ph/0612275}.

\bibitem[{\citenamefont{Arkani-Hamed et~al.}(2001)\citenamefont{Arkani-Hamed,
  Cohen, and Georgi}}]{arkani_01}
\bibinfo{author}{\bibfnamefont{N.}~\bibnamefont{Arkani-Hamed}},
  \bibinfo{author}{\bibfnamefont{A.~G.} \bibnamefont{Cohen}}, \bibnamefont{and}
  \bibinfo{author}{\bibfnamefont{H.}~\bibnamefont{Georgi}},
  \bibinfo{journal}{Phys. Lett. B} \textbf{\bibinfo{volume}{513}},
  \bibinfo{pages}{232} (\bibinfo{year}{2001}), \eprint{hep-ph/0105239}.

\bibitem[{\citenamefont{{Schmaltz} and {Tucker-Smith}}(2005)}]{schmaltz_05}
\bibinfo{author}{\bibfnamefont{M.}~\bibnamefont{{Schmaltz}}} \bibnamefont{and}
  \bibinfo{author}{\bibfnamefont{D.}~\bibnamefont{{Tucker-Smith}}},
  \bibinfo{journal}{Ann. Rev. Nucl. Part. Sci.} \textbf{\bibinfo{volume}{55}},
  \bibinfo{pages}{229} (\bibinfo{year}{2005}), \eprint{arXiv:hep-ph/0502182}.

\bibitem[{\citenamefont{{Cheng} and {Low}}(2003)}]{cheng_03}
\bibinfo{author}{\bibfnamefont{H.-C.} \bibnamefont{{Cheng}}} \bibnamefont{and}
  \bibinfo{author}{\bibfnamefont{I.}~\bibnamefont{{Low}}},
  \bibinfo{journal}{JHEP} \textbf{\bibinfo{volume}{09}}, \bibinfo{pages}{051}
  (\bibinfo{year}{2003}), \eprint{arXiv:hep-ph/0308199}.

\bibitem[{\citenamefont{{Han} et~al.}(2003)\citenamefont{{Han}, {Logan},
  {McElrath}, and {Wang}}}]{han_03}
\bibinfo{author}{\bibfnamefont{T.}~\bibnamefont{{Han}}},
  \bibinfo{author}{\bibfnamefont{H.~E.} \bibnamefont{{Logan}}},
  \bibinfo{author}{\bibfnamefont{B.}~\bibnamefont{{McElrath}}},
  \bibnamefont{and} \bibinfo{author}{\bibfnamefont{L.-T.}
  \bibnamefont{{Wang}}}, \bibinfo{journal}{\prd} \textbf{\bibinfo{volume}{67}},
  \bibinfo{pages}{095004} (\bibinfo{year}{2003}),
  \eprint{arXiv:hep-ph/0301040}.

\bibitem[{\citenamefont{{Birkedal} et~al.}(2006)\citenamefont{{Birkedal},
  {Noble}, {Perelstein}, and {Spray}}}]{birkedal_06}
\bibinfo{author}{\bibfnamefont{A.}~\bibnamefont{{Birkedal}}},
  \bibinfo{author}{\bibfnamefont{A.}~\bibnamefont{{Noble}}},
  \bibinfo{author}{\bibfnamefont{M.}~\bibnamefont{{Perelstein}}},
  \bibnamefont{and} \bibinfo{author}{\bibfnamefont{A.}~\bibnamefont{{Spray}}},
  \bibinfo{journal}{\prd} \textbf{\bibinfo{volume}{74}},
  \bibinfo{pages}{035002} (\bibinfo{year}{2006}),
  \eprint{arXiv:hep-ph/0603077}.

\bibitem[{\citenamefont{{Asano} et~al.}(2007)\citenamefont{{Asano},
  {Matsumoto}, {Okada}, and {Okada}}}]{asano_07}
\bibinfo{author}{\bibfnamefont{M.}~\bibnamefont{{Asano}}},
  \bibinfo{author}{\bibfnamefont{S.}~\bibnamefont{{Matsumoto}}},
  \bibinfo{author}{\bibfnamefont{N.}~\bibnamefont{{Okada}}}, \bibnamefont{and}
  \bibinfo{author}{\bibfnamefont{Y.}~\bibnamefont{{Okada}}},
  \bibinfo{journal}{\prd} \textbf{\bibinfo{volume}{75}},
  \bibinfo{pages}{063506} (\bibinfo{year}{2007}),
  \eprint{arXiv:hep-ph/0602157}.

\bibitem[{\citenamefont{{Dehnen}}(2002)}]{2002ASPC..275..105D}
\bibinfo{author}{\bibfnamefont{W.}~\bibnamefont{{Dehnen}}}, in
  \emph{\bibinfo{booktitle}{Disks of Galaxies: Kinematics, Dynamics and
  Peturbations}}, edited by
  \bibinfo{editor}{\bibfnamefont{E.}~\bibnamefont{{Athanassoula}}},
  \bibinfo{editor}{\bibfnamefont{A.}~\bibnamefont{{Bosma}}}, \bibnamefont{and}
  \bibinfo{editor}{\bibfnamefont{R.}~\bibnamefont{{Mujica}}}
  (\bibinfo{year}{2002}), vol. \bibinfo{volume}{275} of
  \emph{\bibinfo{series}{Astronomical Society of the Pacific Conference
  Series}}, pp. \bibinfo{pages}{105--116}.

\bibitem[{\citenamefont{{Bissantz} and {Gerhard}}(2002)}]{2002MNRAS.330..591B}
\bibinfo{author}{\bibfnamefont{N.}~\bibnamefont{{Bissantz}}} \bibnamefont{and}
  \bibinfo{author}{\bibfnamefont{O.}~\bibnamefont{{Gerhard}}},
  \bibinfo{journal}{Mon. Not. R. Astron. Soc.} \textbf{\bibinfo{volume}{330}},
  \bibinfo{pages}{591} (\bibinfo{year}{2002}), \eprint{arXiv:astro-ph/0110368}.

\bibitem[{\citenamefont{{Athanassoula}}(2007)}]{bar_lia_07}
\bibinfo{author}{\bibfnamefont{E.}~\bibnamefont{{Athanassoula}}},
  \bibinfo{journal}{Mon. Not. R. Astron. Soc.} p. \bibinfo{pages}{339} (\bibinfo{year}{2007}),
  \eprint{arXiv:astro-ph/0703184}.

\bibitem[{\citenamefont{{Athanassoula}
  et~al.}(2005)\citenamefont{{Athanassoula}, {Ling}, and
  {Nezri}}}]{non_spher_halo_lia_fu_manu_05}
\bibinfo{author}{\bibfnamefont{E.}~\bibnamefont{{Athanassoula}}},
  \bibinfo{author}{\bibfnamefont{F.-S.} \bibnamefont{{Ling}}},
  \bibnamefont{and} \bibinfo{author}{\bibfnamefont{E.}~\bibnamefont{{Nezri}}},
  \bibinfo{journal}{\prd} \textbf{\bibinfo{volume}{72}},
  \bibinfo{pages}{083503} (\bibinfo{year}{2005}), \eprint{astro-ph/0504631}.

\bibitem[{\citenamefont{{Boezio} et~al.}(2000)\citenamefont{{Boezio},
  {Carlson}, {Francke}, {Weber}, {Suffert}, {Hof}, {Menn}, {Simon}, {Stephens},
  {Bellotti} et~al.}}]{2000ApJ...532..653B}
\bibinfo{author}{\bibfnamefont{M.}~\bibnamefont{{Boezio}}},
  \bibinfo{author}{\bibfnamefont{P.}~\bibnamefont{{Carlson}}},
  \bibinfo{author}{\bibfnamefont{T.}~\bibnamefont{{Francke}}},
  \bibinfo{author}{\bibfnamefont{N.}~\bibnamefont{{Weber}}},
  \bibinfo{author}{\bibfnamefont{M.}~\bibnamefont{{Suffert}}},
  \bibinfo{author}{\bibfnamefont{M.}~\bibnamefont{{Hof}}},
  \bibinfo{author}{\bibfnamefont{W.}~\bibnamefont{{Menn}}},
  \bibinfo{author}{\bibfnamefont{M.}~\bibnamefont{{Simon}}},
  \bibinfo{author}{\bibfnamefont{S.~A.} \bibnamefont{{Stephens}}},
  \bibinfo{author}{\bibfnamefont{R.}~\bibnamefont{{Bellotti}}},
  \bibnamefont{et~al.}, \bibinfo{journal}{\apj} \textbf{\bibinfo{volume}{532}},
  \bibinfo{pages}{653} (\bibinfo{year}{2000}).

\bibitem[{\citenamefont{{DuVernois} et~al.}(2001)\citenamefont{{DuVernois},
  {Barwick}, {Beatty}, {Bhattacharyya}, {Bower}, {Chaput}, {Coutu}, {de Nolfo},
  {Lowder}, {McKee} et~al.}}]{2001ApJ...559..296D}
\bibinfo{author}{\bibfnamefont{M.~A.} \bibnamefont{{DuVernois}}},
  \bibinfo{author}{\bibfnamefont{S.~W.} \bibnamefont{{Barwick}}},
  \bibinfo{author}{\bibfnamefont{J.~J.} \bibnamefont{{Beatty}}},
  \bibinfo{author}{\bibfnamefont{A.}~\bibnamefont{{Bhattacharyya}}},
  \bibinfo{author}{\bibfnamefont{C.~R.} \bibnamefont{{Bower}}},
  \bibinfo{author}{\bibfnamefont{C.~J.} \bibnamefont{{Chaput}}},
  \bibinfo{author}{\bibfnamefont{S.}~\bibnamefont{{Coutu}}},
  \bibinfo{author}{\bibfnamefont{G.~A.} \bibnamefont{{de Nolfo}}},
  \bibinfo{author}{\bibfnamefont{D.~M.} \bibnamefont{{Lowder}}},
  \bibinfo{author}{\bibfnamefont{S.}~\bibnamefont{{McKee}}},
  \bibnamefont{et~al.}, \bibinfo{journal}{\apj} \textbf{\bibinfo{volume}{559}},
  \bibinfo{pages}{296} (\bibinfo{year}{2001}).

\bibitem[{\citenamefont{{AMS Collaboration} et~al.}(2002)\citenamefont{{AMS
  Collaboration}, {Aguilar}, {Alcaraz}, {Allaby}, {Alpat}, {Ambrosi},
  {Anderhub}, {Ao}, {Arefiev}, {Azzarello} et~al.}}]{2002PhR...366..331A}
\bibinfo{author}{\bibnamefont{{AMS Collaboration}}},
  \bibinfo{author}{\bibfnamefont{M.}~\bibnamefont{{Aguilar}}},
  \bibinfo{author}{\bibfnamefont{J.}~\bibnamefont{{Alcaraz}}},
  \bibinfo{author}{\bibfnamefont{J.}~\bibnamefont{{Allaby}}},
  \bibinfo{author}{\bibfnamefont{B.}~\bibnamefont{{Alpat}}},
  \bibinfo{author}{\bibfnamefont{G.}~\bibnamefont{{Ambrosi}}},
  \bibinfo{author}{\bibfnamefont{H.}~\bibnamefont{{Anderhub}}},
  \bibinfo{author}{\bibfnamefont{L.}~\bibnamefont{{Ao}}},
  \bibinfo{author}{\bibfnamefont{A.}~\bibnamefont{{Arefiev}}},
  \bibinfo{author}{\bibfnamefont{P.}~\bibnamefont{{Azzarello}}},
  \bibnamefont{et~al.}, \bibinfo{journal}{Phys. Rept.}
  \textbf{\bibinfo{volume}{366}}, \bibinfo{pages}{331} (\bibinfo{year}{2002}).

\bibitem[{\citenamefont{{Orito} et~al.}(2000)\citenamefont{{Orito}, {Maeno},
  {Matsunaga}, {Abe}, {Anraku}, {Asaoka}, {Fujikawa}, {Imori}, {Ishino},
  {Makida} et~al.}}]{2000PhRvL..84.1078O}
\bibinfo{author}{\bibfnamefont{S.}~\bibnamefont{{Orito}}},
  \bibinfo{author}{\bibfnamefont{T.}~\bibnamefont{{Maeno}}},
  \bibinfo{author}{\bibfnamefont{H.}~\bibnamefont{{Matsunaga}}},
  \bibinfo{author}{\bibfnamefont{K.}~\bibnamefont{{Abe}}},
  \bibinfo{author}{\bibfnamefont{K.}~\bibnamefont{{Anraku}}},
  \bibinfo{author}{\bibfnamefont{Y.}~\bibnamefont{{Asaoka}}},
  \bibinfo{author}{\bibfnamefont{M.}~\bibnamefont{{Fujikawa}}},
  \bibinfo{author}{\bibfnamefont{M.}~\bibnamefont{{Imori}}},
  \bibinfo{author}{\bibfnamefont{M.}~\bibnamefont{{Ishino}}},
  \bibinfo{author}{\bibfnamefont{Y.}~\bibnamefont{{Makida}}},
  \bibnamefont{et~al.}, \bibinfo{journal}{Physical Review Letters}
  \textbf{\bibinfo{volume}{84}}, \bibinfo{pages}{1078} (\bibinfo{year}{2000}),
  \eprint{arXiv:astro-ph/9906426}.

\bibitem[{\citenamefont{{Maeno} et~al.}(2001)\citenamefont{{Maeno}, {Orito},
  {Matsunaga}, {Abe}, {Anraku}, {Asaoka}, {Fujikawa}, {Imori}, {Makida},
  {Matsui} et~al.}}]{2001APh....16..121M}
\bibinfo{author}{\bibfnamefont{T.}~\bibnamefont{{Maeno}}},
  \bibinfo{author}{\bibfnamefont{S.}~\bibnamefont{{Orito}}},
  \bibinfo{author}{\bibfnamefont{H.}~\bibnamefont{{Matsunaga}}},
  \bibinfo{author}{\bibfnamefont{K.}~\bibnamefont{{Abe}}},
  \bibinfo{author}{\bibfnamefont{K.}~\bibnamefont{{Anraku}}},
  \bibinfo{author}{\bibfnamefont{Y.}~\bibnamefont{{Asaoka}}},
  \bibinfo{author}{\bibfnamefont{M.}~\bibnamefont{{Fujikawa}}},
  \bibinfo{author}{\bibfnamefont{M.}~\bibnamefont{{Imori}}},
  \bibinfo{author}{\bibfnamefont{Y.}~\bibnamefont{{Makida}}},
  \bibinfo{author}{\bibfnamefont{N.}~\bibnamefont{{Matsui}}},
  \bibnamefont{et~al.}, \bibinfo{journal}{Astroparticle Physics}
  \textbf{\bibinfo{volume}{16}}, \bibinfo{pages}{121} (\bibinfo{year}{2001}),
  \eprint{arXiv:astro-ph/0010381}.

\bibitem[{\citenamefont{{Asaoka} et~al.}(2002)\citenamefont{{Asaoka},
  {Shikaze}, {Abe}, {Anraku}, {Fujikawa}, {Fuke}, {Haino}, {Imori}, {Izumi},
  {Maeno} et~al.}}]{2002PhRvL..88e1101A}
\bibinfo{author}{\bibfnamefont{Y.}~\bibnamefont{{Asaoka}}},
  \bibinfo{author}{\bibfnamefont{Y.}~\bibnamefont{{Shikaze}}},
  \bibinfo{author}{\bibfnamefont{K.}~\bibnamefont{{Abe}}},
  \bibinfo{author}{\bibfnamefont{K.}~\bibnamefont{{Anraku}}},
  \bibinfo{author}{\bibfnamefont{M.}~\bibnamefont{{Fujikawa}}},
  \bibinfo{author}{\bibfnamefont{H.}~\bibnamefont{{Fuke}}},
  \bibinfo{author}{\bibfnamefont{S.}~\bibnamefont{{Haino}}},
  \bibinfo{author}{\bibfnamefont{M.}~\bibnamefont{{Imori}}},
  \bibinfo{author}{\bibfnamefont{K.}~\bibnamefont{{Izumi}}},
  \bibinfo{author}{\bibfnamefont{T.}~\bibnamefont{{Maeno}}},
  \bibnamefont{et~al.}, \bibinfo{journal}{Physical Review Letters}
  \textbf{\bibinfo{volume}{88}}, \bibinfo{pages}{051101}
  (\bibinfo{year}{2002}), \eprint{arXiv:astro-ph/0109007}.

\bibitem[{\citenamefont{{Haino} and {et al.}}(2005)}]{2005ICRC....3...13H}
\bibinfo{author}{\bibfnamefont{S.}~\bibnamefont{{Haino}}} \bibnamefont{and}
  \bibinfo{author}{\bibnamefont{{et al.}}}, in
  \emph{\bibinfo{booktitle}{International Cosmic Ray Conference}}
  (\bibinfo{year}{2005}), vol.~\bibinfo{volume}{3} of
  \emph{\bibinfo{series}{International Cosmic Ray Conference}}, pp.
  \bibinfo{pages}{13--+}.

\bibitem[{\citenamefont{{Bullock} et~al.}(2001)\citenamefont{{Bullock},
  {Kolatt}, {Sigad}, {Somerville}, {Kravtsov}, {Klypin}, {Primack}, and
  {Dekel}}}]{bullock_etal_01}
\bibinfo{author}{\bibfnamefont{J.~S.} \bibnamefont{{Bullock}}},
  \bibinfo{author}{\bibfnamefont{T.~S.} \bibnamefont{{Kolatt}}},
  \bibinfo{author}{\bibfnamefont{Y.}~\bibnamefont{{Sigad}}},
  \bibinfo{author}{\bibfnamefont{R.~S.} \bibnamefont{{Somerville}}},
  \bibinfo{author}{\bibfnamefont{A.~V.} \bibnamefont{{Kravtsov}}},
  \bibinfo{author}{\bibfnamefont{A.~A.} \bibnamefont{{Klypin}}},
  \bibinfo{author}{\bibfnamefont{J.~R.} \bibnamefont{{Primack}}},
  \bibnamefont{and} \bibinfo{author}{\bibfnamefont{A.}~\bibnamefont{{Dekel}}},
  \bibinfo{journal}{Mon. Not. R. Astron. Soc.} \textbf{\bibinfo{volume}{321}},
  \bibinfo{pages}{559} (\bibinfo{year}{2001}), \eprint{astro-ph/9908159}.

\bibitem[{\citenamefont{{Diemand}
  et~al.}(2007{\natexlab{b}})\citenamefont{{Diemand}, {Kuhlen}, and
  {Madau}}}]{2007ApJ...667..859D}
\bibinfo{author}{\bibfnamefont{J.}~\bibnamefont{{Diemand}}},
  \bibinfo{author}{\bibfnamefont{M.}~\bibnamefont{{Kuhlen}}}, \bibnamefont{and}
  \bibinfo{author}{\bibfnamefont{P.}~\bibnamefont{{Madau}}},
  \bibinfo{journal}{\apj} \textbf{\bibinfo{volume}{667}}, \bibinfo{pages}{859}
  (\bibinfo{year}{2007}{\natexlab{b}}), \eprint{arXiv:astro-ph/0703337}.

\bibitem[{\citenamefont{{Moskalenko} and {Strong}}(1998)}]{1998ApJ...493..694M}
\bibinfo{author}{\bibfnamefont{I.~V.} \bibnamefont{{Moskalenko}}}
  \bibnamefont{and} \bibinfo{author}{\bibfnamefont{A.~W.}
  \bibnamefont{{Strong}}}, \bibinfo{journal}{\apj}
  \textbf{\bibinfo{volume}{493}}, \bibinfo{pages}{694} (\bibinfo{year}{1998}),
  \eprint{astro-ph/9710124}.

\bibitem[{\citenamefont{{Barwick} et~al.}(1997)\citenamefont{{Barwick},
  {Beatty}, {Bhattacharyya}, {Bower}, {Chaput}, {Coutu}, {de Nolfo}, {Knapp},
  {Lowder}, {McKee} et~al.}}]{1997ApJ...482L.191B}
\bibinfo{author}{\bibfnamefont{S.~W.} \bibnamefont{{Barwick}}},
  \bibinfo{author}{\bibfnamefont{J.~J.} \bibnamefont{{Beatty}}},
  \bibinfo{author}{\bibfnamefont{A.}~\bibnamefont{{Bhattacharyya}}},
  \bibinfo{author}{\bibfnamefont{C.~R.} \bibnamefont{{Bower}}},
  \bibinfo{author}{\bibfnamefont{C.~J.} \bibnamefont{{Chaput}}},
  \bibinfo{author}{\bibfnamefont{S.}~\bibnamefont{{Coutu}}},
  \bibinfo{author}{\bibfnamefont{G.~A.} \bibnamefont{{de Nolfo}}},
  \bibinfo{author}{\bibfnamefont{J.}~\bibnamefont{{Knapp}}},
  \bibinfo{author}{\bibfnamefont{D.~M.} \bibnamefont{{Lowder}}},
  \bibinfo{author}{\bibfnamefont{S.}~\bibnamefont{{McKee}}},
  \bibnamefont{et~al.}, \bibinfo{journal}{Astrophysical Journal Letters}
  \textbf{\bibinfo{volume}{482}}, \bibinfo{pages}{L191+}
  (\bibinfo{year}{1997}), \eprint{arXiv:astro-ph/9703192}.

\bibitem[{\citenamefont{{Beatty} et~al.}(2004)\citenamefont{{Beatty},
  {Bhattacharyya}, {Bower}, {Coutu}, {Duvernois}, {McKee}, {Minnick},
  {M{\"u}ller}, {Musser}, {Nutter} et~al.}}]{2004PhRvL..93x1102B}
\bibinfo{author}{\bibfnamefont{J.~J.} \bibnamefont{{Beatty}}},
  \bibinfo{author}{\bibfnamefont{A.}~\bibnamefont{{Bhattacharyya}}},
  \bibinfo{author}{\bibfnamefont{C.}~\bibnamefont{{Bower}}},
  \bibinfo{author}{\bibfnamefont{S.}~\bibnamefont{{Coutu}}},
  \bibinfo{author}{\bibfnamefont{M.~A.} \bibnamefont{{Duvernois}}},
  \bibinfo{author}{\bibfnamefont{S.}~\bibnamefont{{McKee}}},
  \bibinfo{author}{\bibfnamefont{S.~A.} \bibnamefont{{Minnick}}},
  \bibinfo{author}{\bibfnamefont{D.}~\bibnamefont{{M{\"u}ller}}},
  \bibinfo{author}{\bibfnamefont{J.}~\bibnamefont{{Musser}}},
  \bibinfo{author}{\bibfnamefont{S.}~\bibnamefont{{Nutter}}},
  \bibnamefont{et~al.}, \bibinfo{journal}{Physical Review Letters}
  \textbf{\bibinfo{volume}{93}}, \bibinfo{pages}{241102}
  (\bibinfo{year}{2004}), \eprint{arXiv:astro-ph/0412230}.

\bibitem[{\citenamefont{{AMS-01 Collaboration}
  et~al.}(2007)\citenamefont{{AMS-01 Collaboration}, {Aguilar}, {Alcaraz},
  {Allaby}, {Alpat}, {Ambrosi}, {Anderhub}, {Ao}, {Arefiev}, {Azzarello}
  et~al.}}]{2007PhLB..646..145A}
\bibinfo{author}{\bibnamefont{{AMS-01 Collaboration}}},
  \bibinfo{author}{\bibfnamefont{M.}~\bibnamefont{{Aguilar}}},
  \bibinfo{author}{\bibfnamefont{J.}~\bibnamefont{{Alcaraz}}},
  \bibinfo{author}{\bibfnamefont{J.}~\bibnamefont{{Allaby}}},
  \bibinfo{author}{\bibfnamefont{B.}~\bibnamefont{{Alpat}}},
  \bibinfo{author}{\bibfnamefont{G.}~\bibnamefont{{Ambrosi}}},
  \bibinfo{author}{\bibfnamefont{H.}~\bibnamefont{{Anderhub}}},
  \bibinfo{author}{\bibfnamefont{L.}~\bibnamefont{{Ao}}},
  \bibinfo{author}{\bibfnamefont{A.}~\bibnamefont{{Arefiev}}},
  \bibinfo{author}{\bibfnamefont{P.}~\bibnamefont{{Azzarello}}},
  \bibnamefont{et~al.}, \bibinfo{journal}{Physics Letters B}
  \textbf{\bibinfo{volume}{646}}, \bibinfo{pages}{145} (\bibinfo{year}{2007}),
  \eprint{arXiv:astro-ph/0703154}.

\bibitem[{\citenamefont{{Casadei} and {Bindi}}(2004)}]{2004ApJ...612..262C}
\bibinfo{author}{\bibfnamefont{D.}~\bibnamefont{{Casadei}}} \bibnamefont{and}
  \bibinfo{author}{\bibfnamefont{V.}~\bibnamefont{{Bindi}}},
  \bibinfo{journal}{\apj} \textbf{\bibinfo{volume}{612}}, \bibinfo{pages}{262}
  (\bibinfo{year}{2004}).

\bibitem[{\citenamefont{{Picozza} et~al.}(2008)}]{picozza_blois_08}
\bibinfo{author}{\bibfnamefont{P.}~\bibnamefont{{Picozza}}}
  \bibnamefont{et~al.}, \bibinfo{journal}{Rencontres de Blois 2008, Challenges
  in Particle Astrophysics}  (\bibinfo{year}{2008}).

\bibitem[{\citenamefont{{Cirelli} et~al.}(2008)\citenamefont{{Cirelli},
  {Kadastik}, {Raidal}, and {Strumia}}}]{2008arXiv0809.2409C}
\bibinfo{author}{\bibfnamefont{M.}~\bibnamefont{{Cirelli}}},
  \bibinfo{author}{\bibfnamefont{M.}~\bibnamefont{{Kadastik}}},
  \bibinfo{author}{\bibfnamefont{M.}~\bibnamefont{{Raidal}}}, \bibnamefont{and}
  \bibinfo{author}{\bibfnamefont{A.}~\bibnamefont{{Strumia}}},
  \bibinfo{journal}{ArXiv e-prints}  (\bibinfo{year}{2008}),
  \eprint{0809.2409}.

\bibitem[{\citenamefont{{Casolino} et~al.}(2007)\citenamefont{{Casolino},
  {Picozza}, {Altamura}, {Basili}, {De Simone}, {Di Felice}, {De Pascale},
  {Marcelli}, {Minori}, {Nagni} et~al.}}]{2007arXiv0708.1808C}
\bibinfo{author}{\bibfnamefont{M.}~\bibnamefont{{Casolino}}},
  \bibinfo{author}{\bibfnamefont{P.}~\bibnamefont{{Picozza}}},
  \bibinfo{author}{\bibfnamefont{F.}~\bibnamefont{{Altamura}}},
  \bibinfo{author}{\bibfnamefont{A.}~\bibnamefont{{Basili}}},
  \bibinfo{author}{\bibfnamefont{N.}~\bibnamefont{{De Simone}}},
  \bibinfo{author}{\bibfnamefont{V.}~\bibnamefont{{Di Felice}}},
  \bibinfo{author}{\bibfnamefont{M.~P.} \bibnamefont{{De Pascale}}},
  \bibinfo{author}{\bibfnamefont{L.}~\bibnamefont{{Marcelli}}},
  \bibinfo{author}{\bibfnamefont{M.}~\bibnamefont{{Minori}}},
  \bibinfo{author}{\bibfnamefont{M.}~\bibnamefont{{Nagni}}},
  \bibnamefont{et~al.}, \bibinfo{journal}{ArXiv e-prints}
  (\bibinfo{year}{2007}), \eprint{arXiv:0708.1808}.

\bibitem[{\citenamefont{{Moiseev} et~al.}(2007)\citenamefont{{Moiseev},
  {Ormes}, and {Moskalenko}}}]{2007arXiv0706.0882M}
\bibinfo{author}{\bibfnamefont{A.~A.} \bibnamefont{{Moiseev}}},
  \bibinfo{author}{\bibfnamefont{J.~F.} \bibnamefont{{Ormes}}},
  \bibnamefont{and} \bibinfo{author}{\bibfnamefont{I.~V.}
  \bibnamefont{{Moskalenko}}}, \bibinfo{journal}{ArXiv e-prints}
  \textbf{\bibinfo{volume}{706}} (\bibinfo{year}{2007}), \eprint{0706.0882}.

\bibitem[{\citenamefont{{Diemand} et~al.}(2005)\citenamefont{{Diemand},
  {Moore}, and {Stadel}}}]{2005Natur.433..389D}
\bibinfo{author}{\bibfnamefont{J.}~\bibnamefont{{Diemand}}},
  \bibinfo{author}{\bibfnamefont{B.}~\bibnamefont{{Moore}}}, \bibnamefont{and}
  \bibinfo{author}{\bibfnamefont{J.}~\bibnamefont{{Stadel}}},
  \bibinfo{journal}{\nat} \textbf{\bibinfo{volume}{433}}, \bibinfo{pages}{389}
  (\bibinfo{year}{2005}), \eprint{astro-ph/0501589}.

\bibitem[{\citenamefont{{Gentile} et~al.}(2004)\citenamefont{{Gentile},
  {Salucci}, {Klein}, {Vergani}, and {Kalberla}}}]{2004MNRAS.351..903G}
\bibinfo{author}{\bibfnamefont{G.}~\bibnamefont{{Gentile}}},
  \bibinfo{author}{\bibfnamefont{P.}~\bibnamefont{{Salucci}}},
  \bibinfo{author}{\bibfnamefont{U.}~\bibnamefont{{Klein}}},
  \bibinfo{author}{\bibfnamefont{D.}~\bibnamefont{{Vergani}}},
  \bibnamefont{and}
  \bibinfo{author}{\bibfnamefont{P.}~\bibnamefont{{Kalberla}}},
  \bibinfo{journal}{Mon. Not. R. Astron. Soc.} \textbf{\bibinfo{volume}{351}},
  \bibinfo{pages}{903} (\bibinfo{year}{2004}), \eprint{arXiv:astro-ph/0403154}.

\bibitem[{\citenamefont{{Englmaier} and {Gerhard}}(2006)}]{2006CeMDA..94..369E}
\bibinfo{author}{\bibfnamefont{P.}~\bibnamefont{{Englmaier}}} \bibnamefont{and}
  \bibinfo{author}{\bibfnamefont{O.}~\bibnamefont{{Gerhard}}},
  \bibinfo{journal}{Celestial Mechanics and Dynamical Astronomy}
  \textbf{\bibinfo{volume}{94}}, \bibinfo{pages}{369} (\bibinfo{year}{2006}),
  \eprint{arXiv:astro-ph/0601679}.

\bibitem[{\citenamefont{{Bosma}}(2004)}]{2004IAUS..220...39B}
\bibinfo{author}{\bibfnamefont{A.}~\bibnamefont{{Bosma}}}, in
  \emph{\bibinfo{booktitle}{Dark Matter in Galaxies}}, edited by
  \bibinfo{editor}{\bibfnamefont{S.}~\bibnamefont{{Ryder}}},
  \bibinfo{editor}{\bibfnamefont{D.}~\bibnamefont{{Pisano}}},
  \bibinfo{editor}{\bibfnamefont{M.}~\bibnamefont{{Walker}}}, \bibnamefont{and}
  \bibinfo{editor}{\bibfnamefont{K.}~\bibnamefont{{Freeman}}}
  (\bibinfo{year}{2004}), vol. \bibinfo{volume}{220} of
  \emph{\bibinfo{series}{IAU Symposium}}, pp. \bibinfo{pages}{39--+}.

\end{thebibliography}

\end{document}